\documentclass{aa}
\usepackage{longtable}
\usepackage{graphicx}
\usepackage{txfonts}
\begin{document}
%
   \title{First detection of acceleration and deceleration in protostellar jets?}
   \subtitle{Time variability in the Cha II outflows
   \thanks{Based on observations collected at the European Southern Observatory,
 (Paranal and La Silla, Chile) (55.C-0908, 62.I-0136(A), 63.I-0031(A), 69.C-0269(A), 74.C-0235(A), 77.C-0047(B), 078.C-0089(A))}
       }

   \author{A. Caratti o Garatti \inst{1,2},
   	   J. Eisl\"{o}ffel \inst{1},
	   D. Froebrich \inst{3},
	   B. Nisini \inst{4},
	   T. Giannini \inst{4},
	  \and
	   L. Calzoletti \inst{4}
          }

\offprints{A. Caratti o Garatti \email{alessio@cp.dias.ie}}

   \institute{Th\"uringer Landessternwarte Tautenburg,
              Sternwarte 5, D-07778 Tautenburg, Germany\\
             \email{caratti;jochen@tls-tautenburg.de}
             \and	
             Dublin Institute for Advanced Studies, 31 Fitzwilliam Place, Dublin 2, Ireland\\ 
              \email{alessio@cp.dias.ie}
		\and
	      Centre for Astrophysics and Planetary Science, University of Kent, Canterbury, CT2 7NH, United Kingdom\\
	      \email{df@star.kent.ac.uk}
	      \and
              INAF - Osservatorio Astronomico di Roma, via Frascati 33, I-00040 Monte Porzio, Italy \\
             \email{nisini;giannini;calzoletti@oa-roma.inaf.it}
             }

   \date{Received; accepted }


  \abstract
   {Kinematical and time variability studies of protostellar jets are fundamental for understand the dynamics and the physics of these objects.
    Such studies remain very sporadic, since they require long baselines before they can be accomplished.}
   {We present for the first time a multi-epoch (20 years baseline) kinematical investigation of HH\,52, 53, and 54 at optical and
    near-IR wavelengths, along with medium (optical) and high resolution (NIR) spectroscopic analyses, probing the kinematical
    and physical time variability conditions of the gas along the flows.}
   {By means of multi-epoch and multi-wavelength narrow-band images, we derived proper motions (P.M.s), tangential velocities,
   velocity and flux variability of the knots. Radial velocities and physical parameters of the gas were derived from spectroscopy.
   Finally, spatial velocities and inclination of the flows were obtained by combining both imaging and spectroscopy.}
   {The P.M. analysis reveals three distinct, partially overlapping outflows. Spatial velocities of the knots vary from 50\,km\,s$^{-1}$ to 120\,km\,s$^{-1}$. The inclinations of the three flows are 58$\pm$3$^\circ$, 84$\pm$2$^\circ$, and 67$\pm$3$^\circ$
   (HH\,52, HH\,53, and HH\,54 flows, respectively).
   In 20 years, about 60\% of the observed knots show some degree of flux variability. 
   Our set of observations apparently indicates acceleration and deceleration in a variety of knots
   along the jets. For about 20\% of the knots, mostly coincident with working surfaces or interacting knots along the flows, a relevant variability in both flux and velocity is observed. We argue that both variabilities are related and that all or part of the kinetic energy lost by the interacting knots is successively radiated.
   The physical parameters derived from the diagnostics are quite homogeneous along and among the three outflows. The analysis indicates the presence of very light ($N_{\rm H}\sim$10$^3$\,cm$^{-3}$), ionised ($X_{\rm e}\sim$0.2-0.6), and hot ($T_{\rm e}\sim$14\,000-26\,000\,K) flows,
   impacting a denser medium. Several knots are deflected, especially in the HH\,52 flow. At least for a couple of them (HH\,54 G and G0), the deflection originates from the collision of the two. For the more massive parts of the flow, the deflection is likely the result of the flow collision with a dense cloud or with clumps.
   Finally, we discuss the possible driving sources of the flows.}
   {}


\authorrunning{A.Caratti o Garatti et al.}
\titlerunning{Time-variability in the Cha II outflows}
\maketitle
%

\section{Introduction}
\label{intro:sec}

Protostellar jets and outflows represent fundamental processes in star formation.
These outflows are a direct consequence of the accretion mechanism in young stellar objects (YSOs) during their earliest phase
(see e.\,g. Cabrit et al.~\cite{cabrit}; Shu et al.~\cite{shu}; Reipurth \& Bally~\cite{reipurth01}), when
a fraction of the infalling material is ejected in the form of collimated jets. The interaction between the ejecta and
the circumstellar medium occurs via radiative shocks.

In the very young and still deeply embedded sources, these flows are mostly observed
in molecular (H$_2$) and atomic ([\ion{Fe}{ii}]) lines in the near-infrared (NIR) (see e.\,g. Schwartz et al.~\cite{schwartz87};
Eisl\"{o}ffel et al.~\cite{eis94a}; Davis et al.~\cite{davis94}). At later phases
-- Class I and Class II -- the sources also become visible in the optical,
and their jets can then be traced by optical lines as [\ion{S}{ii}]
(6716, 6730\,$\AA$) and H$\alpha$, as well (see e.\,g. Graham \& Elias~\cite{G83}; Mundt~\cite{mundt}).
Observations of the emission lines from the shock-heated gas allow us to study physical parameters in detail, such
as the density, the temperature, the ionisation fraction, and the abundances (see e.\,g. Hartigan et al.~\cite{hart94}; Bacciotti \& Eisl\"offel~\cite{BE99}; Lavalley-Fouquet et al.~\cite{lavalley}). Moreover, since shocks in outflows trace ejecta that are progressively older with distance from the source, it is also possible to gain important information about their driving source properties,
reconstructing the YSO mass ejection history.

A description of the gas dynamics, however, only becomes available through
measurements of the flow velocity. Spectroscopic radial velocity measurements are necessary,
but not sufficient for deriving the flow speed, since they deliver only the projected
velocities along the line of sight. In order to determine the true
3-dimensional kinematics, i.\,e. the absolute flow speed and the inclination of
the flow with respect to the plane of the sky, proper motion measurements
of the knots are also necessary. With this information in hand, a
number of important questions can be addressed straightforwardly.

i) The sources of the outflows can be unambiguously identified and knots can be correctly
    assigned to their driving sources.

ii) Using multi-wavelength data, molecular, and atomic,
    it is possible to study the different kinematics of the flow by means of tracers that
    indicate various shock conditions and velocities of the gas.

iii) The internal kinematics of the flows can be studied in detail.
     Possibly, acceleration, and deceleration can be derived from a multi-epoch analysis
     of the motions and their origin can be investigated in detail.

iv)  Flux variability can be studied and related to the motion variability in the flow.

v)  Combining the kinematical results and the gas physical parameters, it is possible to infer the mass-loss rate and the energy budget
    along the flow. In turn, we can compare these quantities with the physical properties of the exciting source
    to investigate how it evolves with time.

To address these points, we collected a multi-epoch and multi-wavelength (optical and NIR) data archive of several Herbig Haro (HH) objects and protostellar jets.
The database includes both imaging and spectroscopy (low and high resolution), mainly covering the optical atomic (H$\alpha$, [\ion{S}{ii}])
and NIR molecular emission (H$_2$). For the first time it is possible to study the kinematics of the outflows during an elapse of time up to 20 years at different wavelengths, observing the formation and the evolution of the single structures inside the jets.
This paper represents the first work of our project.

The structure is as follows. In Sect.~2 we shortly describe the investigated region.
In Sect.~3 our observations are presented. Section~4 reports an overview of our results, including
morphology of the flows, their kinematics, flux and velocity variability,
physical parameters of the gas, and mass flux rates. An investigation on the possible exciting sources of the
flows is presented, as well. In Sect.~5 a discussion is
proposed in terms of kinematics vs. flow variability, flow deflection mechanisms,
length, and dynamical age of the flows. Our conclusions are summarised in Sect.~6.
Finally, a detailed description of our results on the single knots is given in the Appendices.

\section{The investigated region}
\label{region:sec}

\begin{figure*}
 \centering
   \includegraphics [width=15.5 cm] {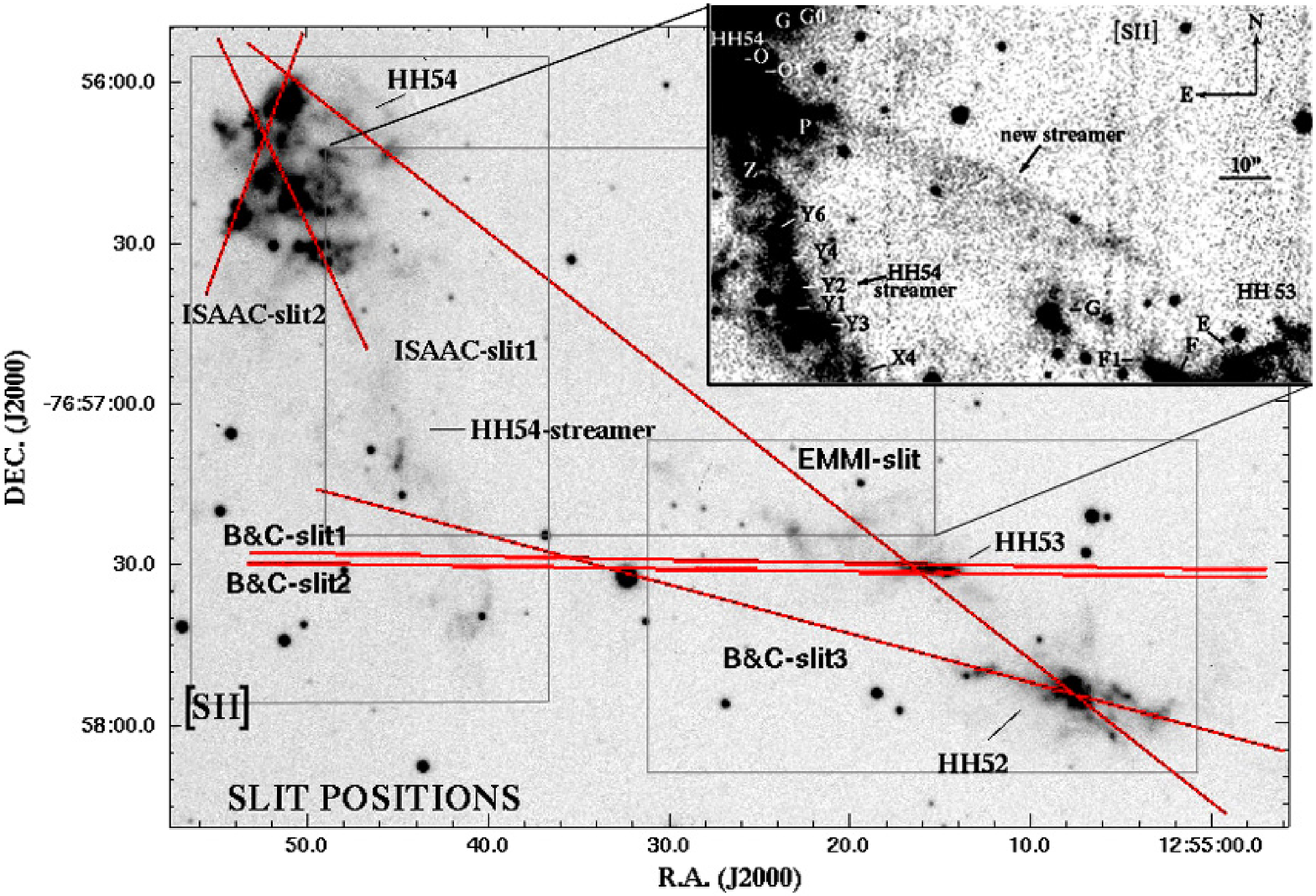}
   \caption{ [\ion{S}{ii}] EMMI 2006 image of the region indicates the locations of the HH objects, along with
   EMMI, B\&C, and ISAAC slit positions. In the upper-right inset [\ion{S}{ii}] deep image (from 1993) we show the newly detected HH\,52 streamer.
   A rectangular box (in the centre of the image) shows its exact location. Additionally, left and right boxes
   approximatively indicate the fields displayed in Figures~\ref{HH52-3-4Hfig:fig},~\ref{HH52-3-pms:fig}, and~\ref{HH54-pms:fig}.
\label{slit-positions:fig}}
\end{figure*}

HH\,52, 53, and 54 are located at the northern edge of the Chamaleon II (Cha II) cloud (Schwartz~\cite{schwartz}) at a distance of
178$\pm$18\,pc\footnote{Hereafter we refer to this value to compute spatial distances and tangential velocities of
the knots.} (Whittet et al.~\cite{whittet}). As shown in Fig.~\ref{slit-positions:fig}, the three HH objects are roughly placed along a straight
path, moving from SW to NE, with a P.A. of $\sim$55$\degr$.
The morphology of the HH objects is approximatively the same in the H$\alpha$ and [\ion{S}{ii}] filters.
HH\,52 has a bow shock shape pointing towards HH\,54. About 20$\arcsec$ NE, a chain of bright knots (HH\,53\,C, A, and B),
roughly aligned E-W, is observed. Moreover, farther NE ($\sim$1$\arcmin$), a cluster of several bright knots (HH\,54) is detected.
This group spans some tens of arcseconds and elongates towards SSW, delineating the so-called HH\,54 streamer (see Sandell et al.~\cite{sandell}). In this work, they described the morphology
of the HH objects\footnote{The nomenclature used in this paper mainly refers to Sandell et al.~(\cite{sandell}).} in some details,
also for the first time reporting a contour map of the molecular hydrogen emission in the NIR.
The HHs were spectroscopically observed in the optical by Schwartz \& Dopita~(\cite{SD80}) and Graham \& Hartigan~(\cite{GH88}),
who found only blue-shifted radial velocities ranging from -40 to -100\,km\,s$^{-1}$. Knee~(\cite{knee}) detected two blue-shifted CO
outflows, coinciding with HH\,52 and 53, and HH\,54 emissions, and only one red-shifted lobe, located approximately between the two. H$_2$ and [\ion{Fe}{ii}]
imaging together with low-resolution spectra of HH\,54 were published by Gredel~(\cite{gredel}). The morphology in the H$_2$ and [\ion{Fe}{ii}] filters appears similar to the optical images. More low-resolution spectra in the NIR, with a detailed study of the excitation of the gas in HH\,54, have been published by Caratti o Garatti~(\cite{caratti06a}) and Giannini et al.~(\cite{giannini06}). From these authors H$_2$ medium-resolution spectra of HH\,54 and radial velocities are also available. These data cannot explain, however, the kinematics of the object.
An optical P.M. analysis of the HHs with a seven year baseline was attempted by Schwartz et al.~(\cite{schwartz84})
without conclusive results, because of small P.M. values and relatively large errors.

Finally, the exciting sources of the HHs still remain unclear,
even after a proper motion analysis of the infrared knots (Caratti o Garatti et al.~\cite{caratti06b}), the publication of MIPS maps and
YSO catalogues for this region (Young et al.~\cite{young}; Porras et al.~\cite{porras}; Alcal\'{a} et al.~\cite{alcala}).

\section{Observations, data reduction \& analysis}
\label{observations:sec}

\begin{table*}
\caption[]{ Journal of observations - Imaging
    \label{obs_imaging:tab}}
\begin{center}
\begin{tabular}{ccccccc}
\hline \hline\\[-5pt]
Date of obs. &      Telescope/  &  Filter  & Resolution & seeing & Exp. Time  & Notes                      \\
(d,m,y)   &         Instrument  &  Band & ($\arcsec$/pixel) & ($\arcsec$) & (s) & Objects Observed \\
 \hline\\[-5pt]
14.05.2006  & ESO-NTT/EMMI  & H$\alpha$ & 0.166 & 0.8 & 3600 & HH52,HH53,HH54 \\
12.05.1995  & ESO-NTT/EMMI  & H$\alpha$ & 0.266 & 1.1 & 3600 & HH52,HH53,HH54 \\
31.03.1993  & ESO/MPG2.2-m/EFOSC2  & H$\alpha$ & 0.35 & 0.9 & 4500 & HH52,HH53,HH54 \\
08.05.1992  & ESO/MPG2.2-m/EFOSC2  & H$\alpha$ & 0.35 & 1.4 & 5400 & HH52,HH53,HH54 \\
15.01.1990  & ESO/MPG2.2-m/CCD  & H$\alpha$ & 0.35 & 1.0 & 6000 & HH54 \\
03.02.1989  & ESO/MPG2.2-m/CCD  & H$\alpha$ & 0.35 & 1.0 & 4200 & HH54 \\
04.03.1987  & ESO/MPG2.2-m/CCD  & H$\alpha$ & 0.35 & 1.2 & 3600 & HH54 \\
            &  & & &        &                    &\\
14.05.2006  & ESO-NTT/EMMI   & [\ion{S}{ii}] & 0.166 & 0.7 & 3600 & HH52,HH53,HH54 \\
31.03.1993  & ESO/MPG2.2-m/EFOSC2  & [\ion{S}{ii}] & 0.35 & 1.0 & 4200 & HH52,HH53,HH54 \\
09.02.1991  & ESO/MPG2.2-m/EFOSC2  & [\ion{S}{ii}] & 0.35 & 1.1 & 1800 & HH52,HH53,HH54 \\
14.01.1990  & ESO/MPG2.2-m/CCD  & [\ion{S}{ii}] & 0.35 & 0.8 & 4800 & HH52,HH53 \\
01.02.1989  & ESO/MPG2.2-m/CCD  & [\ion{S}{ii}] & 0.35 & 1.2 & 3600 & HH52,HH53 \\
28.02.1987  & ESO/MPG2.2-m/CCD  & [\ion{S}{ii}] & 0.35 & 1.5 & 3600 & HH52,HH53 \\
            &  & & &        &                    &\\
01.01.2005  & ESO-VLT/ISAAC      & H$_2$ & 0.148 & 0.5 & 40 & HH54 \\
02.07.2002  & ESO-VLT/ISAAC      & H$_2$ & 0.148 & 0.5 & 60 & HH54 \\
06.06.1999  & ESO-NTT/SofI       & H$_2$ & 0.29 & 1.0 & 300 & HH52,HH53,HH54 \\
31.03.1999  & ESO-NTT/SofI       & H$_2$ & 0.29 & 0.9 & 600 & HH52,HH53,HH54 \\
15.03.1995  & ESO/MPG2.2-m/IRAC2b       & H$_2$ & 0.51 & 1.1 & 500 & HH52,HH53,HH54\\
10.04.1993  & ESO/MPG2.2-m/IRAC2b       & H$_2$ & 0.51 & 1.7 & 200 & HH54 \\
03.04.1993  & ESO/MPG2.2-m/IRAC2b       & H$_2$ & 0.51 & 1.7 & 200 & HH52,HH53 \\
            &  & & &        &                    &\\
01.09.2007  & ESO-NTT/SofI       & H$_2$ & 0.29 & 1.0 & 1080 & IRAS\,12500-7658 \\
01.09.2007  & ESO-NTT/SofI       & K$_s$ & 0.29 & 1.0 & 270 & IRAS\,12500-7658 \\
\hline \hline
\end{tabular}
\end{center}
\end{table*}

\begin{table*}
\caption[]{ Journal of observations - Spectroscopy
    \label{obs_spec:tab}}
\begin{center}
\begin{tabular}{cccccccc}
\hline \hline\\[-5pt]
Telescope/  & Date of obs.  & Wavelength &  t$_{int}$ & P.A. & $\mathcal R$ & Encompassed Knots \\
Instrument  & (d,m,y) &  ($\AA$) & (s) &  ($\degr$)  &  & \\
 \hline\\[-5pt]
ESO-NTT/   & 14.05.2006 & 6200--7000    & 2700 & 52    & 3000 & HH52:\,A3-A4,A1,D4; HH53:\,C-C1,E1;\\
EMMI       &            &               &      &       &      & HH54:\,G-G0,G1-G3,C2,C3 \\
ESO-VLT/ISAAC   & 01.01.2005 & 16400         & 3600 & 29.5  & 10000 & HH54:\,Z,A3,A1,B,J-J1,H3,H2 \\
ESO-VLT/ISAAC   & 01.01.2005 & 16400         & 3600 & 160.5 & 10000 & HH54:\,C3,C1,H3,J-J1,E,K \\
ESO-2.2-m/      & 20.02.1987 & 6200--6800    & 3600 & 89.5  & 2000  & HH53:\,B,A,C,F2; HH54:\,X4\,C,X1\,D  \\
Boller \& Chivens (B\&C)      & 21.02.1987 & 6200--6800    & 3600 & 89.5  & 2000  & HH53:\,B,A,C; HH54:\,X3,X1\,A \\
                & 25.02.1987 & 6200--6800    & 3600 & 75   & 2000& HH52:\,B,A,D3-D2; HH54:\,X4\,B-A \\
\hline \hline
\end{tabular}
\end{center}
\end{table*}

Our multi-wavelength database is mainly composed of optical and
NIR observations, collected with the ESO facilities during the
past twenty years. The relevant information for both imaging and
spectroscopy is reported in Tables~\ref{obs_imaging:tab} and
\ref{obs_spec:tab}, respectively.
These data were used to derive the kinematics and the physical properties of the studied
outflows. The advantage of using a multi-epoch image dataset is quite straightforward:
on one hand, we can reduce the errorbars of the P.M.s and check for velocity variations in the flows;
on the other, we can study their flux variability and their evolution with time.

Additionally, this paper makes use of {\it Spitzer Space Telescope} public data from the ``Cores to Disks'' (C2d)
Legacy project (Evans et al.~\cite{evans}), already published and analysed (Young et al.~\cite{young};
Porras et al.~\cite{porras}; Alcal\'{a} et al.~\cite{alcala}) in order to infer the exciting sources
and map the colder gas component of the outflows.

\subsection{Imaging}

\subsubsection{Optical}
Our optical narrow-band images (see Table~\ref{obs_imaging:tab}) are
centred on H$\alpha$ and [\ion{S}{ii}] emissions.
A first set of data was obtained at the ESO/MPG\,2.2-m telescope (between 1987 and 1993) with three different detectors:
an RCA 30\,$\mu$m CCD chip (1987) (2 images) and an RCA 15\,$\mu$m CCD chip (1989-1990) (4 images),
both mounted on the Cassegrain focus, and EFOSC2 (Buzzoni et al.~\cite{buzzoni}) (1991-1993)
(4 images). A second set of images was retrieved at the New Technology Telescope
(1995 and 2006) with EMMI (Dekker et al.~\cite{dekker})(3 images).
Because of the relatively small FoV of the CCDs used
between 1987 and 1990, HH\,52 and HH\,53, and HH\,54 were observed
separately with the [\ion{S}{ii}] ($\lambda\lambda$ 6716, 6731) and
with the H$\alpha$ filters, respectively. Starting from 1991, all
the three objects were observed in both filters within a single
frame. The exposure times range from 1800 to 6000\,s, resulting in
deep images with a high signal-to-noise ratio (S/N). The raw data were
reduced by using standard procedures in \emph{IRAF}\footnote{IRAF (Image
Reduction and Analysis Facility) is distributed by the National
Optical Astronomy Observatories, which are operated by AURA, Inc.,
cooperative agreement with the National Science Foundation.},
subtracting bias, flat-fielding and removing bad pixels. Cosmic
rays were removed by using the \emph{L.A.\,Cosmic} algorithm (van Dokkum
\cite{dokkum}), which was extremely efficient in cleaning
cosmic-ray crowded images. The measured seeing on our images
ranges between 0\farcs7 and 1\farcs5 (Table~\ref{obs_imaging:tab}),
with the best values obtained in the most recent ones.

Only images from 2006 have their own photometric standard star,
observed with both H$\alpha$ and [\ion{S}{ii}] filters.
The selected standard is LTT\,3218 (Hamuy et al.~\cite{hamuy}), for which a
flux calibrated spectrum also exists. At the required wavelengths, the flux of the
star was obtained by convolving its flux density with the profiles of the ESO filters.

To flux-calibrate the earlier epoch images, 30 bright
field stars were selected and photometry in both filters
performed on them. For each epoch in each filter, the star fluxes
were cross-checked, and those showing variations less than 3\% were
chosen as secondary standard stars to calibrate the fields.

\subsubsection{NIR}
\label{observations_NIR:sec}

Our narrow-band H$_2$ (2.12\,$\mu$m) images were collected during different runs between 1993 and 2005,
at the ESO/MPG\,2.2-m telescope with IRAC2b (Lidman et al.~\cite{lidman}) (3 images), at the
ESO-NTT with SofI (Moorwood et al.~\cite{moor1}) (2 images),
and at ESO-VLT with ISAAC (Moorwood et al.~\cite{moor2}) (2 images) (see Table~\ref{obs_imaging:tab}).
Some data on HH\,54 have been presented in previous papers (Giannini et al.~\cite{giannini06}; Caratti o Garatti et al.~\cite{caratti06b}),
where P.M.s were derived comparing the SofI image of June 1999 with the ISAAC one of January 2005.
Here, we enlarge the baseline of our observations including IRAC2b data from 1993 and 1995, and an ISAAC image from 2002 (covering only HH\,54) that, on one hand, allow us to also derive H$_2$ P.M.s for HH\,52 and HH\,53, and, on the other, add more data points to HH\,54, reducing the uncertainties on P.M.s and P.A.s of the knots.
The first set of IRAC2b data (1993) covers HH\,52 and HH\,53, and HH\,54, separately, exhibiting a poor seeing of 1\farcs7.
The second (1995) consists of a mosaic of the region ($\sim$7$\arcmin\times$7$\arcmin$) and has a seeing of 1\farcs1.
In addition, more H$_2$ SofI data from the ESO science archive facility\footnote{http://archive.eso.org/} (taken less then 3 months apart from those of June 1999, see Table~\ref{obs_imaging:tab}) were retrieved and coadded to create a deeper SofI H$_2$ map of the region covering $\sim$11$\arcmin\times$11$\arcmin$.

Finally, additional H$_2$ and K$_s$ images around $IRAS\,12500-7658$, one of the possible driving sources of HH\,54 (see also Sects.\,\ref{observations_spitzer:sec} and \ref{excitingsources:sec}), were taken in Jan. 2007 to detect a possible H$_2$ jet from that source.

As for the optical, all the raw data were reduced by using \emph{IRAF} packages applying standard procedures for sky subtraction, dome
flat-fielding, bad pixel and cosmic ray removal and image mosaicing.
Flux calibration by means of a photometric standard star was only possible for the June 1999 SofI image.
As for the optical, the remaining H$_2$ images were calibrated by selecting a few field stars that showed flux variations smaller than 10\%.

\subsubsection{Spitzer data}
\label{observations_spitzer:sec}

\begin{figure*}
 \centering
   \includegraphics [width= 13.5 cm] {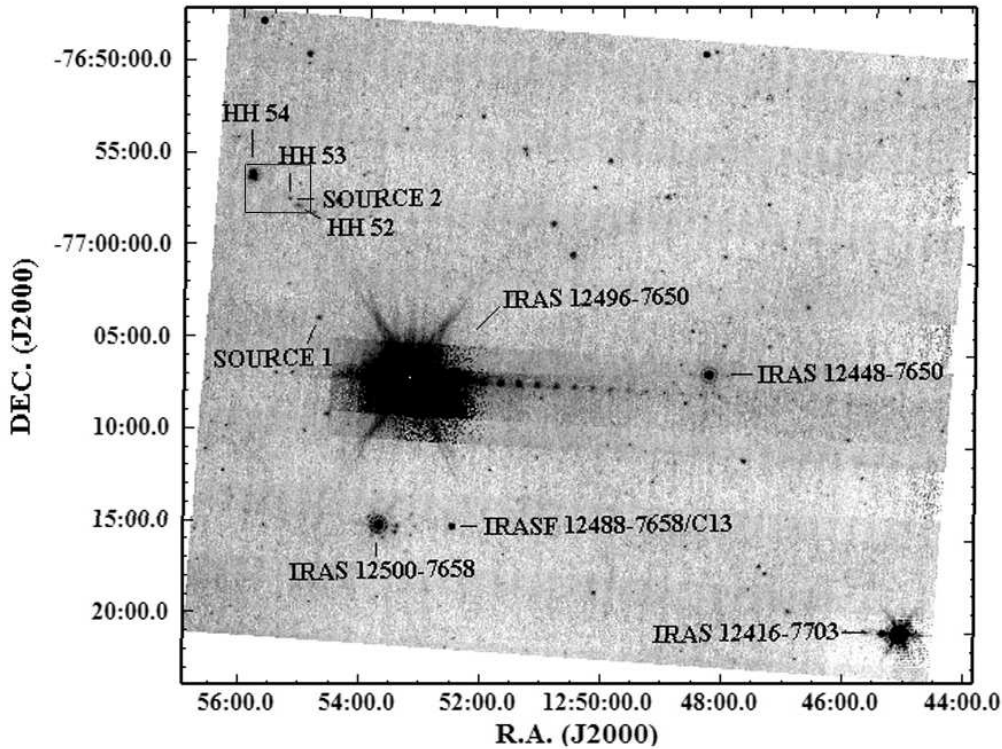}
   \caption{\emph{Spitzer-MIPS} 24\,$\mu$m map of the investigated Cha II region. The candidate exciting sources and the HH objects are reported.
   A rectangular box indicates the region displayed in Fig.~\ref{slit-positions:fig}.
\label{spitzer:fig}}
\end{figure*}

Reduced \emph{Spitzer} data have been obtained by the public catalogue c2d (third delivery).
They consist of large mosaics of the Cha II cloud
observed with the Infrared Array Camera (IRAC) in four channels (at 3.6, 4.5, 5.8, and 8.0 $\mu$m) ($\sim$1.04\,deg$^2$) and with \emph{MIPS}
(at 24, 70 $\mu$m) ($\sim$1.5\,deg$^2$). In Fig.~\ref{spitzer:fig} we report part of the \emph{Spitzer-MIPS} map (at 24\,$\mu$m), indicating
the location of the studied HH objects and the possible driving sources.

\subsubsection{Knot identification and flux analysis}
\label{flux_analysis}

To identify the knots along the flows, we used the best image in each filter (i.\,e. with highest S/N and best resolution).
This resulted to be the last epoch image (i.\,e. EMMI 2006 image in H$\alpha$ and [\ion{S}{ii}],
SofI 1999 image for HH\,52 and HH\,53 in H$_2$, and ISAAC 2005 image for HH\,54 in H$_2$).
These were also used as reference images for the P.M. analysis (see Sect.~\ref{pm_analysis}).
Each structure was then identified and labelled (in each filter) within a 5\,$\sigma$ contour around the local maximum.
These contours were also used to measure the total flux of the knots in each epoch and filter, and
to study the flux variability as well.

\subsubsection{P.M. analysis}
\label{pm_analysis}

Particular attention was paid to align the images. In fact, an accurate
determination of the proper motions requires that the
multi-epoch images have to be scaled and then registered to a subpixel
accuracy in each filter. Our observed region contains several field stars:
therefore, the image registration was easy to implement, and the
distortions introduced by the different optical systems were
corrected by means of \emph{geomap} (with a polynomial fit of 3rd
order) and \emph{geotran} routines in \emph{IRAF}. Because of the large
baseline of the observations, few stars showed a significant proper
motion and were excluded from the sample. Each image in each filter was aligned
to a common (scaled) reference frame (see Sect.~\ref{flux_analysis}).
The resulting error is given by the residuals of the fit.
Statistical errors in the optical were typically about $\pm$0.1-0.2
pixels (i.\,e. $\sim$0\farcs03-0\farcs07), depending on the seeing and on the resolution of the image.
In the NIR, due to the reduced FoV, ISAAC, and IRAC2b (1993) images have an exiguous number of stars
($\le$20) that diminishes the accuracy in the alignment, not allowing the entire area to be sampled homogeneously.
As a consequence, here the errors were up to $\pm$0.3-0.4 pixels.

Knot shifts were determined between image pairs (i.\,e. last epoch and previous epoch images)
using a cross-correlation method. A
rectangle was defined round each knot, enclosing its 5\,$\sigma$ contour. The 1st-epoch image was then
shifted with subpixel accuracy (0.1 pix) with respect to the last epoch image
and then multiplied by it. For each shift (x,y), the total integrated flux (f)
in the rectangle around each knot was measured in the product image. As the final shift
for each knot, we used the position of the maximum of f(x,y), determined
via a Gaussian fit. For each structure, an initial guess of the shift vector was estimated
visually by blinking the image pairs, defining a maximum shift for each rectangle.
This served as input into our routine for the cross-correlation.
For several features, the analysis routine converged rapidly to a solution, independent of the initial guess.
A more accurate estimate (within a few pixels) of the size and direction of the proper motion was needed to avoid the code
to correlate spurious pairs in case of diffuse knots located close to brighter knots or stars.
To prove the code accuracy, we tested it on both nebulous and stellar objects
by shifting the images and recovering the offsets with the code.

To quantify the systematic errors for the shift measurements, we varied the size and shape of the rectangle that
defines the object.
The resulting range of values gives a systematic error, which depends on the S/N, shape, and time
variability of the knot.
This error is comparable to or even larger than the alignment error.
Thus, each single shift error was derived combining both the alignment error of the two epoch images and
the uncertainty in the centring routine. Usually, for each pair, the final error was around 10\% of the P.M. value.
However, for slow knots with a low S/N, the error was up to 100\%.

\begin{figure}
 \centering
   \includegraphics [width= 10 cm] {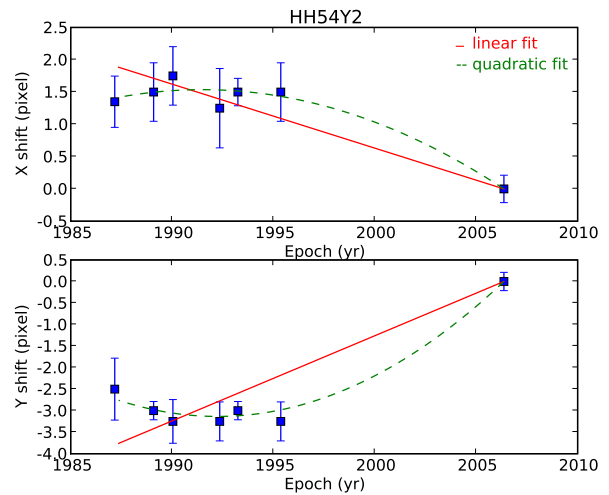}
   \caption{Example of linear and quadratic fits performed on the dataset to derive
            proper and accelerated motions. Here the fits on HH\,54\,Y2 H$\alpha$ data are reported.
\label{fits:fig}}
\end{figure}

The final P.M. value and error for each knot in each filter were obtained from a
weighted least square fit of the shifts, fitting the motion in X (R.\,A.) and
Y (Dec.) simultaneously (see Fig.~\ref{fits:fig}), using the equations of both linear and accelerated motions: $[x(t),y(t)]=p_{x,y}+v_{x,y}t+(\frac{1}{2}a_{x,y}t^2$).
Both linear and quadratic fits (for knots with a number of epochs $\ge 3$)
were performed to detect any possible acceleration or deceleration of the knot, as well.
The resulting errors on proper and accelerated motions were derived from the fits.
An example of our fits is shown in Fig.~\ref{fits:fig}, where an accelerated motion is detected.
The upper and lower panels report the X and Y shifts with errors for the H$\alpha$ filter
of HH\,54\,Y2, respectively. Straight and dashed lines indicate linear and quadratic fits.
In this particular case both X and Y shifts are better modelled by a quadratic fit.

\subsection{Spectroscopy}

\subsubsection{Optical}

\begin{figure*}
   \fbox{\includegraphics [width=6.95 cm] {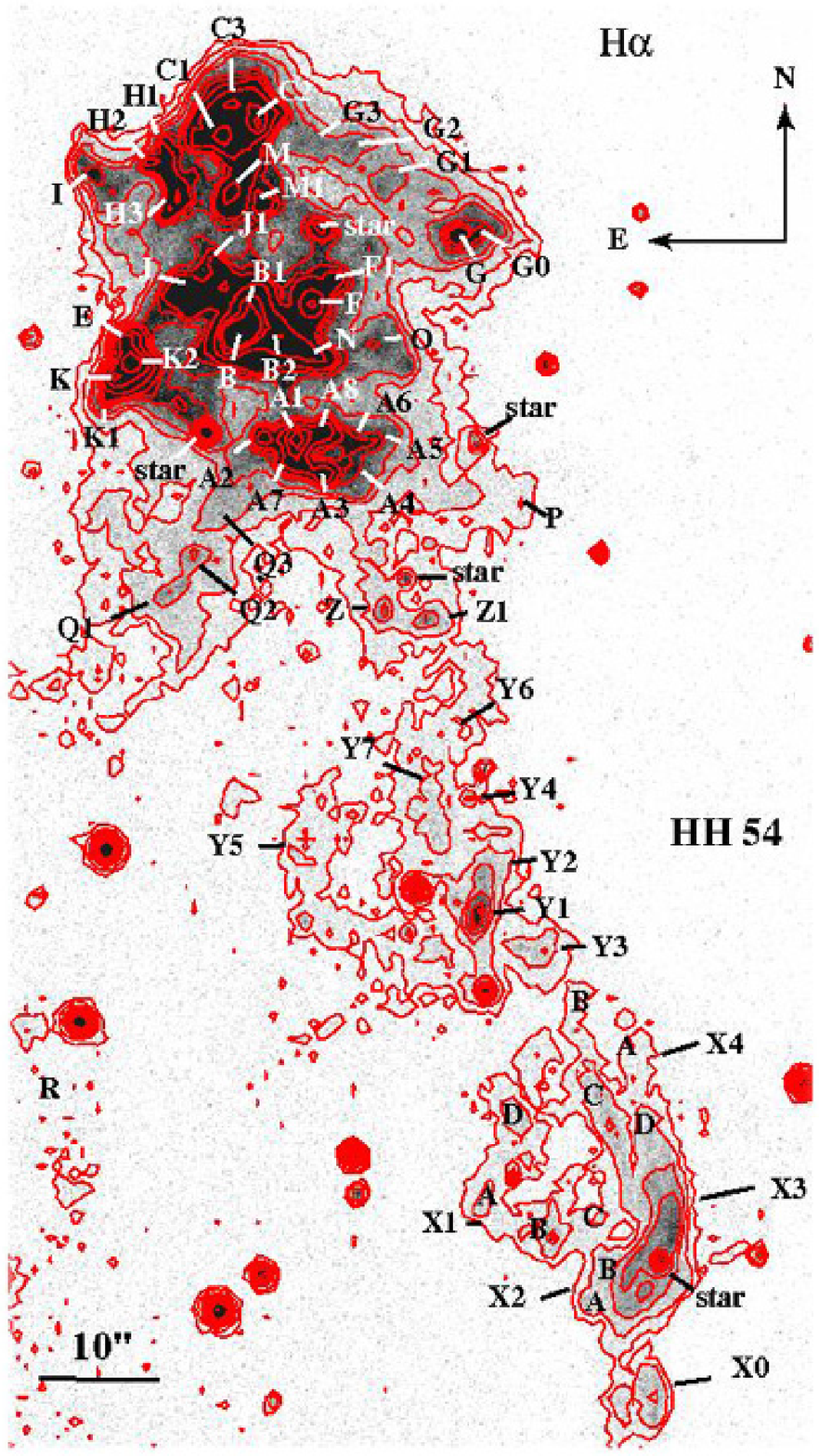}}
   \fbox{\includegraphics [width=11.0 cm] {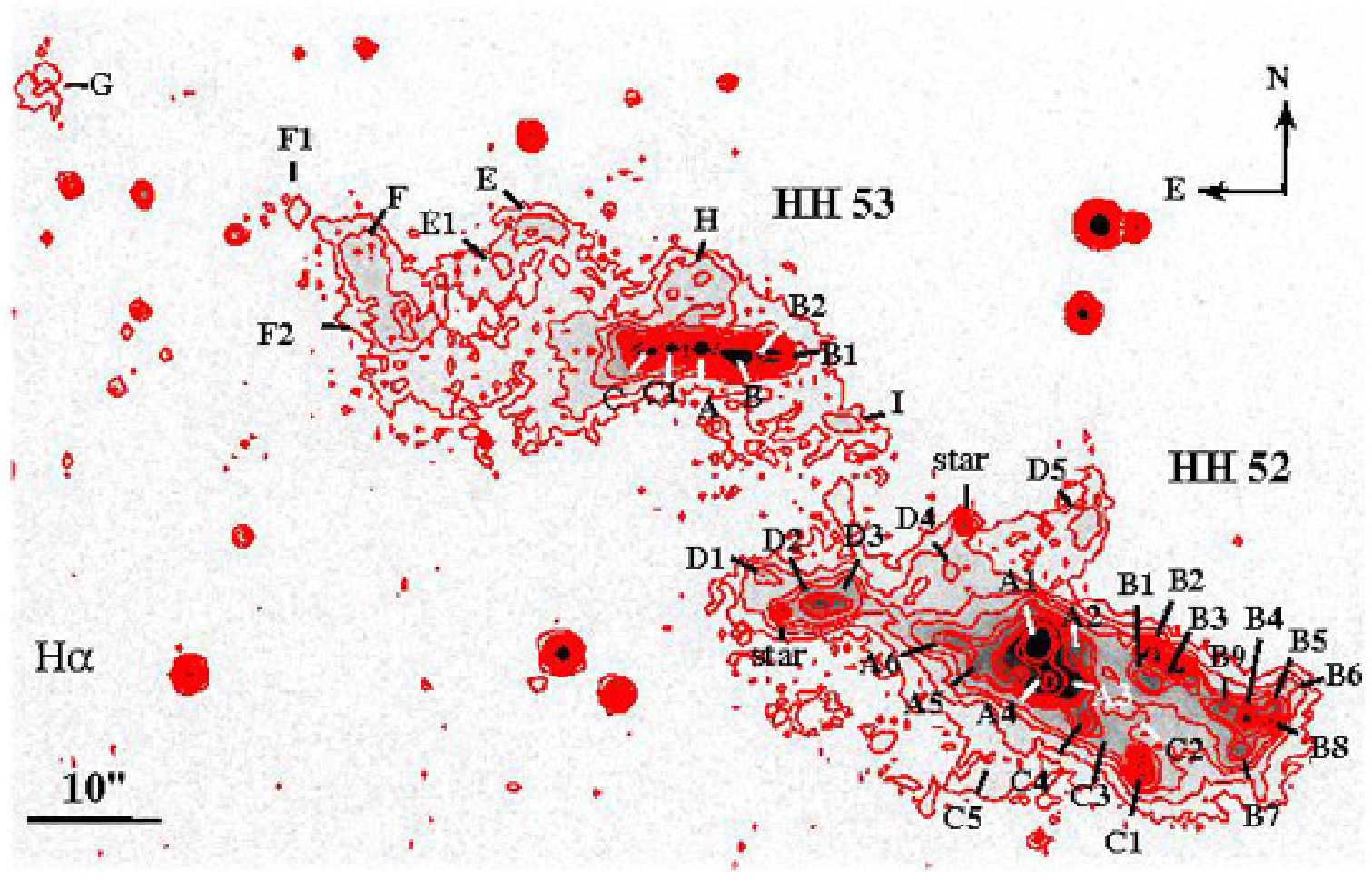}}\\
   \caption{ H$\alpha$ images (EMMI 2006) of HH\,54 ({\bf left panel}), HH\,52 and HH\,53 ({\bf right panel}). The labels indicate the position of the knots, including the newly detected ones. The contour levels are 3, 10, 20, 30, 40, 50, 60, 70, 80, 90, 100$\times$ the standard deviation to the mean background ($\sigma\sim$~4~$\times$~10$^{-17}$~erg~s$^{-1}$~cm$^{-2}$~arcsec$^{-2}$).
\label{HH52-3-4Hfig:fig}}
\end{figure*}

Four optical medium-resolution spectra ($\mathcal R\sim$2000-3000) were collected with a Boller \& Chivens (B\&C) spectrograph at the ESO/MPG\,2.2-m telescope (1987) and with EMMI at the NTT (2006). They range from 6200 to 7000\,$\AA$  and cover the spectral region containing bright lines from [\ion{O}{i}]
(6300\,$\AA$ and 6363\,$\AA$), [\ion{N}{ii}] (6548\,$\AA$ and 6583\,$\AA$), H$\alpha$ (6563\,$\AA$), and [\ion{S}{ii}] (6716\,$\AA$ and 6731\,$\AA$), widely used for the diagnostics of the shocked gas (see e.\,g. Bacciotti \& Eisl\"{o}ffel~\cite{BE99}, hereafter BE99).
The observed targets and the slit P.A. are listed in
Table~\ref{obs_spec:tab}, together with the date of the observations and the total integration time. B\&C slit~1 and 2 are positioned on the three HH\,53 bright knots (A, B and C) (see also Fig.~\ref{slit-positions:fig}),
and intersect the HH\,54 streamer. They have the same P.A. (89\fdg5) and almost the same positions, with slit~1 shifted slightly northward
and also encompassing knots F2 in HH\,53 and X4\,C and X1\,D in HH\,54, while slit~2 covers knots X3 and X1\,A of HH54 streamer
(see Fig.~\ref{HH52-3-4Hfig:fig} and Appendix~\ref{appendixA:sec} for a detailed description of the flow morphology).
B\&C slit~3 (P.A. = 75$\degr$) is placed on HH\,52 A, B and D2-D3, and also includes knots X4\,A and B in HH\,54 streamer.
Finally, EMMI slit (P.A. = 52$\degr$) covers HH\,52\,B, A, D3-D2, HH\,53\,C and C1, HH\,54\,G, and C.
The total integration time for B\&C and EMMI spectra are 3600\,s and 2700\,s, respectively.
We used the same procedure as for the optical imaging to reduce the data.
A first rough wavelength calibration was retrieved from helium and argon lamps. A second, more accurate
calibration was done using several atmospheric OH lines (Osterbrock et al.~\cite{osterbrock}) present in each frame with a final
accuracy of 4\,km\,s$^{-1}$ in the EMMI spectrum.
LTT\,7379 was the standard star observed to flux calibrate the EMMI spectrum.
The B\&C data were calibrated using our narrow-band images.

\subsubsection{NIR}

Two medium-resolution spectra around the 1.64\,$\mu$m [\ion{Fe}{ii}] line ($\lambda_{vac}$ = 1.64399\,$\mu$m;
Johansson~\cite{johansson}) were obtained with the ISAAC IR camera, with a 0\farcs3 slit,
corresponding to a nominal resolving power of $\sim$10\,000.
Targets and P.A.s of the slits are reported in Table~\ref{obs_spec:tab}, while the slit positions are shown in
Fig.~\ref{slit-positions:fig} (ISAAC slit~1 and 2). The two slits encompass only HH\,54.
The first slit (P.A. = 29\fdg5) encompasses knots A3, A1, B3, B1, partially J and J1, H3, and H2.
The second slit (P.A. = 160\fdg5) is centred on knots K1, K, and E, but partially encompasses knots J, J1, H3,
C1, and C3.

To perform our spectroscopic measurements, we adopted the usual ABB'A'
configuration, for a total integration time of 3600\,s. Each observation was flat-fielded, sky-subtracted, and corrected
for the curvature derived by longslit spectroscopy, while
atmospheric features were removed by dividing the spectra by
a telluric standard star (O8 spectral type), normalised to the
blackbody function at the stellar temperature and corrected for
hydrogen recombination absorption features intrinsic to the star.
The wavelength calibration was performed using atmospheric OH lines (Rousselot et al.~\cite{rousselot}) with an
accuracy of 3\,km\,s$^{-1}$. The instrumental profile in the dispersion direction, measured from
Gaussian fits to the sky OH emission lines, was 38.0\,km\,s$^{-1}$.

\section{Results}
\label{results:sec}

\subsection{Morphology}
\label{morphology:sec}

An overview of the HH\,52, 53, and 54 regions is given in Fig.~\ref{slit-positions:fig}.
Additionally, in Figures~\ref{HH52-3-4Hfig:fig}, \ref{HH52-3-fig:fig}, and \ref{HH54-fig:fig}\footnote{Figs.~\ref{HH52-3-fig:fig} and \ref{HH54-fig:fig} are presented in Appendix~\ref{appendixA:sec} as online material only, available in electronic form at http://www.edpsciences.org.} we present H$\alpha$, [\ion{S}{ii}], and H$_2$ images of HH\,52, HH\,53, and HH\,54 with labels indicating the position of the knots, including the newly detected ones. Their coordinates are reported in Columns~2 and 3 of Tab.~\ref{PM52:tab}-\ref{PM54_h2:tab}\footnote{Tables~\ref{PM52:tab}-\ref{PM54_h2:tab} are presented in Appendix~\ref{appendixB:sec} as online material only.}. Note that some emissions are not visible in all the three filters. The atomic and molecular emissions only partially overlap, sometimes showing a different spatial distribution of the gas in each filter that indicates various excitational conditions along the flows.

A first glance to Fig.~\ref{slit-positions:fig} reveals at least two flows converging towards HH\,54 brightest region, which
shows a chaotic structure (see also Fig.~\ref{HH52-3-4Hfig:fig}). A first, well-known outflow, delineated by the HH\,54 streamer, is oriented
NNE with a position angle of $\sim$22$\degr$ (see also Fig.~\ref{HH52-3-4Hfig:fig}). A second outflow follows an NE direction with
a position angle of 50$\degr$-55$\degr$, grouping HH\,52, 53, and 54. In our images we detect a new streamer (\textit{hereafter} HH\,52
streamer, see upper-right panel in Fig.~\ref{slit-positions:fig}), which connects the HH\,52 bow shock to the main body of HH\,54 and it is
superimposed on HH\,53. The structure is barely visible in H$_2$, but well delineated in our deepest [\ion{S}{ii}] and H$\alpha$ images. It
is detected in the \emph{Spitzer} images as well. It appears as continuous slightly curved emission between HH\,54 and 53 knots, while it
is fragmented in single knots between HH\,53 and 52. Both the narrow-band images and the EMMI spectrum seem to indicate that the HH\,52
streamer only presents line emission.

We do not detect any other additional emission in the large H$_2$ SofI map of the region down to the 3$\sigma$ limit of
$\sim$3$\times$10$^{-16}$~erg~s$^{-1}$~cm$^{-2}$~arcsec$^{-2}$.
Neither do we detect any jet or molecular emission in the 2.12\,$\mu$m image around $IRAS\,12500-7658$ (a possible driving source of HH\,54 - see also Fig.~\ref{spitzer:fig} and Sect.\,\ref{excitingsources:sec}) down to a 3$\sigma$ limit of
$\sim$1.5$\times$10$^{-15}$~erg~s$^{-1}$~cm$^{-2}$~arcsec$^{-2}$.
Moreover, we do not observe any further emission in the \emph{Spitzer} large field images, with the exception of a faint nebulosity in IRAC band 2 (4.5\,$\mu$m)
located south $\sim$2$\arcmin$.8 SSE of HH\,52 (at about $\alpha=12^h54^m50^s$ and $\delta$ =-77$\degr$00$\arcmin$37\farcs2) and
extending approximatively $\sim$1$\arcmin$ towards $IRAS\,12496-7650$, located $\sim$8$\arcmin$ farther SSE. It is not clear,
however, whether this emission is correlated with the outflows.

Finally, in the Appendix~\ref{appendixA:sec}, for each object, a detailed description of the morphology is reported.

\subsection{Proper motions}
\label{PMs:sec}

\begin{figure*}
 \centering
 \fbox{\includegraphics [width=7.0 cm] {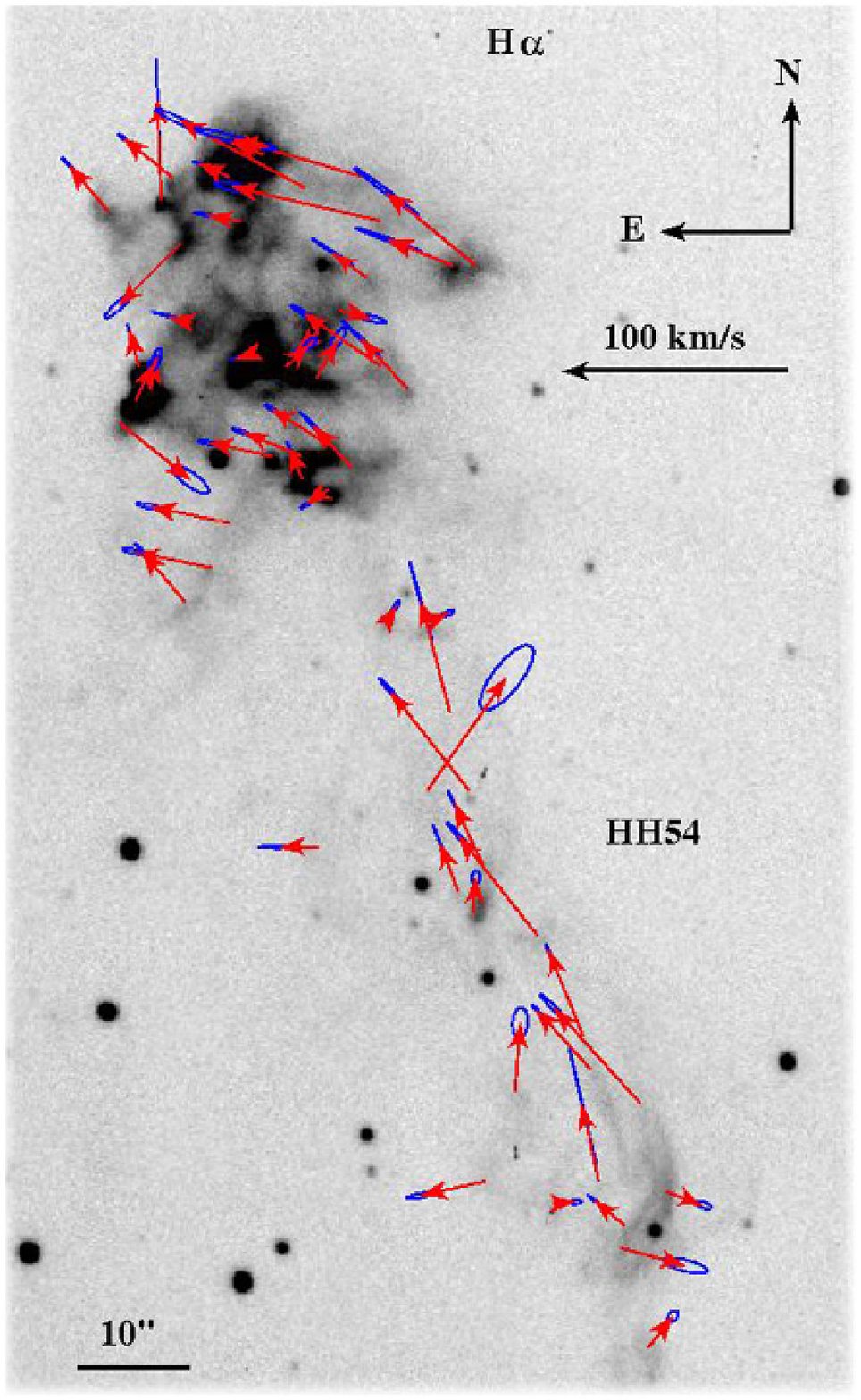}}
 \fbox{\includegraphics [width=10.8 cm] {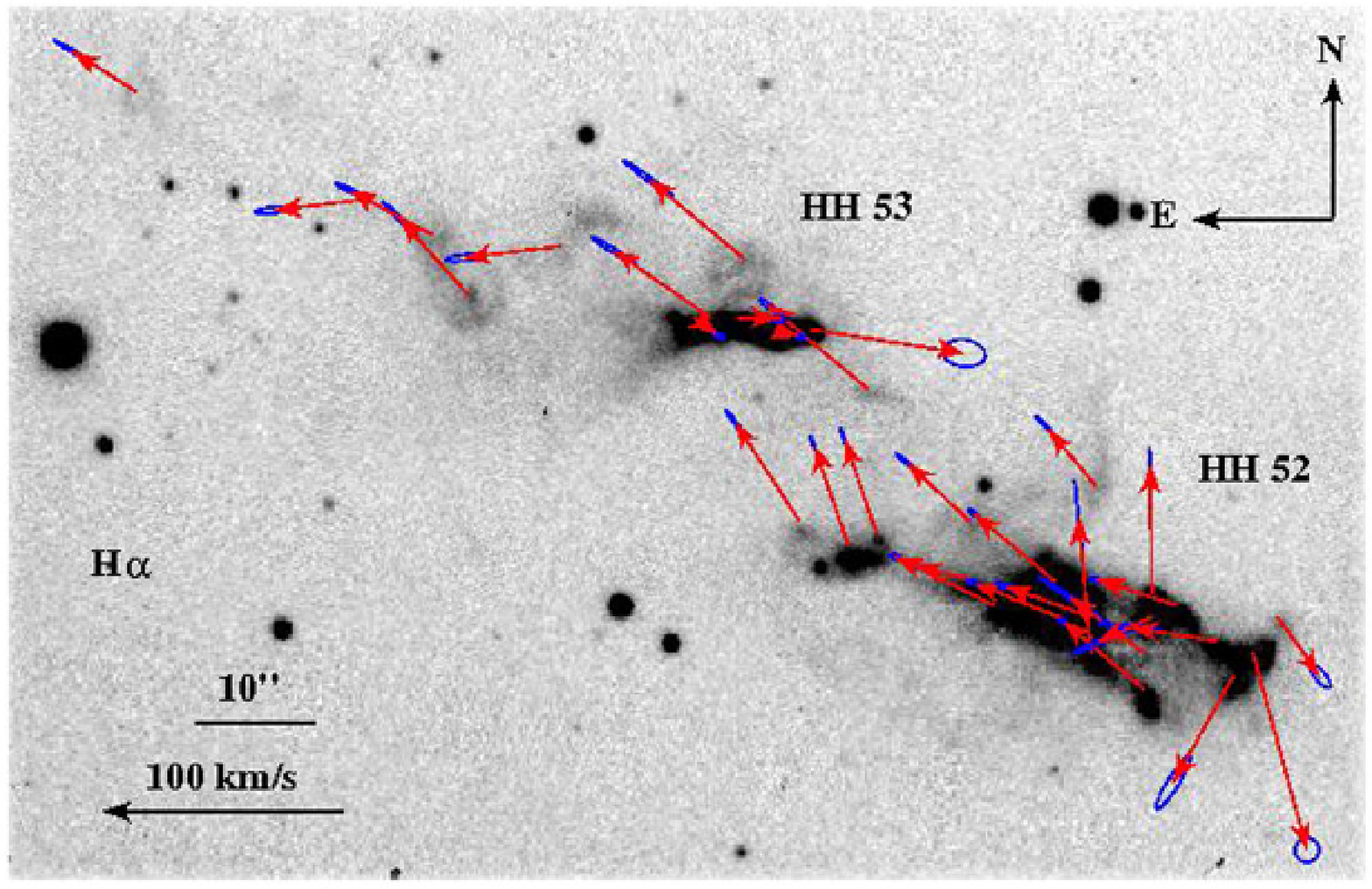}}
   \caption{ Flow charts of HH\,54 ({\bf left panel}), HH\,52 and 53 ({\bf right panel}) in the H$\alpha$ filter.
    Proper motions and their error bars are indicated by arrows and ellipses, respectively.
\label{HH52-3-pms:fig}}
\end{figure*}

\begin{figure*}
 \centering
 \fbox{\includegraphics [width=7.0 cm] {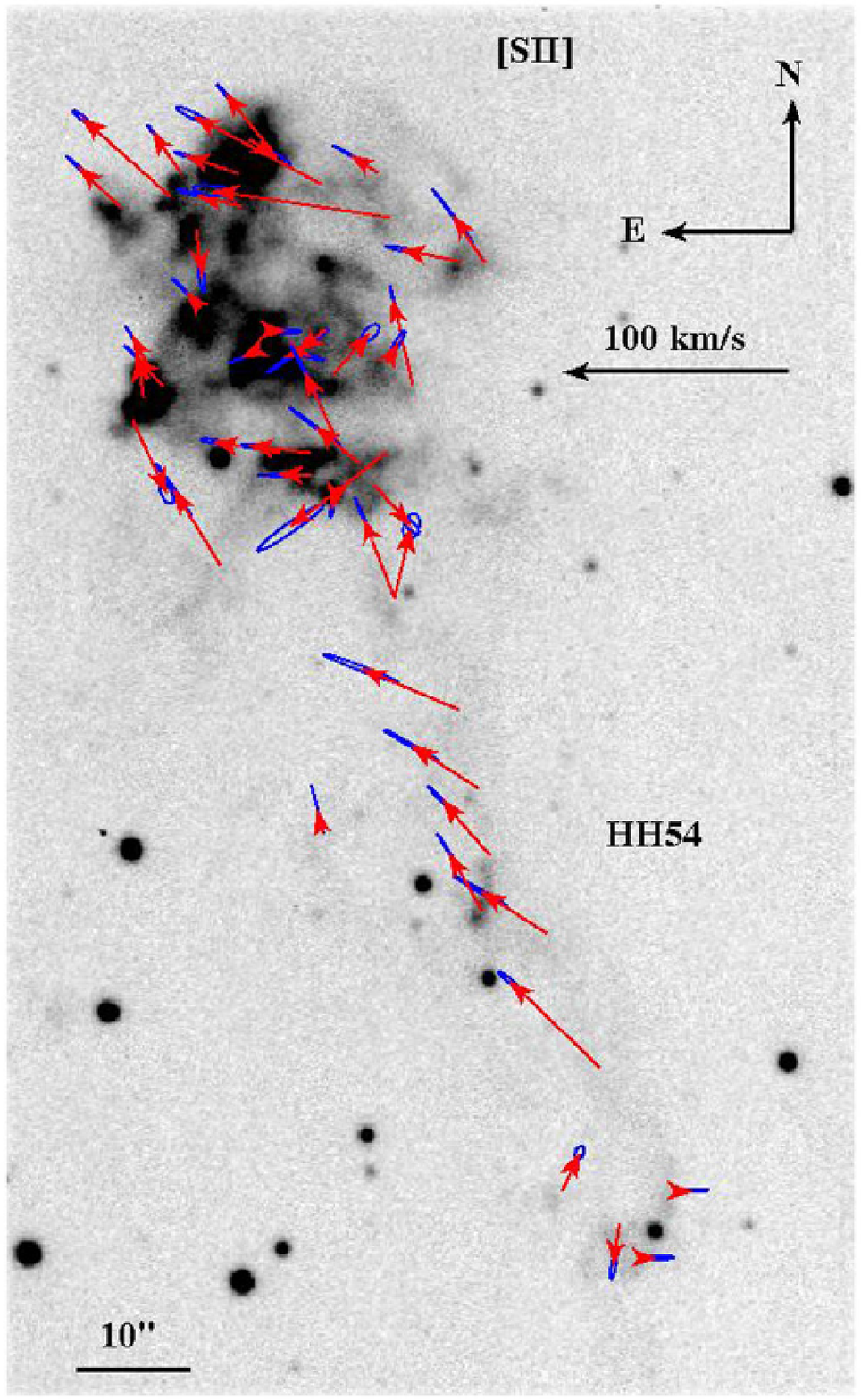}}
 \fbox{\includegraphics [width=10.8 cm] {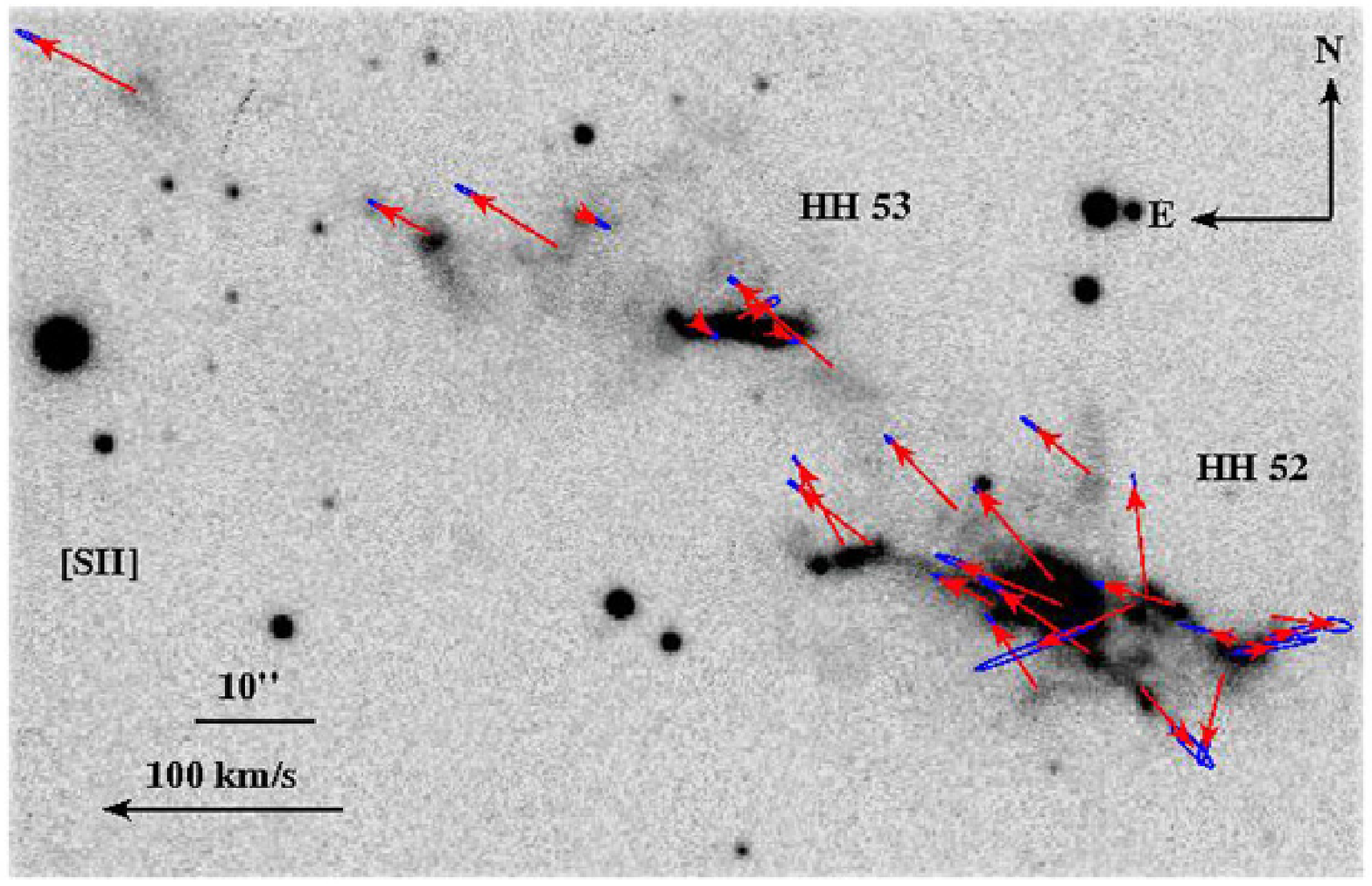}}
   \caption{ Flow charts of HH\,54 ({\bf left panel}), HH\,52 and 53 ({\bf right panel}) in the [\ion{S}{ii}] filter. 
    Proper motions and their error bars are indicated by arrows and ellipses, respectively.
\label{HH54-pms:fig}}
\end{figure*}

In this section we report the general results of our proper motion analysis.
A detailed description for each HH object is given separately in Appendix~\ref{appendixB:sec}.
During the time interval of our observations, some objects showed high degrees of variability, considerably changing in velocity, direction, shape, and flux. Sometimes it was possible to observe the formation or the vanishing of structures along the flows. Therefore, we leave to Sects.~\ref{acc_motion:sec}, \ref{variability:sec}, and to Appendix~\ref{appendixC:sec} the detailed report of these phenomena.

In Figs.~\ref{HH52-3-pms:fig}, \ref{HH54-pms:fig}, and \ref{HH52-H2-pms:fig}\footnote{Fig.~\ref{HH52-H2-pms:fig} is presented in Appendix~\ref{appendixB:sec} as online material only.}, we show the flow charts of the two regions for the three
filters (H$\alpha$, [\ion{S}{ii}], and H$_2$, respectively). 
The P.M.s and their error bars are indicated by arrows and ellipses, respectively.
The results of our P.M. measurements (from the linear fits) are given as online material in Appendix~\ref{appendixB:sec}.
Table~\ref{PM52:tab} (HH\,52 and 53) and Table~\ref{PM54:tab} (HH\,54) report the values of the optical lines, and Table~\ref{PM52_h2:tab} (HH\,52 and 53) and \ref{PM54_h2:tab} (HH\,54) those of the H$_2$ line.
Each table lists the knot ID, the coordinates, the measured proper motion from the linear fit,
the derived tangential velocity ($v_\mathrm{tan}$) at a distance of 178\,pc, and the filter used.
Where P.M.s could not be measured only the knot ID, the coordinates, and the filter (in which the knot was detected) are reported.
Those data which are better modelled by a quadratic fit are labelled with an `a', and the results of the quadratic fits are listed
in Table~\ref{acc:tab}\footnote{Table~\ref{acc:tab} is presented in Appendix~\ref{appendixC:sec} as online material only.} (see Sec.~\ref{acc_motion:sec}).

With few exceptions, the derived P.M.s have similar values in the optical filters, with
values ranging between 0.007 and 0.097 $\arcsec$\,yr$^{-1}$, corresponding to a $v_\mathrm{tan}$
between 6 and 82\,km\,s$^{-1}$. On average, the H$_2$ P.M.s have lower values, ranging between
0.007 and 0.041 $\arcsec$\,yr$^{-1}$ (or 6 and 35\,km\,s$^{-1}$), and the uncertainties are usually greater,
especially for the HH\,52 and 53 regions, due to the smaller number of epochs, poor seeing, lower resolution
of the used instruments, and less accuracy in the alignment (see also Sect.~\ref{observations_NIR:sec}).

A quick inspection of HH\,52 and 53 (right panel of Figs.~\ref{HH52-3-pms:fig} and \ref{HH54-pms:fig}) reveals the presence of two different outflows.
A first flow (\textit{hereafter} HH\,52 flow) is composed of the majority of the knots (those of HH\,52 and several of HH\,53, i.\,e. from C1 to G). It moves NE with a position angle of $\sim$55$^\circ$ (towards HH\,54), roughly following the path delineated by the HH\,52 streamer and  partially superposing on a second flow (\textit{hereafter} HH\,53 flow). This one, made of HH\,53\,A, B, B1, and C, has approximatively an E-W direction and does not seem to be related to the HH\,52 flow.
The features of the HH\,52 flow are moving at a $v_\mathrm{tan}$ between $\sim$30 and 60\,km\,s$^{-1}$, where the different velocities do not seem to be associated with the position of the objects along the flow, but rather they could originate from local interactions inside the fluid. On the contrary, HH\,53\,A, B and C, in the HH\,53 flow, exhibit smaller P.M.s (0.01-0.02$\arcsec$\,yr$^{-1}$ or 9-18\,km\,s$^{-1}$) with P.A.s around 270$\degr$, indicating that they are probably part of a different outflow.

The left panels of Figs.~\ref{HH52-3-pms:fig} and \ref{HH54-pms:fig} reveal the complex kinematics of HH\,54.
The trajectories of the knots indicate that at least two distinct outflows are present in this region
and they appear to be superimposed on the main body of HH\,54. A first flow is generated by the HH\,54 streamer (\textit{hereafter} HH\,54 flow) and is moving NNE (P.A.$\sim$20$^\circ$). It includes knot groups X, Y and Z (HH\,54 streamer) and likely groups E, K and Q (in the main body), with the $v_\mathrm{tan}$ that decreases from $\sim$60 to $\sim$10\,km\,s$^{-1}$ moving from the stream to the main body.
A second outflow with an NE direction (P.A.$\sim$55$^\circ$) is the extension of the HH\,52 flow and contains groups G, C, and, possibly, A,
which appears deflected (P.A.$\sim$80$^\circ$, $v_\mathrm{tan}\sim$30\,km\,s$^{-1}$).
Indeed, for several knots the affiliation with one of the two flows cannot be certain, and the
presence of other flows, even if unlikely, cannot be ruled out (see also Appendix~\ref{appendixC:sec} for a detailed discussion).
This is particularly evident in the central part of the HH object, where the overlap of the flows and the low P.M. values make it difficult to distinguish between real and apparent motions.

\subsection{Accelerated motions}
\label{acc_motion:sec}

As anticipated in the previous section, Table~\ref{acc:tab} reports those datasets that are better modelled
by a quadratic fit. In particular, the table lists, the knot ID; the goodness of the chi-square for the quadratic fit ($P_\mathrm{a}$($\chi^2/\nu$), i.\,e. the 1-tail probability value associated with the provided chi-square value and degrees of freedom $\nu$); the proper and accelerated motions derived from the fit; the P.A.s of the P.M. and of the acceleration vector; the filter and the notes.
The data have been selected by following criteria:
\textit{a)} $P(\chi^2/\nu)$ of the quadratic fit is greater than that of the linear fit;
\textit{b)} $P_\mathrm{a}$($\chi^2/\nu$)$\ge$0.05; \textit{c)} the measured accelerated motion is $\ge$2$\sigma$.
For datasets with only three measurements, conditions \textit{a} and \textit{b} are always verified with $P_\mathrm{a}$($\chi^2/\nu$)=1.
In this case we report the result of the quadratic fit only to support or reject the evidence of an accelerated motion measured in other filters. It is worth to note that same or opposite directions of proper and accelerated motion (i\,e. P.A. values of \vec{v} and \vec{a}, respectively) indicate acceleration or deceleration of the knot.
As a result of this selection, Table~\ref{acc:tab} lists 33 (out of 108) knots that seem to
show some velocity variability in at least one filter. However, not all of these results seem to be genuine.
For example, a few knots (i.\,e. HH\,52\,A5, 53\,G, 54\,G2, 54\,H3, see Table~\ref{acc:tab}) have inconsistent
P.A.(\vec{a}) values in different filters, whereas we would expect similar dynamics for the same knot
even with different filters (i.\,e. it is unlikely that the same knot is accelerated in a filter and decelerated in the other).
Moreover the P.A.(\vec{v}) of some other knots appears to be inconsistent with
respect to the P.A. measured from the linear analysis, and with respect to the P.A. that is expected from the geometry of the flow.
We labelled all these knots as `no' in the notes of Table~\ref{acc:tab}.
Indeed for these knots the quadratic fits do not represent a reliable physical solution.
After this further selection, only 17 knots are left. They indeed appear to be genuine accelerated motions.
Among them, 12 objects accelerate and 5 decelerate. For most of them, both modules and P.A.s of proper
and accelerated motions are identical (inside the errors) for the two optical filters.

\subsection{Flux, direction, and velocity variability}
\label{variability:sec}

In this section we report some examples of the observed time variability in flux, velocity, and direction.
In Appendix~\ref{appendixC:sec} we provide details on those most significant.

On average, the analysed knot fluxes are variable with time. About
60\% of the analysed knots (62 out of 108) showed a variability above the 3\,$\sigma$ error bars.
However, such variations often ($\sim$60\%) do not have a particular trend, but rather appear
as temporary fluctuations around a mean value.
For these objects, the P.M.s, inside the error bar, are usually constant with time.
On the other hand, in 20 years about 20\% of the knots (i.\,e. 19 out of 108) showed velocity variability,
together with flux variations.

The structures that show the highest variability in morphology, flux, and velocity can be mostly identified with \textbf{(1)} working surfaces or \textbf{(2)} interacting knots, where the faster fluid particles ejected at later times eventually catch up with the slower flow ejected earlier from the source. Usually, the fast gas from behind is decelerated, while the slow gas in front is accelerated. This phenomenon is mostly accompanied by flux variability. In particular, we identify four regions that clearly illustrate such a phenomenon: HH\,52\,B and C, HH\,54\,G, and HH\,54\,H.

\begin{figure}
   \includegraphics [width=9.1 cm] {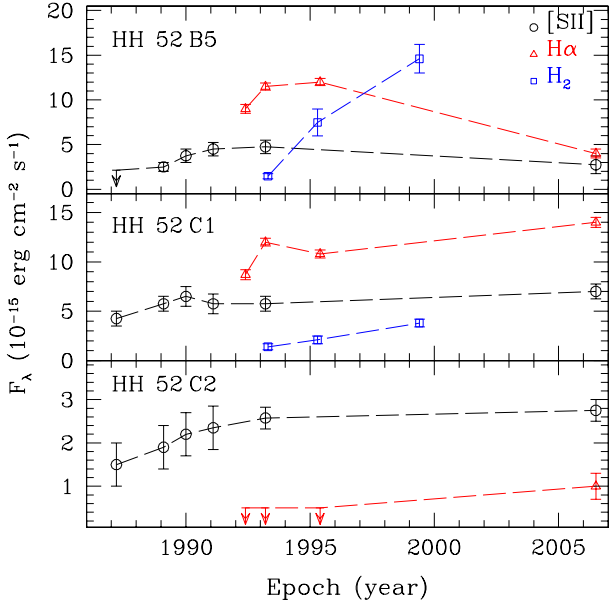}
   \caption{ Measured fluxes (uncorrected for the extinction) in knots HH\,52\,B5 ({\bf top panel}), C1 ({\bf central panel}), and C2 ({\bf bottom panel}) in [\ion{S}{ii}] (circles), H$\alpha$ (triangles), and H$_2$ (squares) filters.
\label{HH52B5-flux:fig}}
\end{figure}

For the analysed knots, the largest flux variations have been usually detected in the H$\alpha$ and H$_2$ emissions, while the [\ion{S}{ii}] usually presents a lower degree of flux variability. Also the timing and the duration of such a variability seem to be different in the three filters. In particular, the optical component seems to excite first and then raises the molecular emission, which
usually seems to exhibit a longer cooling time ($\ge$ 10 years). This is, however, a very qualitative analysis, since we do not have H$_2$ measurements from the beginning of our time series. Moreover, such a trend can be modified by a subsequent shock front.
All this can be seen in Fig.~\ref{HH52B5-flux:fig}, where three examples of the observed knot flux variability are shown.
In addition more examples of flux variability are shown in Appendix~\ref{appendixC:sec}.
The measured flux in the different filters ([\ion{S}{ii}] - circles, H$\alpha$ - triangles, and H$_2$ - squares) is plotted as a function of time.
The top panel of Fig.~\ref{HH52B5-flux:fig} shows the appearance in 1989 of knot HH\,52\,B5, which rapidly increased its brightness and then faded at optical wavelengths, while the H$_2$ emission seems to rise later.
Central and bottom panels of the figure illustrate two more examples of variability in HH\,52. HH\,52\,C1 has a trend similar to B5, but the collision with a fast overtaking shock between 1995 and 2006 (see also Appendix~\ref{appendixC:sec}) newly increases the H$\alpha$ and [\ion{S}{ii}] luminosity. Finally, knot HH\,52\,C2 (bottom panel) represents a quite rare case of [\ion{S}{ii}] variability, where the knot appears visible in H$\alpha$ only in 2006.

\begin{figure}
   \includegraphics [width=8.8 cm] {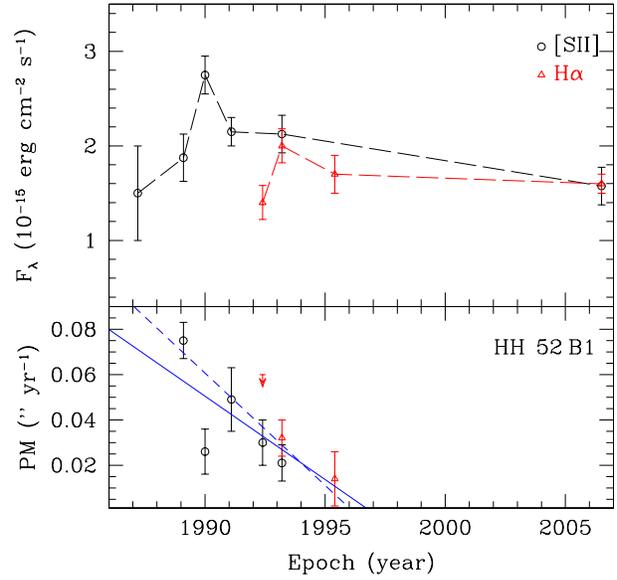}
   \caption{ Variability in HH\,52\,B1. {\bf Bottom panel} shows the P.M.s as a function of time in both [\ion{S}{ii}] (circles), and H$\alpha$ (triangles). The continuous and dashed lines are the best fits of the data points with and without 1990 [\ion{S}{ii}] point, respectively. The slope of the fit gives the deceleration of the knot.
   In the {\bf top panel} the knot flux (in both filters) (uncorrected for the extinction) as a function of time is indicated.
\label{HH52B1-fig:fig}}
\end{figure}

In Fig.~\ref{HH52B1-fig:fig} an example of P.M. and flux variability is indicated.
The bottom panel gives the P.M. values for the optical filters (no molecular emission is detected)
of HH\,52\,B1 plotted against time. As described in Sect.~\ref{pm_analysis}.
The P.M. value decreases in both the [\ion{S}{ii}] and H$\alpha$ filters
(see Table~\ref{acc:tab}, column 4), with a deceleration of 0.022$\pm$0.003 and 0.02$\pm$0.01\,$\arcsec$\,yr$^{-2}$, respectively.
We also note that the P.A.s of the linear motion and of the accelerated motion (Table~\ref{acc:tab}, columns 5 and 6)
are identical in both species (within the errors).
In addition the opposite direction of the accelerated motion indicates
deceleration of the knot for both species. This is also visible in our linear fit of
Fig.~\ref{HH52B1-fig:fig} (bottom panel), where we consider both species
deriving an average deceleration value of
0.017$\pm$0.005$\arcsec$\,yr$^{-2}$ or 0.022$\pm$0.002$\arcsec$\,yr$^{-2}$ without considering the 1990 [\ion{S}{ii}] data point, apparently discordant.
Also the flux of the knot varies with time, with a delay of about 3 years between the [\ion{S}{ii}] and H$\alpha$
lines (as shown in Fig.~\ref{HH52B1-fig:fig}, top panel).

\begin{figure}
\includegraphics [width=8.5 cm] {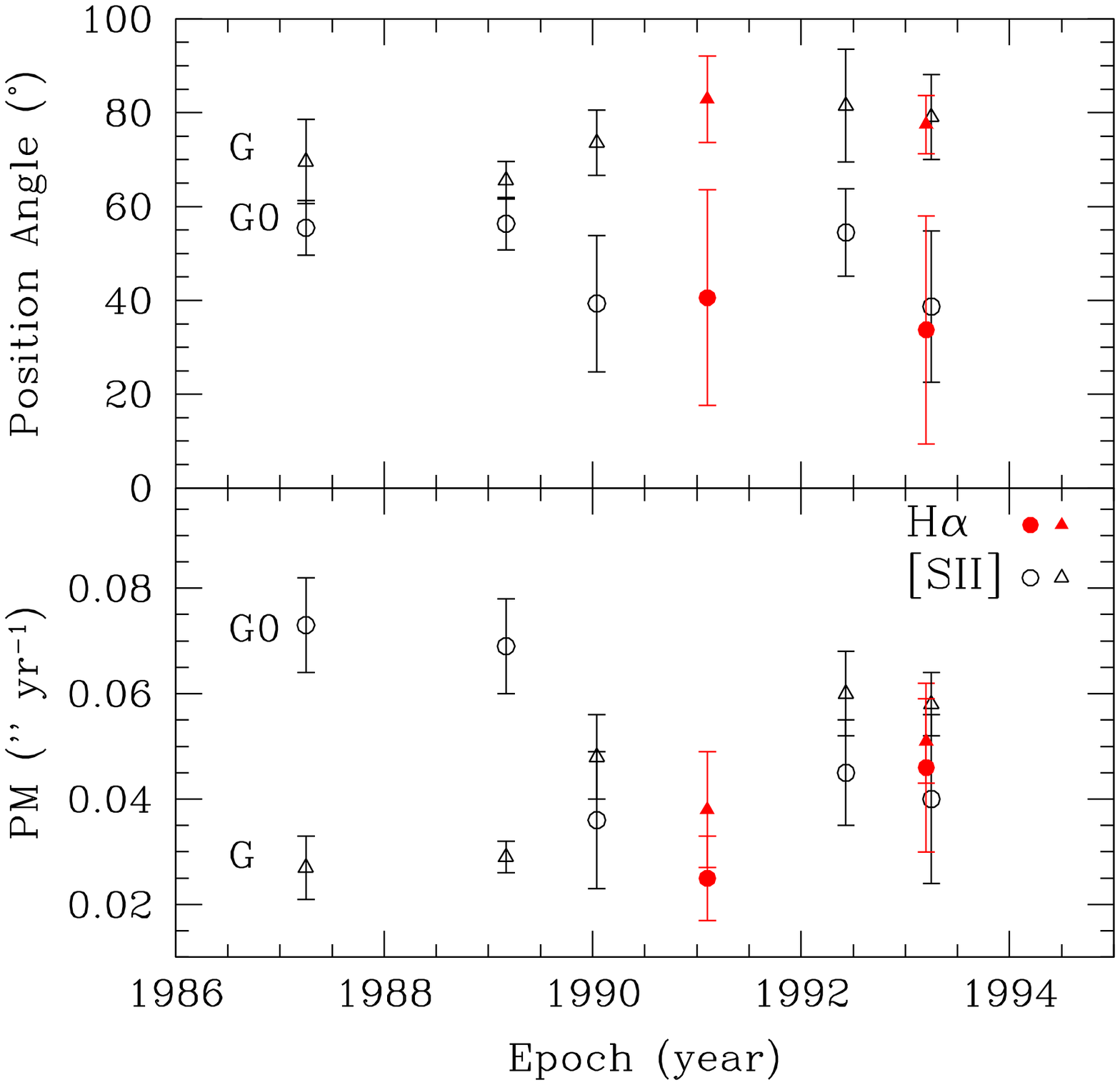}
\includegraphics [width=8.5 cm] {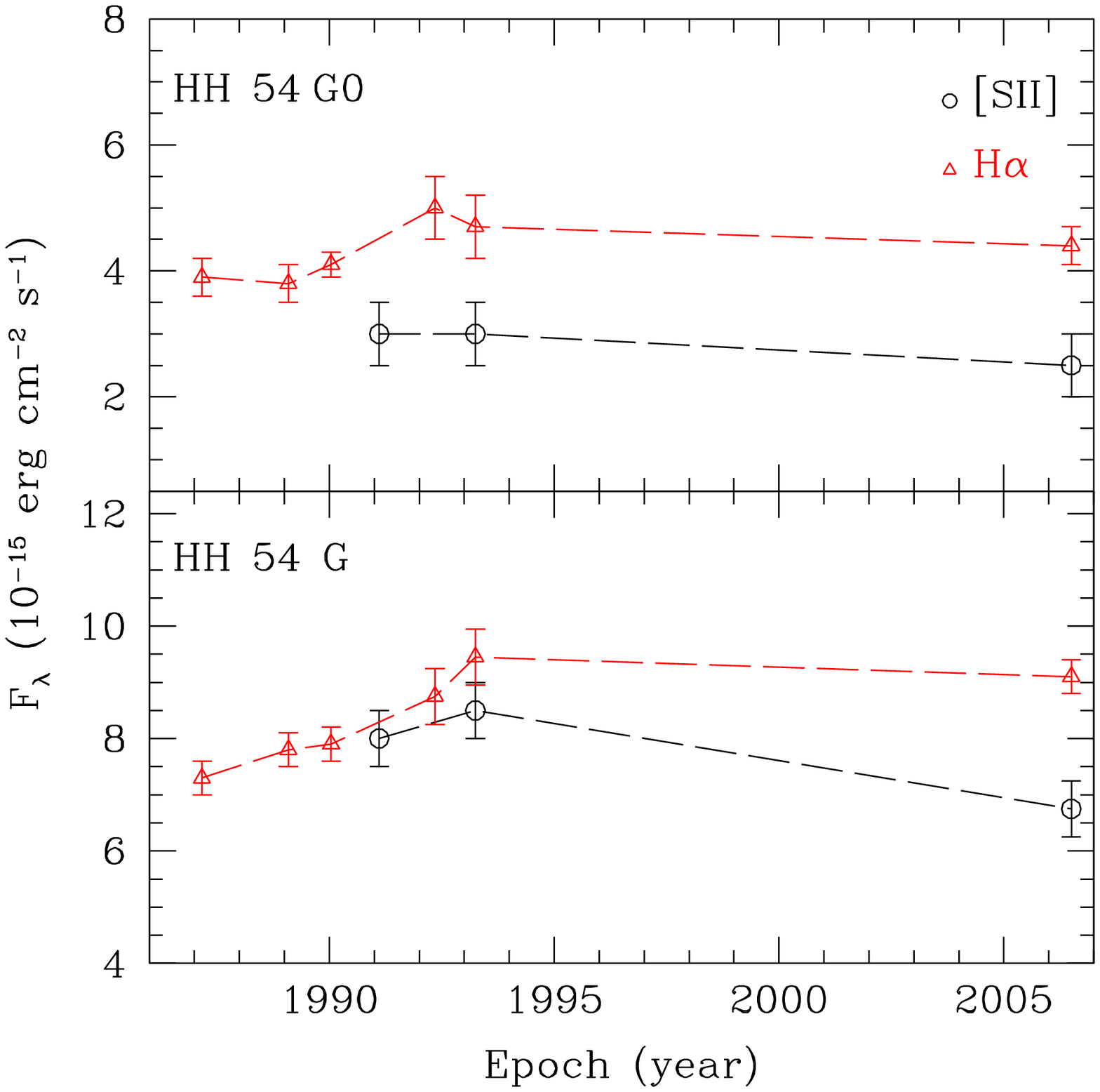}
\caption{ {\bf Upper figure:} changes of proper motions ({\bf bottom panel}) and directions ({\bf top panel}) in HH\,54\,G (triangles) and G0 (circles) with time. Open and filled marks indicate measurements from H$\alpha$ and [\ion{S}{ii}], respectively.
    {\bf Lower figure:} measured (uncorrected for the extinction) fluxes in HH\,54\,G ({\bf bottom panel}) and G0 ({\bf top panel})
   in [\ion{S}{ii}] (circles), H$\alpha$ (triangles), and H$_2$ (squares) filters
\label{HH54G:fig} }
\end{figure}

In Fig.~\ref{HH54G:fig}, as an example of the observed variability in velocity, direction, and flux,
the behaviour of optical knots HH\,54\,G0 and G is presented.
As for the previous knot, our analysis of the accelerated motions of these objects can be found in Table~\ref{acc:tab}.
These knots are part of the HH\,52 flow (P.A:$\sim$55$\degr$) in the NE region of the main body of HH\,54 (see also Fig.~\ref{HH52-3-4Hfig:fig}, Fig.~\ref{HH54-fig:fig}, and Appendices A and B), where the fast flow is suddenly decelerated as it collides with the leading material, which is partially deflected sideways (P.A.$\sim$80$\degr$). G0 is part of such a fast flow, that shocks at the end a slow moving component (knot G), changing its direction.
Bottom panel of the upper figure indicates the measured P.M.s of G0 (circles) and G (triangles) at different epochs in
both optical filters (open and filled marks are H$\alpha$ and [\ion{S}{ii}] data, respectively).
The top panel (upper figure) shows the variations of P.A.s with time for both knots in both filters.
Finally, the lower figure reports the flux variability of the knots. The
P.M. in G0 decreases from $\sim$0.07 to $\sim$0.04 $\arcsec$\,yr$^{-1}$. (i.\,e. from a $v_\mathrm{tan}\sim$65\,km\,s$^{-1}$ to a $v_\mathrm{tan}\sim$35\,km\,s$^{-1}$) as it interacts with knot G ahead. At the same time, its P.A. also exhibits an abrupt change from $\sim$55$\degr$ to $\sim$39$\degr$. Accordingly, P.M. of knot G increases from $\sim$0.03 to $\sim$0.06\,$\arcsec$\,yr$^{-1}$. (i.\,e. from $\sim$25 to $\sim$53\,km\,s$^{-1}$ in $v_\mathrm{tan}$), while its P.A. turns from $\sim$65$\degr$ to $\sim$80$\degr$.
Indeed such a mechanism could at least partially explain the different P.A.s observed along the flows.
As for the other knots, we observe a flux variation in both knots as well (Fig.~\ref{HH54G:fig}, lower figure).

\subsection{Radial velocities}
\label{radialvelocities:sec}

In Tables~\ref{HH52vrad1:tab}-\ref{FeIIvrad:tab}\footnote{Tables~\ref{HH52vrad1:tab}-\ref{FeIIvrad:tab} are presented in Appendix~\ref{appendixD:sec} as online material only.}, the results from our spectroscopical analysis are reported. From Table~\ref{HH52vrad1:tab}
to \ref{HH54vrad1:tab}, we list, for each object and optical instrument (EMMI and B\&C),
the detected lines, the vacuum wavelength, radial velocities (\emph{v$_{rad}$}), and calibrated fluxes.
Finally, Table~\ref{FeIIvrad:tab} indicates the [\ion{Fe}{ii}] (1.64\,$\mu$m) radial velocities
along the ISAAC slits. No flux calibration is available for the NIR spectra.
Each $v_{rad}$ has been computed with respect to the \emph{local standard of rest (LSR)},
assuming a cloud speed of 2\,km\,s$^{-1}$ (see Knee~\cite{knee}).
When detected, two different velocity components are reported.

At variance with B\&C, the EMMI data allowed us to study
single substructures inside the HH objects, thanks to the higher spatial and spectral resolution, and often 
two distinct velocity components were detected.
Moreover, errors in the EMMI spectra are considerably smaller than in the B\&C (4 and 20\,km\,s$^{-1}$, respectively).
However, for those regions encompassed by slits of both instruments, there is excellent agreement among
the values measured with the two instruments. It is also worth noting that all the measured velocities of the knots in this
region are blueshifted, as mentioned by Graham \& Hartigan~(\cite{GH88}).
With a few exceptions, our results are in excellent agreement with theirs. Due to reduced resolution, however, both their spectra and our B\&C spectra  only show one velocity component.

Our measured radial velocities range between -20 and -110 km\,s$^{-1}$ with variations similar to those detected in the P.M.s.
The correlation between radial velocities and P.M.s implies that knot velocity variations are real.
With a few exceptions, radial velocities derived from the detected (atomic) lines have similar values.
The highest values (around -110\,km\,s$^{-1}$) are observed in the three knots of the HH\,53 flow (A, B, C) (see Table~\ref{HH53vrad1:tab})
and along the two streamers (around -100\,km\,s$^{-1}$).
A change in the radial velocity (from $\sim$100\,km\,s$^{-1}$ up to $\sim$40\,km\,s$^{-1}$) is observed ahead of the HH\,52 streamer in groups HH\,54\,G and C, where the fast flow impacts slow-moving material.

The line-profile analysis of the two velocity components indicates that they originate from single bow-shock structures (as in HH\,52\,A), from two different flows (as in HH\,53\,C and C1), or from a flow at different speeds (e.\,g. in HH\,54\,G) (see details in Appendix~\ref{appendixD:sec}).

\begin{figure}
 \centering
   \includegraphics [width=7.3 cm] {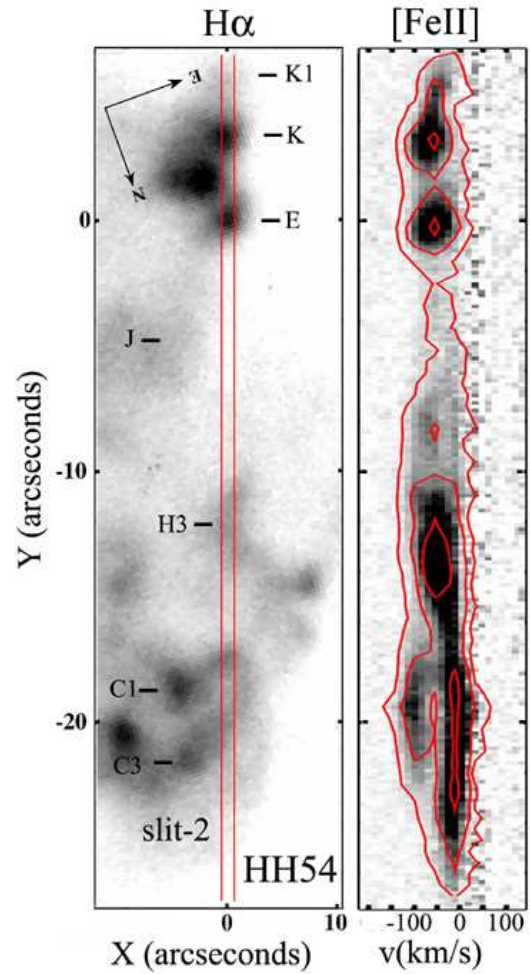}
   \caption{ P-V diagram of HH\,54 slit~2. In the {\bf left} panel an H$\alpha$ image with the slit superimposed is reported.
   In the {\bf right} the [\ion{Fe}{ii}] (1.64\,$\mu$m) spectrum is reported.
\label{HH54-PV2_FeII:fig}}
\end{figure}

Figures~\ref{HH54-PV1_FeII:fig} and \ref{HH54-PV2_FeII:fig} show the position velocity (PV) diagrams of the two ISAAC spectra.
Since we do not have any [\ion{Fe}{ii}] image, we use an H$\alpha$ image (EMMI 2006) to indicate the position of the slits, enlarged to match the spatial resolution of ISAAC. The accordance between the H$\alpha$ and [\ion{Fe}{ii}] emissions is quite satisfactory.
The observed velocity values range from about -10 to -110\,km\,s$^{-1}$. The velocity structure of the region appears complex, and again, often two velocity components are detected. At variance with the EMMI spectra,it is more difficult here  to tell whether these components originate
from single bow shock structures or from the overlapping of two different flows (i.\,e. HH\,52 and 54).

Sometimes the two velocities are well-separated, as in B3 or close to B (Fig.~\ref{HH54-PV1_FeII:fig}), or as in C1 and C3
(Fig.~\ref{HH54-PV2_FeII:fig}). For these objects, the superposition of the two different flows is very likely.
The fast component would then originate from the HH\,54 flow, which shows in Z a v$_{rad}$ of $\sim$-100\,km\,s$^{-1}$
(see Fig.~\ref{HH54-PV1_FeII:fig}).
In other cases, such as in A1, B, H2, and H3 along slit~1 (see Fig.~\ref{HH54-PV1_FeII:fig}), the two velocity components are not well-separated and the line profiles resemble those of a bow shock. For HH\,54\,B, however, the highest intensity in the spectral profile is detected in the lower velocity component, as expected in a cloudlet geometry (see, e.\,g., Schwartz~\cite{schwar78}) and not in the higher velocity component, as observed in a bullet model (e.\,g. Hartigan et al.~\cite{hart87}; Davis et al.~\cite{davis01}). In the first case the observed P.M.s should be close to zero, and, indeed, in knot B we observe a $v_\mathrm{tan}$ close to zero.

\begin{figure}
 \centering
   \includegraphics [width=7.1 cm] {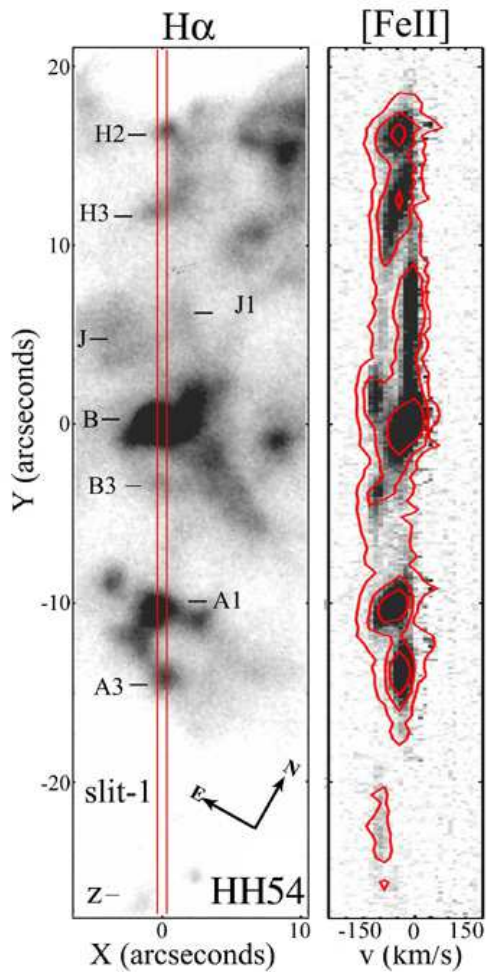}
   \caption{ P-V diagram of HH\,54 slit~1. In the {\bf left} panel, an H$\alpha$ image with the slit superimposed is reported.
   In the {\bf right} the [\ion{Fe}{ii}] (1.64\,$\mu$m) spectrum is reported.
\label{HH54-PV1_FeII:fig}}
\end{figure}

\subsection{Inclination and spatial velocity}
\label{inclination:sec}

\begin{figure*}
 \centering
   \includegraphics [width=13 cm] {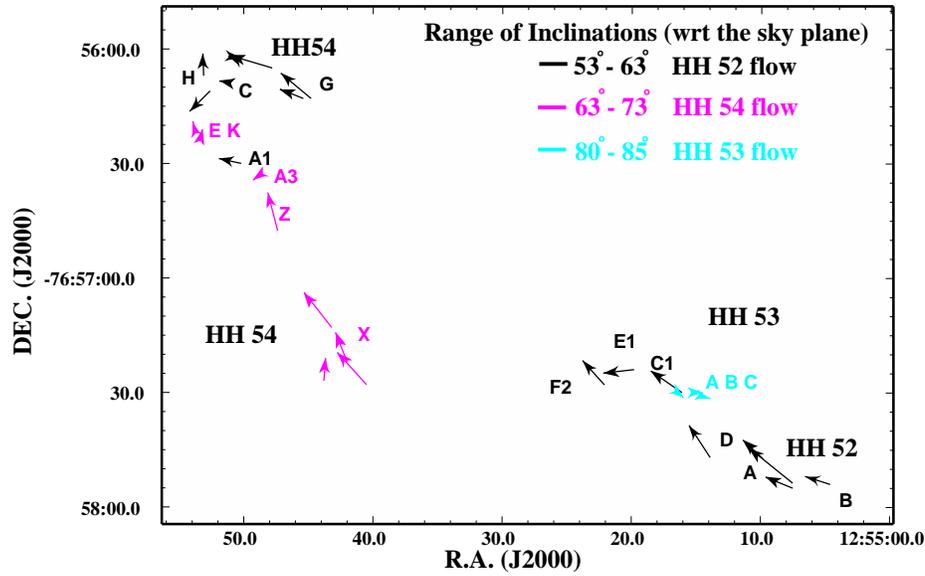}
   \caption{Inclination of the flows with respect to the plane of the sky.
The arrows show the direction of the knots. Their different colours indicate different range of inclination values with respect to the sky plane.
\label{incl:fig}}
\end{figure*}

By combining tangential and radial velocities it is possible to infer an upper limit of the inclination of the single knots and flows (see e.\,g.~\cite{eis94}).
In order to calculate the inclination of each structure and to reduce the uncertainties, we have inferred from our optical data an average value of both the tangential and radial velocities. In Table~\ref{incl:tab}\footnote{Tab.~\ref{incl:tab} is presented in Appendix~\ref{appendixE:sec} as online material only.} the inclination and spatial velocity derived for each knot are reported (averaging both H$\alpha$ and [\ion{S}{ii}] values). A detailed analysis is given in Appendix~\ref{appendixE:sec}.
Inferred spatial velocities range from 25 to 125\,km\,s$^{-1}$.
We obtained three different averaged values of inclination for the flows (58$\pm$3$^\circ$, 84$\pm$2$^\circ$, and 67$\pm$3$^\circ$ for HH\,52, 53, and 54 flows, respectively), which support the proposed scenario of three distinct outflows. The outcome is illustrated in Fig.~\ref{incl:fig}, where the measured knots are spatially plotted with their P.M.s, P.A.s, and inclinations (different colours represent different range of inclinations with respect to the plane of the sky, a few more knots of HH\,54 have been included combining [\ion{Fe}{ii}] and optical data - see next paragraph).

We also attempted to estimate the inclination angle for the structures encompassed by the [\ion{Fe}{ii}] ISAAC slits,
in order to define their proper outflow. To this aim we tentatively combined the [\ion{Fe}{ii}] radial velocities
with the atomic tangential velocities derived from the optical lines, keeping in mind that these species can trace different velocities and regions of the shock.
The results are reported in Fig.~\ref{incl:fig} as well. For knot Z in the HH\,54 streamer we obtain an inclination angle of 72$\degr\pm$2$\degr$, in good agreement with the optical results. Both K and E ($i$=73$\degr\pm$2$\degr$) are also part of this flow.
Knot A1 could be part of the HH\,52 flow ($i$=62$\degr\pm$4$\degr$), while the association of knot A3 is not clear ($i$=68$\degr\pm$8$\degr$). Also J, H2, and H3 are likely parts of this flow with the inclination estimates
of 58$\degr\pm$6$\degr$, 54$\degr\pm$5$\degr$, and 53$\degr\pm$3$\degr$, respectively.

\subsection{Diagnostic results from spectroscopy}
\label{diagnostic:sec}

From the ratio of the detected lines of our optical spectra, it is also possible to
derive the physical parameters of the encompassed knots. The selected optical transitions from [\ion{O}{i}],
[\ion{N}{ii}], and [\ion{S}{ii}] were used to determine the electron density ($N_{\rm e}$), electron temperature ($T_{\rm e}$),
and ionisation fraction ($X_{\rm e}$) of the shocked gas by adopting the so-called BE technique (e.\,g. BE99, Nisini et al.~\cite{nisini05}; Podio et al.~\cite{podio}), which in turn allows us to derive the total hydrogen density $N_{\rm H}$ ($N_{\rm e}$/$X_{\rm e}$) as well.
The assumed chemical abundances are those from Esteban et al.~(\cite{esteban}), extensively discussed and also used
in Podio et al.~(\cite{podio}).
The results are presented in Table~\ref{phys_param:tab} (Col. 2-5). The observed lines have
been dereddened using the optical extinction law of Cardelli et al.~(\cite{cardelli}) assuming an average
$A_V$ value of 2$\pm$1\,mag for all the objects,
as retrieved in previous works on HH\,54 from the NIR analysis (Gredel~\cite{gredel}; Caratti o Garatti et al.~\cite{caratti06a}; Giannini et al.~\cite{giannini06}). The visual extinction in HH\,52 and 53 should have similar values as the extinction map in Porras et al.~(\cite{porras}) suggests (see their Figure\,9). The errors of the physical parameters in Table~\ref{phys_param:tab} were derived by combining the uncertainties on both the line fluxes and extinction.

The values appear quite homogeneous along and among the three flows, indicating the presence of
very light and ionised jets, as sometimes observed in other HH flows, where shocks propagate in a medium of relatively low pre-shock density
(see e.\,g. Podio et al.~\cite{podio}). Actually, both ionisation fraction  and electron temperature have high values, ranging between 0.2-0.6 and 14\,000-26\,000\,K, respectively. On the other hand, both hydrogen and electron densities are extremely
low with values of 2$\times$10$^{2}$-2$\times$10$^{3}$\,cm$^{-3}$ and 50-790\,cm$^{-3}$,
respectively. Remarkably, such a low jet density for HH\,54 was predicted by the shock models of Giannini et al.~(\cite{giannini06}), supported by H$_2$ and FIR observations. There, the emission in HH\,54 was interpreted as
originated by a J-type shock with a magnetic precursor
of $N_{\rm H}\sim$10$^{3}$\,cm$^{-3}$, which interacts with a relatively denser ambient medium ($N_{\rm H}\sim$10$^{4}$\,cm$^{-3}$).

\begin{table*}
\caption[]{ Measured physical parameters and mass fluxes of knots in HH\,52, HH\,53, and HH\,54.
\label{phys_param:tab}}
\begin{center}
\begin{tabular}{llllllll}\\ [+5pt] \hline \hline
knot   &  Log~($N_{\rm e}$) & $X_{\rm e}$ ($N_{\rm e}$/$N_{\rm H}$) & $T_{\rm e}$ &  Log~($N_{\rm H}$)&  $r_{\rm knot}$ & $\dot{M}$ (A)$^a$ & $\dot{M}$ (B)$^b$ \\
       &  (cm$^{-3}$) &  & K & (cm$^{-3}$) &  $\arcsec$ & (10$^{-8}$ $M_{\sun}$\,yr$^{-1}$) & (10$^{-8}$ $M_{\sun}$\,yr$^{-1}$)\\
\hline
HH\,52\,A1    &  2.82$\pm$0.02 & 0.43$\pm$0.02 & 17500$\pm$700 & 3.18$\pm$0.16&  1.5   & 8.2$\pm$0.7  & 1.1$\pm$0.1 \\
HH\,52\,A3-4  &  2.46$\pm^{0.26}_{0.36}$ & 0.35$\pm$0.06 & 23800$\pm$2600 & 2.9$\pm$0.5 &  1.3   & 1.5$\pm$0.4  & 0.3$\pm$0.03\\
HH\,52\,D4   &  2.2$\pm$0.2 & 0.5$\pm$0.05 & 21000$\pm$1000 & 2.5$\pm$0.1 &  1.6   & 2.1$\pm$0.2  & 0.2$\pm$0.03 \\
HH\,52\,B   &  2.88$\pm$0.03 & 0.40$\pm$0.03 & 21500$\pm^{900}_{800}$ & 3.28$\pm^{0.26}_{0.25}$ &  3.3   & 16$\pm$2  & 3$\pm$0.3\\
HH\,52\,D2-D3   &  2.45$\pm^{0.18}_{0.21}$ & 0.23$\pm^{0.04}_{0.06}$ & 14200$\pm^{2400}_{1800}$ & 3.09$\pm^{0.53}_{0.76}$ &  2.4   & 7$\pm$2  & 1$\pm$0.1\\
HH\,53\,A  &  2.88$\pm$0.03 & 0.43$\pm$0.03 & 21600$\pm$800 & 3.15$\pm$0.20 &  0.8   & 1.1$\pm$0.2  & 0.2$\pm$0.02\\
HH\,53\,B  &  2.60$\pm^{0.18}_{0.22}$  & 0.31$\pm^{0.05}_{0.09}$ & 17300$\pm^{2500}_{1900}$ & 2.92$\pm$0.03 &  0.8   & 0.7$\pm$0.1  & 0.3$\pm$0.02\\
HH\,53\,C   &  2.46$\pm$0.11 & 0.34$\pm$0.04 & 18600$\pm^{1200}_{1100}$ & 2.9$\pm$0.3 &  0.6   & 0.3$\pm$0.03  & 0.07$\pm$0.01\\
HH\,53\,C1   &  2.59$\pm^{0.13}_{0.15}$ & 0.32$\pm$0.09 & 24200$\pm$3700 & 3.08$\pm^{0.85}_{0.89}$ &  0.6   & 1.3$\pm$0.1  & 0.3$\pm$0.06\\
HH\,54\,C2   &  2.76$\pm^{0.06}_{0.05}$ & 0.49$\pm^{0.03}_{0.04}$ & 22600$\pm^{2800}_{1600}$ & 3.06$\pm$0.20 &  0.7   & 0.3$\pm$0.05 & 0.3$\pm$0.05 \\
HH\,54\,C3    &  2.66$\pm^{0.15}_{0.16}$ & 0.50$\pm$0.07 & 26000$\pm$7000 & 2.9$\pm$0.4&  0.7   & 0.2$\pm$0.05 & 0.1$\pm$0.05 \\
HH\,54\,G-G0      &  2.65$\pm$0.01 & 0.6$\pm^{0.05}_{0.1}$ & 19700$\pm^{800}_{1700}$ & 2.85$\pm^{0.09}_{0.01}$  &  0.7   & 0.6$\pm$0.03 & 0.05$\pm$0.01\\
HH\,54\,G1-G3   &  2.00$\pm^{0.21}_{0.27}$ & 0.35$\pm^{0.09}_{0.11}$ & 23000$\pm$3000 & 2.15$\pm^{0.56}_{0.65}$ &  0.7   & 19$\pm$6     & 3$\pm$0.5 \\
HH\,54\,X3   &  1.70$\pm$0.7 & 0.34$\pm$0.07 & 25400$\pm^{3600}_{3800}$ & 2.17$\pm^{0.51}_{0.48}$ &  1.3  & 0.3$\pm$0.08  & 0.4$\pm$0.2\\
HH\,54\,X4A-X4B &  2.6$\pm$0.3 & 0.45$\pm$0.05 & 26000$\pm$2000 & 3.0$\pm$0.2&  2.1 & 0.8$\pm$0.1  & 0.6$\pm$0.2 \\
[+5pt] \hline \hline
\end{tabular}
\end{center}
Notes:\\
$^{a}$ $\dot{M}$ derived with method A (see text).\\
$^{b}$ $\dot{M}$ measured from [\ion{S}{ii}], [\ion{O}{i}], and [\ion{N}{ii}] line luminosities (average value)
multiplied by a correction factor that takes the entire area of the knot into account (method B, see text).\\

\end{table*}

\subsection{Mass flux, mass, and luminosity}
\label{massflux:sec}

The physical and kinematical parameters derived in the previous sections allow us to
evaluate the mass flux rate of the knots ($\dot{M}$) in the
studied HH objects. In particular, together with the direction retrieved
from the P.M. analysis, $\dot{M}$ can help us to identify which exciting
source is driving the outflow, comparing the mass flux rates of
the outflow with the theoretical ejection rate of the candidate sources
(also taking their evolutionary stage and bolometric luminosity into account).

\begin{figure}
   \includegraphics [width=9.0 cm] {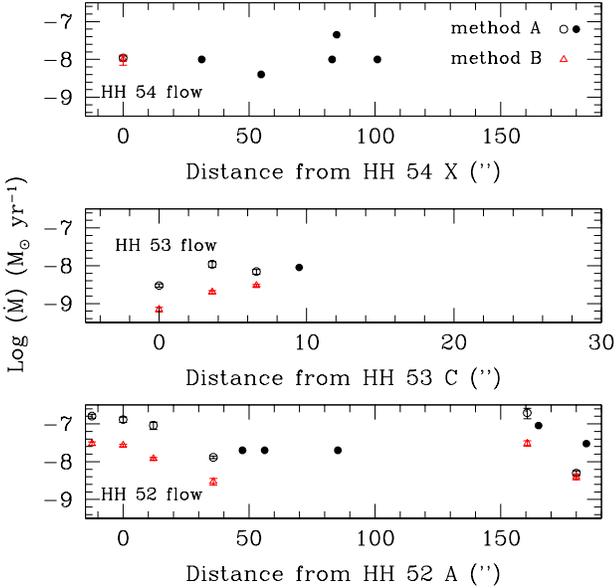}
   \caption{Mass fluxes derived for HH\,52 (\textbf{bottom panel}), HH\,53 (\textbf{central panel}), and HH\,54 flows (\textbf{upper panel}). Circles and triangles refer to $\dot{M}$ values inferred from method A and B, respectively. Filled circles refer to $\dot{M}$ estimates obtained with method A, assuming an $N_{H}$ average value of 10$^{3}$\,cm$^{-3}$.
\label{mdot:fig}}
\end{figure}

We used two different methods to infer $\dot{M}$
(for a detailed discussion see Nisini et al.~\cite{nisini05}; Podio et al.~\cite{podio}).
In the first method (A), the mass flux is estimated as: $\dot{M}_k =  \mu m_H N_H \pi r_k^2 v_k$,
where $\mu$ is the average atomic weight, $m_\mathrm{H}$ is the proton
mass, $N_\mathrm{H}$ the hydrogen density, and $v_\mathrm{k}$ and $r_\mathrm{k}$ are the
velocity and radius of the knot, respectively. For this
we have taken half of the length of the knot, derived from a 3\,$\sigma$ contour on the images.
Method A assumes that the knot is uniformly filled at the derived density,
providing us with an upper limit of the mass flux.
The second method (B) makes use of the luminosities of
the optical lines ([\ion{S}{ii}], [\ion{O}{i}], and [\ion{N}{ii}]
doublets), the hydrogen density, and the tangential velocity:
$\dot{M}_k = \mu m_H (N_H V) v_t / l_t $, with
$N_H V = L_X (h \nu A_i f_i \frac{X^{i}}{X} \frac{[X]}{[H]})^{-1}$,
where L$_\mathrm{X}$ is the luminosity of the element X, for the selected
transition, $A_\mathrm{i}$ and $f_\mathrm{i}$ are the radiative rate and the
fractional population of the upper level of the transition,
$\frac{X^{i}}{X}$ is the ionisation fraction of the considered
species with a total abundance of $\frac{[X]}{[H]}$ with respect
to the hydrogen. Line fluxes were dereddened according to
Sect.~\ref{diagnostic:sec}. The results from each doublet in each knot were
averaged, obtaining a final value.
The values of the mass flux derived from this method can be
considered as lower limits (e.\,g. Nisini et al.
\cite{nisini05}; Podio et al.~\cite{podio}), because method B only
measures gas that is sufficiently heated to radiate in the observed
lines. A correction factor was applied when the slit width was less than the encompassed knot.
The [\ion{S}{ii}] images allowed us to derive such a quantity,
however, for the other two atomic species. The dimension of the emitting region could not be inferred, and
the same correction factor has been adopted, as well.
In Table~\ref{phys_param:tab} (Cols. 7-8) the values of the mass flux derived from both methods are reported.
The resulting mass fluxes are presented in Fig.~\ref{mdot:fig}, as well, where for each flow, $\dot{M}$ of each group of knots
is plotted as a function of the distance from the first knot of the flow.
As expected $\dot{M}(A)$ is, on average, almost an order of magnitude greater than $\dot{M}(B)$ (in 2 out of 3 flows).
Along the HH\,52 flow (bottom panel of Fig.~\ref{mdot:fig}), $\dot{M}(A)$ oscillates between $\sim$2$\times$10$^{-7}$
and $\sim$10$^{-8}$\,$M_{\sun}$\,yr$^{-1}$ (between $\sim$3$\times$10$^{-8}$ and $\sim$3$\times$10$^{-9}$\,$M_{\sun}$\,yr$^{-1}$
with method B). Mass flux along HH\,53 flow ($M_{\sun}(A)\le$10$^{-8}$\,$M_{\sun}$\,yr$^{-1}$, $M_{\sun}(B)\le$3$\times$10$^{-9}$\,$M_{\sun}$\,yr$^{-1}$ ) is more than an order of magnitude lower than in HH\,52. This could indicate a less massive and/or an older exciting source.
The value in HH\,54\,X (along the streamer) is almost the same with both methods ($\sim$10$^{-8}$\,$M_{\sun}$\,yr$^{-1}$).

Using method A and considering an average hydrogen density of 10$^{3}$\,cm$^{-3}$ (see also Table~\ref{phys_param:tab}), we also
estimated the mass flux for most of the other knot groups. The values are also reported in Fig.~\ref{mdot:fig} as filled circles.
$\dot{M}(A)$ has almost the same value in HH\,52\,E, F, and G ($\sim$2$\times$10$^{-8}$\,$M_{\sun}$\,yr$^{-1}$)
Along the HH\,54 streamer, the estimate of $\dot{M}(A)$ ranges between 3$\times$10$^{-9}$ and $\sim$10$^{-8}$\,$M_{\sun}$\,yr$^{-1}$ in Y and Z, respectively. On the main body of HH\,54, we can evaluate a mass flux rate of about 9$\times$, 4.5$\times$, 3$\times$,
1$\times$10$^{-8}$\,$M_{\sun}$\,yr$^{-1}$ for A, E-K, H, and I, respectively.

It is worth noting that such values are in excellent agreement with the mass flux rates
estimated in Giannini et al.~(\cite{giannini06}) and with the linear momentum ($\dot{P}$) obtained from CO
in Knee~(\cite{knee}), considering the velocities and the mass flux rates measured in this paper.
This agreement between atomic and molecular momentum flux can only be explained if the jet is strongly under-dense with respect to
the ambient medium (10-100 times). Then the efficiency of the momentum transfer from the jet (HH) to the medium (CO) is close to one.

Moreover, to measure the energy budget of the flows, we computed mass and luminosity of the knots
(for the detected species, i.\,e. HI, [\ion{S}{ii}], and H$_2$).
To derive the mass, we assumed $M_k = \mu m_H N_H V$, where $\mu$ is the average atomic weight, $m_\mathrm{H}$ is the proton
mass, $N_\mathrm{H}$ the hydrogen density, and $V$ the volume of the knot. For those knots where the hydrogen density
was not measured, we assumed an $N_\mathrm{H}$ average value of 10$^3$\,cm$^{-3}$ (see also Table~\ref{phys_param:tab}).
The luminosity of each element is $L=4\pi d^2 I$, where $d$ is the distance and $I$ the total intensity.
The total intensity is derived from the dereddened line fluxes measured from narrow-band imaging.
As in Sect.\,\ref{diagnostic:sec} the adopted extinction value is $A_\mathrm{V}$=2$\pm$1\,mag.
For the [\ion{S}{ii}] emission an NLTE model has been employed to derive the line ratio with respect to the sulphur doublet
(see e.\,g Nisini et al.~\cite{nisini05}; Caratti o Garatti et al.~\cite{caratti06a}).
The input parameters are $T_{\rm e}$, $N_{\rm e}$, $X_{\rm e}$, and $A_\mathrm{V}$.
Where $T_{\rm e}$, $N_{\rm e}$, $X_{\rm e}$ have not been measured,
average values of 20\,000\,K, 10$^{-3}$\,cm$^{-3}$, and 0.4, have been adopted, respectively.
The HI intensity was obtained from the H$\alpha$ line, assuming that the radiated energy is emitted under Case B recombination,
and using emissivity values from Storey \& Hummer~(\cite{storey}).
Finally, for the molecular hydrogen, when observed, we consider an LTE gas at $T$=2\,000\,K (see e.\,g Caratti o Garatti et al.~\cite{caratti06a}).

\subsection{The candidate driving sources}
\label{excitingsources:sec}

\begin{figure}
 \centering
   \includegraphics [width= 7.8 cm] {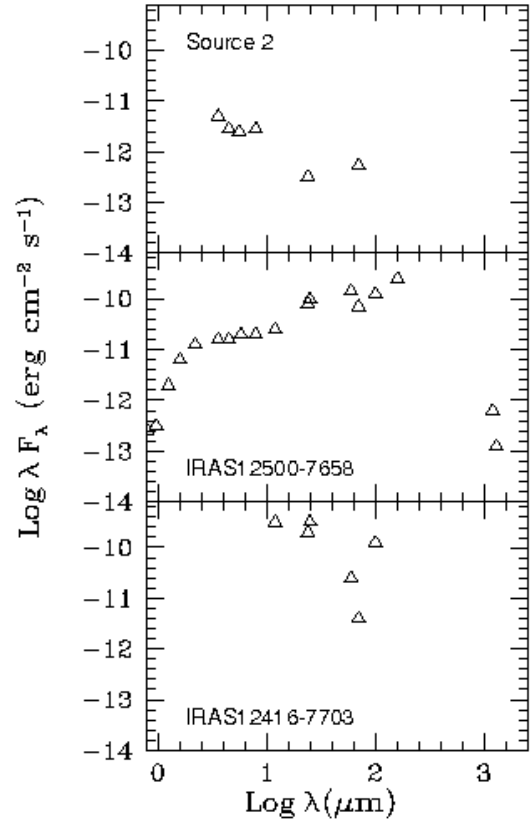}
   \caption{SEDs of the discussed candidate driving sources.
\label{lbol:fig}}
\end{figure}

As seen in Sect.~\ref{morphology:sec}, our H$_2$ maps do not reveal any molecular jet close to the possible driving sources, therefore we do not have any direct indication of which are the driving sources of these flows.
We thus selected the possible candidates on the basis of several YSO catalogues (i.\,e. \emph{IRAS}, \emph{2MASS}, and \emph{MSX}) and publications (see e.\,g. Young et al.~\cite{young}; Spezzi et al.~\cite{spezzi}; Alcal\'{a} et al.~\cite{alcala};
Spezzi et al.~\cite{spezzi2}), on the IR colours of the YSO in the \emph{Spitzer} images, and on their positions with respect to the HH\, objects.
In Fig.~\ref{spitzer:fig}, the 24\,$\mu$m \emph{Spitzer-MIPS} map of the region indicates the selected candidates and the location of the studied HH objects. There are five \emph{IRAS} sources in the region with a position barely compatible with the derived P.A. of the HH flows. These young sources have been deeply analysed in the literature (see i.\,e. Spezzi et al.~\cite{spezzi}; Alcal\'{a} et al.~\cite{alcala}; Spezzi et al.~\cite{spezzi2}), where the authors report their main characteristics and spectral energy distributions (SEDs).
From the \emph{Spitzer} IR colours, we recognised two more possible candidates that have been added to our list.
We note, however, that these two objects were not included in previous Spitzer studies of the region, because they do not fit
some of the selection criteria used (see Spezzi et al.~\cite{spezzi}). In particular the SED analysis in the C2D catalogue indicates that Source\,1 (see Fig.~\ref{spitzer:fig}) could be a galaxy.
The sources, together with short comments, are reported below. The SEDs of the eligible candidates are shown in Fig.~\ref{lbol:fig}. They have been obtained combining \emph{2MASS}, \emph{Spitzer}, \emph{IRAS}, and literature data.

$IRAS\,12496-7650$ (DK Cha) is a Herbig Ae star with a high velocity and weak CO outflow (P.A.$\sim$10$\degr$-15$\degr$, Knee,~\cite{knee}), apparently not related with HH\,52, 53, or 54 outflows (Knee,~\cite{knee}), even if its estimated
mass accretion rate ($\dot{M}_{acc}\sim$10$^{-6}$\,$M_{\sun}$\,yr$^{-1}$, Alcal\'{a} et al.~\cite{alcala}) would make it an excellent candidate (see discussion in Alcal\'{a} et al.~\cite{alcala}).
Our P.M. measurements of the three flows indicate, however, that this cannot be the driving source of any of these flows.

$IRAS\,12448-7650$ is a Class III YSO with a $\dot{M}_{acc}\sim$10$^{-8}$-10$^{-9}$\,$M_{\sun}$\,yr$^{-1}$ (Alcal\'{a} et al.~\cite{alcala}) and located $\sim$68$\degr$ with respect to HH\,52 flow, i.\,e. 10$\degr$-15$\degr$ of misalignment. Moreover,
its optical spectrum (Alcal\'{a} and Spezzi private communication) shows no \ion{Li}{i} absorption feature,
indicating that the source has an age $\ge$2$\times$10$^6$\,yr (see e.\,g. Magazz\'{u} et al~\cite{maga}).

$IRAS\,12416-7703$, a Class II YSO (Young et al.~\cite{young}, Alcal\'{a} et al.~\cite{alcala}), has the correct
position angle ($\sim$56$\degr$) to drive the HH\,52 flow. The SED analysis reveals that the source has a bolometric
luminosity of about 8\,L$_{\sun}$ (see Fig.~\ref{lbol:fig}) (a similar value is by Spezzi et al.~\cite{spezzi2}).
In theory, this would be the most reliable candidate for the HH\,52 flow. Nonetheless, its optical spectrum (Alcal\'{a} and Spezzi private communication) has a weak H$\alpha$ emission but no \ion{Li}{i} absorption feature, thus $IRAS\,12416-7703$ seems too old
to be the driving source. It is puzzling, however, that no other YSOs are present in this region (see Figs.~\ref{spitzer:fig} and 9 in Alcal\'{a} et al.~\cite{alcala}).

$IRAS\,12500-7658$ is a Class I (Chen et al.~\cite{chen}) source with an $L_{bol}$ of 0.5\,L$_{\sun}$ (Young et al.~\cite{young}) with
an estimated $\dot{M}_{acc}$ of 2$\times$10$^{-6}$\,$M_{\sun}$\,yr$^{-1}$ (Alcal\'{a} et al.~\cite{alcala}). Indeed this is
the best candidate for the HH\,54 flow ($\dot{M}\sim$10$^{-8}$\,$M_{\sun}$\,yr$^{-1}$, see Fig.~\ref{mdot:fig}) because of the P.A. of $\sim$20$\degr$ with respect to the HH object and the $\dot{M}$/$\dot{M}_{acc}$ (0.1-0.01) ratio, compatible with a Class\,I source.

$IRAS\,F12488-7658/C13$, located two arcminutes west of the previous source, is a Class III YSO (Voung et al.~\cite{voung}) with a $\dot{M}_{acc}$4$\times$10$^{-8}$\,yr$^{-1}$ (Alcal\'{a} et al.~\cite{alcala}) and
has a P.A.$\sim$30$\degr$ with respect to HH\,54. The misalignment, the age, and the $\dot{M}_{acc}$ of the source seem to indicate that this is not the driving source.

Finally, we detect in the \emph{Spitzer} images an MIR source almost coincident with HH\,53
(indicated as Source\,2 in Fig.~\ref{spitzer:fig}, $\alpha=12^h55^m17^s$ and $\delta$ =-76$\degr$57$\arcmin$29\farcs3),
appearing as a faint star in our optical, H$_2$, and \emph{2MASS} images. A faint emission is also visible
in the IRAC images (partially detached with respect to the HH\,53 knots), while this object is not resolved
in the MIPS images and part or all the measured flux could originate in the HH itself.
The coordinates of this object are almost the same as the infrared source HH\,53*1
reported by Sandell et al.~(\cite{sandell}) and indicated as the possible driving source of HH\,53.
On the contrary, Graham \& Hartigan~(\cite{GH88}) firmly rejected this hypothesis because their
observations showed a blueshifted emission both to the west and to the east of the star
(we know, however, that such emission could originate in the HH\,52 flow) and because it
did not show any unusual strong emission feature in the star spectrum.
Furthermore, the optical photometry of Spezzi and Alcal\'{a} (private communication) indicates that the object
would not belong to Cha\,II and could be a background star.
Nonetheless, our SED analysis (Fig.~\ref{lbol:fig}, from the Spitzer photometry and under the uncertain hypothesis
that the MIR comes entirely from the object) would indicate that it is a Class II YSO,
with a $L_{bol}\sim$10$^{-2}$\,L$_{\sun}$. We tentatively speculate that this could be the exciting source of HH\,53.
The mass ejection rate measured on the knots (3$\times$10$^{-9}$-10$^{-8}M_{\sun}$\,yr$^{-1}$) appears, however, too high
compared to the Class and the luminosity of the presumed source.
Indeed a more in-depth spectroscopical analysis is needed to solve this conundrum.
Nevertheless, it is worth noting that no other candidates are detected east of HH\,53.
In conclusion, two of the three identified flows (i.\,e. HH\,52 and 53) still have not a reliable driving source candidate.

\section{Discussion}
\label{discussion:sec}

\subsection{Variability}

\subsubsection{Flux and velocity variability of knots}

Our analysis indicates that flux and velocity variability is observed in knots along the flows mostly in coincidence
with interacting knots or working surfaces, where pairs of knots interact by accelerating and decelerating the gas in front and behind,
respectively, or where the jet/streamer hits the slow moving material ahead.
On the other hand, knots with a constant or slightly variable flux do not show relevant changes
in velocity. It is therefore reasonable to investigate if and how such a flux variability is connected to
the observed velocity variations. We can argue that part of the kinetic energy of the impacting knot is transferred to
the preceding knot or to the surrounding medium, and then converted in radiated energy.
However, it is worth noting that here we do not consider how the energy is transformed and which processes are involved.
In a forthcoming paper (De Colle \& Caratti o Garatti, in prep.), we report how the observed variability can be reproduced
by MHD simulations, when knots interact along the flow.

In order to check how much of the kinetic energy is radiated during this process, we therefore computed the energy budget for
those knots that are interacting and that clearly show both flux increment and velocity decrement.
We estimate both the amount of kinetic energy lost in the interaction ($\Delta K$) and the variation in the radiated energy during the time interval of our observations ($\Delta E_{rad}$). The kinetic energy lost by the knot can be expressed as
\begin{equation}
\label{kinematics:eq}
\Delta K_k = 1/2 M_k \Delta v^2
\end{equation}
where $M_k$ is the mass of the knot and $\Delta v^2$ is the difference between the initial and final squared velocities.

On the other hand, $\Delta E_{rad}$ can be written as,

\begin{equation}
\label{Erad:eq}
\Delta E_{rad}= \sum_{species} (\int_{t1}^{t2} \Delta L dt)
\end{equation},
which is the integral of the luminosity variations over the time interval of observations, summed for all the detected species
(i.\,e. HI, [\ion{S}{ii}], and H$_2$). Knot luminosities, and masses were estimated as described in Sect.~\ref{massflux:sec}.
The results are reported in Table~\ref{discussion:tab}, where $\Delta E_{rad}$ of each element, and
the total $\Delta E_{rad}$ are indicated for each analysed knot $\Delta K$.

\begin{table*}
\caption[]{ Estimated kinetic energy lost and variations in radiated energy
 for individual knots in HH\,52 and HH\,54 . \label{discussion:tab}}
\begin{center}
\begin{tabular}{cccccc}\\ [+5pt] \hline \hline
knot     &  $\Delta K$ & $\Delta E_{rad}(HI)$ &  $\Delta E_{rad}(SII)$ & $\Delta E_{rad}(H_2)$ & $\Delta E_{rad}(tot)$ \\
  ID     &   (10$^{38}$\,erg ) &   (10$^{38}$\,erg ) &   (10$^{38}$\,erg ) &   (10$^{38}$\,erg ) &   (10$^{38}$\,erg ) \\
\hline
HH\,52\,B1   &  70       & 1.1  & 0.08  & 0.2-0.4  &  1.3-1.5 \\
HH\,52\,C1   &  17       & 0.8  & 0.18  & 0.4  &  1.3 \\
HH\,52\,C2   &  3.5      & 0.35  & 0.1  & -  &  0.4 \\
HH\,54\,H1   &  8.8      & 2.9  & 0.2    &  -       &  3.0     \\
HH\,54\,H2   &  9.6      & 2.8  & 0.12   &  0.9     &  3.8     \\
HH\,54\,H3   &  57       & 2.7  & 0.2    &  2     &  4.8     \\
HH\,54\,G    &  -18$^a$  & 4.0  & 0.02   & -  & 4.0  \\
HH\,54\,G0   &  26       & 0.8  & $<$0.01  & -  & 0.8  \\
\hline

\hline \hline
\end{tabular}
\end{center}
Notes:\\
$^{a}$ here the kinetic energy is increased from the interaction with G0.\\
\end{table*}


As a result $\Delta E_{rad}$ is always smaller than $\Delta K$.
This is possibly due to the longer cooling time of the gas with respect to the interval of our observations.
Moreover this is also because we are not taking into account
the $\Delta E_{rad}$ contribution from other species like FeII, OI, or CO, for example.
We can give a rough upper limit estimate of the FeII contribution, if any, for example.
The brightest [\ion{Fe}{ii}] emission in HH\,52-54 complex is observed in HH\,54\,B, where a flux of
$\sim$8$\times$10$^{-15}$\,erg\,s$^{-1}$\,cm$^{-2}$ (1.64\,$\mu$m line) has been measured (see Gredel~\cite{gredel}, Giannini et al.~\cite{giannini06},
Caratti o Garatti~\cite{caratti06a}). This corresponds to a luminosity of $L_{FeII}\sim$4$\times$10$^{29}$\,erg\,s$^{-1}$ (adopting the same physical parameters used for the other ions). This means that even assuming a luminosity variability equal to the knot luminosity
the radiated energy in FeII would be 10$^{36}$-10$^{37}$\,erg, which is one or two orders of magnitude less than the HI contribution.
The contribution from [\ion{S}{ii}] is usually smaller, down to two orders of magnitude (see Table~\ref{discussion:tab}). Therefore, among the considered atomic lines, HI radiates most of the energy observed in such a variability.
This is well visible from the flux variability in Figures~\ref{HH52B1-fig:fig}-\ref{HH54G:fig} (see also Figures in Appendix~\ref{appendixC:sec}), where H$\alpha$ usually exhibits the largest flux variability. This is possibly due to the different cooling times of the two atomic species (a few years for HI and $\sim$30 years for [\ion{S}{ii}], see e.\,g. Hartigan et al.~\cite{hart01},~\cite{hart05}). As a consequence, usually H$\alpha$ traces morphological changes better and faster than [\ion{S}{ii}] does.

Conversely, the H$_2$ emission, when observed, partially contributes to the gas cooling, but with longer timescales.
We note, however, that only part of the H$_2$ variability was estimated (i\,.e form the warm component $\sim$2000\,K traced by the 2.12\,$\mu$m line), and we are likely missing the contribution of the cold H$_2$ component that can contribute up to 50\% to the whole radiative cooling (Caratti o Garatti et al.~~\cite{caratti08}).

It is not possible to properly evaluate the remaining molecular contribution, as for the CO. However it should not be greater
than the atomic one, since the observed medium is partially ionised.
Therefore it is reasonable to conclude that a large part
of the kinetic energy lost in between two working surfaces is converted in radiated energy.
This also means that the energy is rapidly converted, the shock dissociates the H$_2$, and we observe the HI
emission. After a few years, the temperature drops, and the H$_2$ emission is then detected.

However, in some cases, the kinetic energy is not completely radiated but can be partially transferred to the medium or to the knots ahead of the flow, as shown in the interaction between knots HH\,54\,G0 and G, for example. Here, the kinetic energy lost by the bullet (G0)
is mostly ceded to the target (G) and only partially lost ($\sim$20\%) in the radiative process. In fact, the energy balance of the system ($\Delta K(G0)$-$\Delta K(G)$-$\Delta E_{rad}(G0+G)$) is close to zero, as can be derived from Table~\ref{discussion:tab}.

Finally, there are some knots that show flux variability but no change in velocity (inside the error bars). In some cases, where the flux decreases, we are observing the radiative cooling of the knot, but the momentum of the knot is conserved. In other cases, where the flux increases (see e.\,g. HH\,54\,A1 and A8 in Appendix~\ref{appendixC:sec}), we have estimated the amount of kinetic energy needed to produce the observed $\Delta E_{rad}$. As a result the variation in velocity that produces such a $\Delta K$ is well below the error bars that we measure in our P.M.s.

\subsubsection{Cooling time and velocity variability}

Our set of observations indicate apparent acceleration and deceleration for several knots, but just a few of them seem 
to be genuine. Often the observed velocity changes do not reflect true fluid velocity variations, 
but merely \textit{phase velocity} effects (see e.\,g. Bally et al.~\cite{bally}), 
wherein some portions of the post-shock gas fade and others brighten. This is particularly true when the elapsed time
between two examined images is longer than the post-shock cooling time of the analysed knot. 

We can thus compare the timescale of the observed variability with the expected atomic and molecular cooling times.
The cooling time is 
\begin{equation}
\label{cooling:eq}
 t_{\rm c} = \frac{T}{|dT/dt|} \sim \frac{nk_{\rm B}T}{\Lambda}
\end{equation}
where $T$ is the temperature, $k_{\rm B}$ the Boltzmann constant, and $\Lambda$ the cooling 
rate (in erg\,cm$^{-3}$\,s$^{-1}$)\footnote{alternatively the cooling coefficient $\Lambda_c$ (in erg\,cm$^{3}$\,s$^{-1}$) can be used
and Eq.~\ref{cooling:eq} becomes: $t_c \sim \frac{k_{\rm B}T}{n \Lambda_c}$}. To estimate $t_{\rm c}$,
we use the H$_2$ cooling rates from Shull \& Hollenbach~(\cite{shull}), and the atomic cooling coefficients from Cox \& Tucker~(\cite{cox}) and 
Gnat \& Sternberg~(\cite{gnat}). For the H$_2$ we consider a range of densities (from 10$^3$-10$^5$\,cm$^{-3}$) and
temperatures (from 2000 to 3000 \,K). As a result, the H$_2$ cooling time ranges from $\sim$1 to 9\,yr, where the shorter
value is obtained for higher temperatures and densities. For 
the atomic component, we consider an average $N_\mathrm{H}$ of 10$^3$\,cm$^{-3}$ and $T_\mathrm{e}$ ranging from 10\,000 to 25\,000\,K.
We thus obtain 1$\leq t_{\rm c}(H\alpha)\geq$6\,yr. 
The cooling time of the [\ion{S}{ii}] is almost one order of magnitude longer than $t_{\rm c}(H\alpha)$.

The complex proper motion variations may then indicate the combined effects of photometric and morphological changes, occurring on timescales comparable to the cooling time in the post-shock gas, combined with the detection of collisions of fast, fragmented ejecta
overrunning slower moving fragmented ejecta. 
The highly fragmented morphology of these shocks and the very short acceleration/deceleration timescales ($\sim$10\,yrs)
indicate that various types of instabilities (cooling instabilities, Vinsniac instabilities, etc)
may have caused the flow elements to fragment into tiny, dense clumps surrounded by lower density plasma or gas.
As a fast shrapnel over-takes slower shrapnel, both accelerating and decelerating post-shock fluid elements are observed.
Such timescales could be actually good measurements of the size-scales of the clumps in the colliding fluids.

\subsubsection{Direction variability and flow deflection}

Our P.M. analysis clearly indicates that part of the HH\,52 flow is deflected.
Some knots in the HH\,54\,G group have been observed to change their direction {\it `in real time'}, due to the interaction with
other colliding knots moving faster along the flow. In this case, most of the kinetic energy of the impacting knot (bullet)
is transferred to the knot ahead (target), as shown in the previous section. To cause the deflection, the bullet trajectory
is slightly off-axis with respect to the target. All this implies a time variability in the ejection velocity
and a small precession of the flow, as well (see e.\,g. Raga et al.~\cite{raga93}, V\"{o}lker et al~\cite{volker}).

Other knots, such as HH\,54\,A, C, M, or HH\,52\,G, F, exhibiting different P.A.s of 20$\degr$-40$\degr$
with respect to the main flow, were probably deflected earlier, and it is not evident how this
bending originated. However, it is not clear whether this `bullet-target' mechanism would remarkably deflect massive features like HH\,54\,A.
Such a deflection is more likely the result of a collision of the flow with a dense cloud or dense clumps in the medium,
as also observed in HH\,110 (Reipurth et al.\cite{reip96}) and in several numerical simulations (see e.\,g. de Gouveia Dal Pino~\cite{gouveia},
Raga et al.~\cite{raga02}, Baek et al.~\cite{baek}). In this scenario the deflected beam initially describes a C-shaped trajectory around the curved jet/cloud contact
discontinuity; later, when the jet has penetrated most of the cloud/clump extension, the flow resumes its original direction of
propagation. This could be the reason we observe both a deflected (HH\,54\,A) beam and a continuation of the flow (HH\,54\,G and C).

The observed velocities in the deflected beam are between 30\% and 60\% lower than those detected in HH\,52. This is
consistent with the values predicted by the simulations (see e.\,g. de Gouveia Dal Pino~\cite{gouveia}, Baek et al.~\cite{baek}) with
$v_{fin}$=$v_{in}$cos$\theta$, where $\theta$ is the deflection angle, $v_{fin}$ and $v_{in}$ are the deflected and incident velocities,
respectively. Also the predicted ratio between jet and ambient medium ($n_j$/$n_m\le$10$^{-2}$) agrees with our
observations (see Sects.~\ref{diagnostic:sec} and \ref{massflux:sec}).
Therefore the chaotic structure and P.M.s measured in HH\,54 should be interpreted not only as the simple superposition of two distinct outflows but also as the result of a deflection in the HH\,52 flow.

\subsection{Length, dynamical age of the outflows, counterflows}
\label{taudyn:sec}

\begin{figure*}
 \centering
   \includegraphics [width= 14 cm] {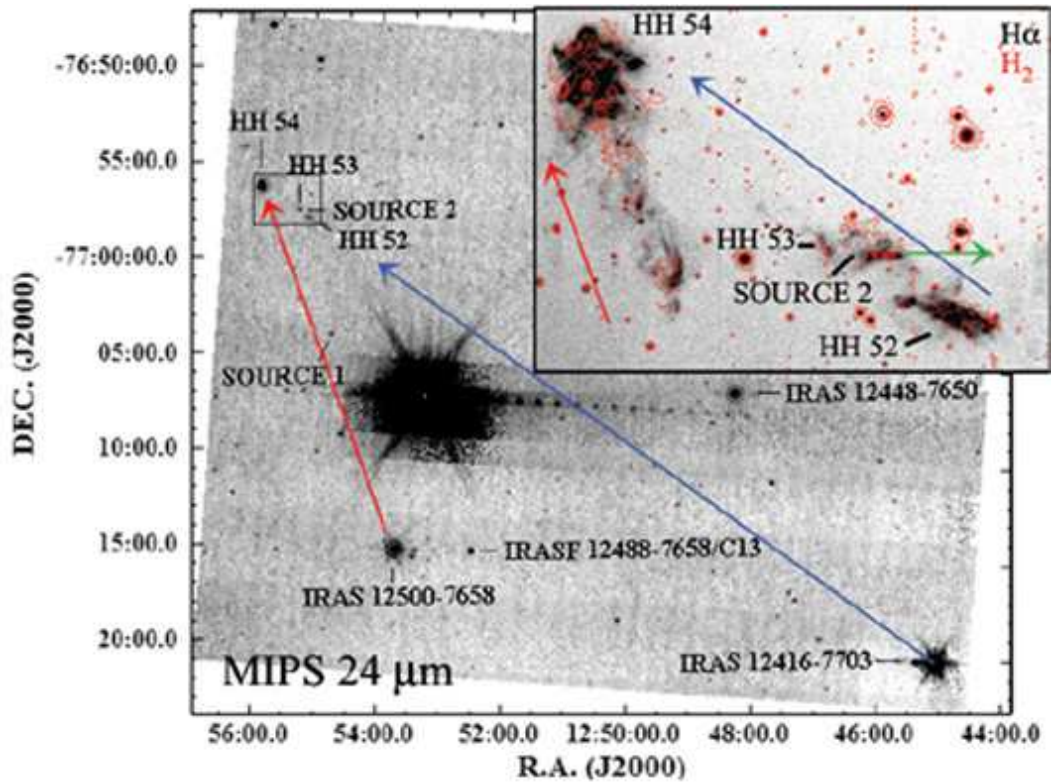}
   \caption{ Summary figure. \emph{Spitzer-MIPS}\,24\,$\mu$m map of the investigated Cha II region, indicating the candidate
   exciting sources, the outflows, and their orientations. 
   A rectangular box indicates the position of the upper-right inset, where the H$\alpha$ EMMI 2006 image along with the H$_2$ SofI 199 
   image contours of the three outflows are reported. 
\label{riass:fig}}
\end{figure*}

Assuming that $IRAS\,12416-7703$, Source\,2, and $IRAS\,12500-7658$ are the driving sources of HH\,52, HH\,53, and HH\,54, respectively, we can infer two important parameters of the flows from our kinematical analysis: the spatial length and the dynamical age ($\tau_{\rm dyn}$), which is the ratio between the distance of the knot(s) from the YSO (corrected for the inclination angle $i$) and the total velocity.
The inferred results are reported in Table~\ref{tau:tab}. Additionally, Fig.~\ref{riass:fig} summarises the results,
showing the candidate exciting sources, the outflows, and their orientations.
Two out of the three outflows, namely HH\,52 and HH\,54, would be parsec scale outflows,
with a length of about 4\,pc and 2.2\,pc, respectively. The dynamical age of the flows would then be
up to 6$\times$10$^4$\,yr (HH\,52 flow, knots HH\,54\,A and C), and up to 4$\times$10$^4$\,yr (HH\,54 flow, knots E, K).
On the other hand, if we assume Source\,2 as the HH\,53 driving source (but this is highly questionable), here the flow has a smaller extension 0.1-0.2\,pc and a shorter dynamical age ($<$10$^3$\,yr).

The detection of parsec scale outflows from Class\,0 \& I (e.\,g. Reipurth et al.~\cite{reip97}, Eisl\"{o}ffel~\cite{eis00})
as from Class\,II YSOs (McGroarty \& Ray~\cite{mcgroarty}) is quite common. It is also not surprising that these flows have 4-6$\times$10$^4$ years of age. What is indeed peculiar is that no further emission is detected between the parsec scale flows and the possible exciting sources. This could mean that the ejection (and therefore also the accretion) activity in two of the YSOs suddenly stopped or at least was reduced several thousand years ago. A different explanation could be that part of the flow is not emitting
(i.\,e. it is neutral), and it only becomes visible when it shocks the slow material moving ahead. This seems to be the case
of the HH\,52\,B and C regions, to the rear of the HH\,52 flow, where we observe the formation of new knots.

Another puzzling question is the absence of a counterjet in all three flows.
The red-shifted CO emission between HH\,52 and HH\,54 (see Knee~\cite{knee}) could be the HH\,53 missing counterflow. However, we do not detect any jet in our images. Also for the two parsec scale flows, no emission is observed along the possible counterflow axes in \emph{Spitzer}, \emph{2MASS}, or \emph{DSS} images. Alcal\'{a} et al.~(\cite{alcala}) report the discovery of a faint H$\alpha$ emission, located
$\sim$20$\arcmin$ to the SW of $IRAS\,12496-7650$, but it is not aligned to the HH\,54 flow.
Finally, the \emph{Spitzer} maps do not cover the hypothetical coordinates of HH\,52 and HH\,54 counterflows.
According to CO maps of the Cha II cloud (see e.\,g. Hayakawa et al.~\cite{haya}), both these coordinates are close to the
southern cloud edge. Thus, because of the very low medium density, it is possible that the ejected matter cannot produce any shock.

\begin{table}
\caption[]{ Lengths and dynamical age of the flows. \label{tau:tab}}
\begin{center}
\begin{tabular}{cccc}\\ [+5pt] \hline \hline
 knot   & Possible &  Length &  $\tau_{\rm dyn}$  \\
  ID	& Exciting Source   &   (pc)  &	 (10$^3$\,yrs) \\
\hline
HH\,52\,A   &  IRAS\,12416-7703      & 3.9  & 39  \\
HH\,54\,C   &  IRAS\,12416-7703      & 4.2  & 63         \\
HH\,54\,G   &  IRAS\,12416-7703      & 4.2  & 41    \\
HH\,54\,A   &  IRAS\,12416-7703      & 4.2  & 63    \\
HH\,53\,A-C &  Source\,2           & 0.1-0.2 & 0.3-0.8   \\
HH\,54\,Y   &  IRAS\,12500-7658      & 2.1  &  18  \\
HH\,54\,K-E &  IRAS\,12500-7658      & 2.2  &  43  \\
\hline

\hline \hline
\end{tabular}
\end{center}
\end{table}


\section{Conclusions}

We have presented a detailed multi-epoch (20 years baseline) kinematical investigation of HH\,52, 53, and 54 at optical and
near-IR wavelengths, along with medium (optical) and high-resolution (NIR) spectroscopical analysis, probing the kinematical
and physical conditions of the gas along the flows. We investigated flux and velocity variability with time of the flows,
reporting the first detections of acceleration and deceleration of protostellar jets.
We find that there is a correlation, at least in some cases, between the flux and the velocity variabilities
observed along the flows. Finally, we discussed the possible flow exciting sources. The main results of this work are the following:

\begin{enumerate}

\item[-] The P.M. analysis reveals that there are three distinct outflows, partially overlapping. The first outflow (HH\,54 flow) is oriented NNE with a position angle of $\sim$22$^\circ$, delineated by the HH\,54 streamer, and includes some of the knots in the main body of HH\,54. The second (HH\,52 flow) follows an NE direction with a position angle of $\sim$55$^\circ$, grouping HH\,52 and part of HH\,53 and 54 knots.
The third (HH\,53 flow), delineated by the three brightest knots of HH\,53, is moving in an E-W direction.
The HH\,52 and 53 outflows partially overlap and seem to converge on the HH\,54 bow shock.

\item[-] With a few exceptions, the derived P.M.s have similar values in the the optical filters (H$\alpha$, [\ion{S}{ii}])
ranging between 0.007\,$\arcsec$yr$^{-1}$ and $\sim$0.097\,$\arcsec$yr$^{-1}$, corresponding to a tangential velocity between 6 and 82\,km\,s$^{-1}$ at a distance of 180\,pc. On average the H$_2$ P.M.s have lower values (from 0.007 to 0.041,$\arcsec$yr$^{-1}$).

\item[-] Measured radial velocities of the atomic components (both optical and NIR) range between -20 and -120 km\,s$^{-1}$
with variations similar to the P.M.s. Spatial velocities of the knots vary from 50\,km\,s$^{-1}$ to 120\,km\,s$^{-1}$.
The inclinations of the three flows are 58$\pm$3$^\circ$, 84$\pm$2$^\circ$, and 67$\pm$3$^\circ$ (HH\,52, HH\,53, and HH\,54, respectively).

\item[-] In 20 years, about 60\% of the observed knots in the flows show flux variability. 
Moreover, our set of observations indicates acceleration and deceleration in several measurements,
but just some of them seem to be reliable. These knots are
working surfaces or interacting knots. In this case
a relevant flux variability is observed as well. We argue that both flux and velocity variability are related and that part of or
all the kinetic energy lost by the decelerating knots is successively radiated.
However, the complex proper motion variations may also indicate the combined effects of photometric and morphological changes, occurring on timescales comparable to the cooling time in the post-shock gas, combined with the detection of collisions of fast, fragmented ejecta overrunning slower moving fragmented ejecta.

\item[-] The physical parameters from the diagnostic are quite homogeneous along and among the three flows. The analysis indicates the presence of
very light ($N_{\rm H}\sim$10$^3$\,cm$^{-3}$), ionised (X$_{\rm e}\sim$0.2-0.6), and hot ($T_{\rm e}\sim$14\,000-26\,000\,K) jets,
impacting against a denser medium.

\item[-] The measured mass flux rates of the knots span from 10$^{-9}$\,$M_{\sun}$\,yr$^{-1}$ to some 10$^{-7}$\,$M_{\sun}$\,yr$^{-1}$,
where the highest values are detected in the HH\,52 flow.

\item[-] Several knots are deflected, at least in the HH\,52 flow. At least for a couple of them (HH\,54 G and G0), the deflection originates
in the collision of the two (what we call a bullet-target deflection mechanism). However, for the more massive parts of the flow, the deflection
is more likely the result of the flow collision with a denser cloud or clump.

\item[-] We investigated the possible driving sources of the flows,
comparing knot P.M.s and mass flux rates with the position, Class and bolometric luminosity of the source candidates.
We indicated three possible candidates, $IRAS\,12416-7703$, Source\,2, and $IRAS\,12500-7658$ for HH\,52, HH\,53, and HH\,54, respectively.
Only $IRAS\,12500-7658$ is, however, a reliable candidate.
\end{enumerate}

\begin{acknowledgements}
      We thank Reimhard Mundt for putting the first optical imaging and spectroscopy at our disposal.
      We are grateful to Juan Alcal\'{a} and Loredana Spezzi for fruitful discussions and for
      providing us with photometric data of the YSOs candidates. 
      Finally we would like to thank the referee, John Bally, for his helpful comments, which really improved the manuscript.
      The present work was supported in part by the European Community's Marie Curie Actions-Human
      Resource and Mobility within the JETSET (Jet Simulations, Experiments and Theory) network under
      contract MRTN-CT-2004 005592. 
      This publication has made use of data from the ``From Molecular Cores to Planet-forming Disks'' (c2d) Legacy project.
      This research has also made use of NASA's Astrophysics Data System Bibliographic Services and the SIMBAD database, operated
      at the CDS, Strasbourg, France, and the 2MASS data, obtained as part of the Two Micron All Sky Survey,
      a joint project of the University of Massachusetts and the Infrared Processing and Analysis Center/California Institute of Technology,
      funded by the National Aeronautics and Space Administration and the National Science Foundation.
\end{acknowledgements}

\Online


\begin{appendix}

\section{Morphology}
\label{appendixA:sec}

\subsection{HH\,52 - 53}
\label{appendix_HH52:sec}

\begin{figure*}
 \centering
\fbox{\includegraphics [width=12.1 cm] {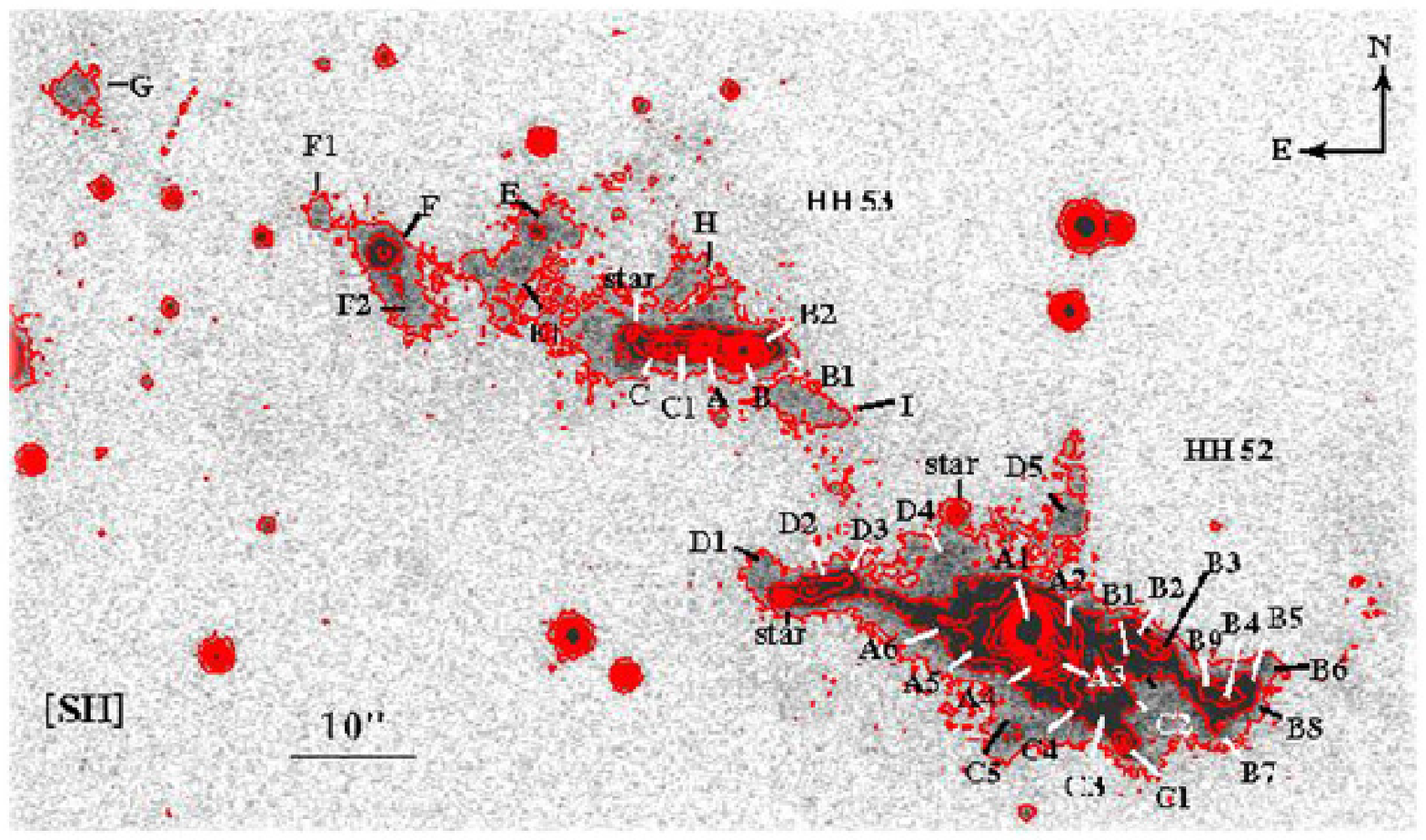}}\\
\fbox{\includegraphics [width=12.1 cm] {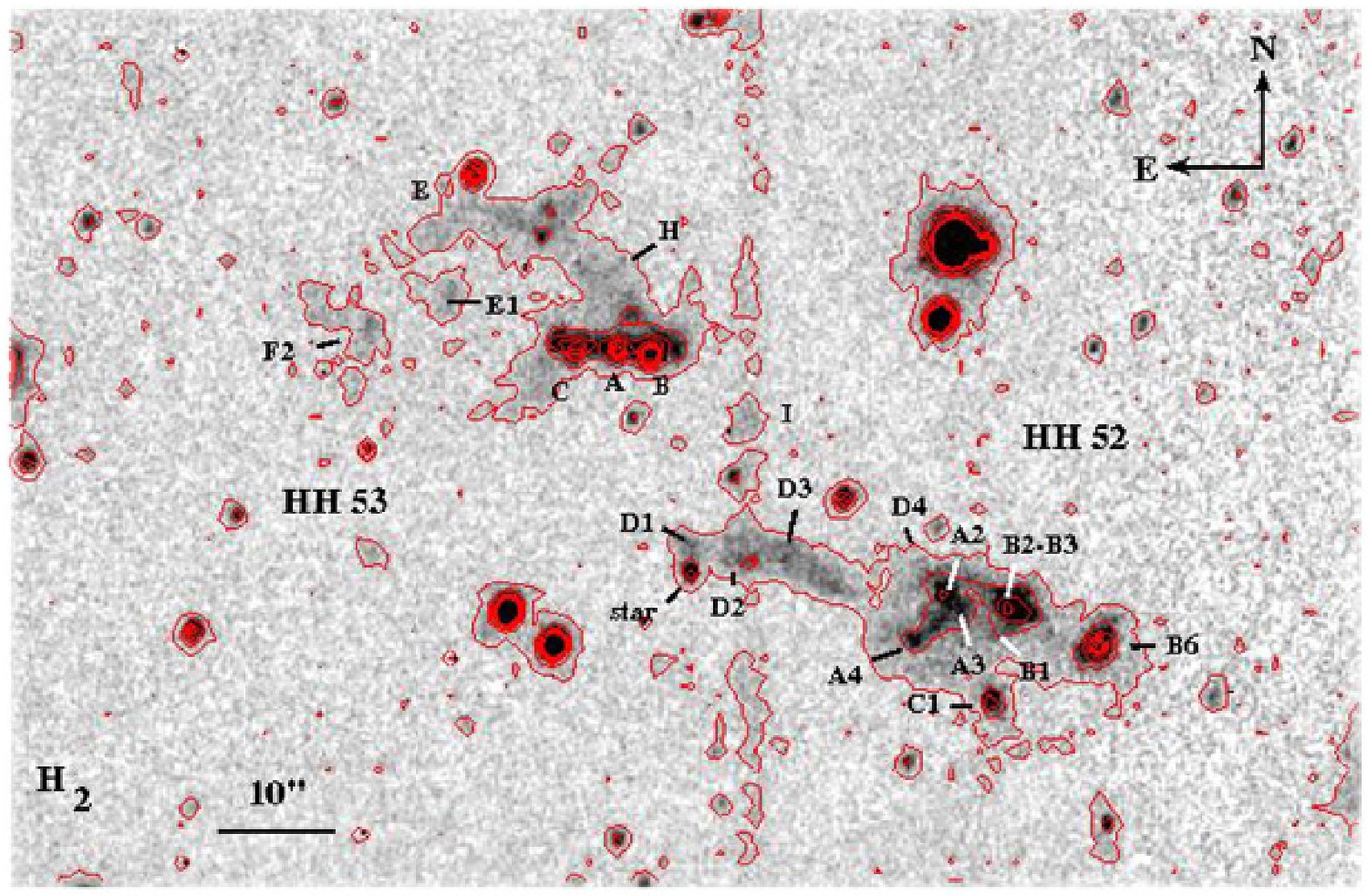}}\\
   \caption{ HH\,52 and HH\,53 regions with [\ion{S}{ii}] (EMMI 2006) ({\bf top}), and H$_2$ (SofI 1999) ({\bf bottom}) filters.
   The labels indicate the position of the knots, including the newly detected ones.
   The contour levels of the [\ion{S}{ii}] are 3, 10, 20, 30, 50, 60, 80, 90, 100, 150, 200$\times$ $\sigma$
   ($\sim$~4~$\times$~10$^{-17}$~erg~s$^{-1}$~cm$^{-2}$~arcsec$^{-2}$). H$_2$ levels are 3, 10, 20, 30, 40, 50, 60$\times$ $\sigma$
   ($\sim$10$^{-16}$~erg~s$^{-1}$~cm$^{-2}$~arcsec$^{-2}$).
\label{HH52-3-fig:fig}}
\end{figure*}

Our high resolution images allow us to disentangle the structure
of the studied HH objects. Figure~\ref{HH52-3-4Hfig:fig} (right panel) and \ref{HH52-3-fig:fig} show the
HH\,52 and 53 region in the three different filters. In H$\alpha$ and
[\ion{S}{ii}] HH\,52 exhibits a
roughly bow shape, where knots labelled A can be identified as the
head of the shock (nucleus). The brightest knot (A1) is surrounded by
several smaller structures, with knots A3 and A4 placed behind, while
A5 and A6 are in front, elongated towards a group of
separated knots (D1, D2, and D3) roughly E-W aligned.
A faint emission (knot D4) is placed just in front of the nucleus.
Groups of knots B and C constitute the right and left wings of the bow,
respectively. The right wing appears particularly fragmented with
knots B1-B3 representing a first condensation close to the
nucleus. B4-B9 are located behind.

The H$_2$ emission (Fig.~\ref{HH52-3-fig:fig}, bottom) shows a more amorphous shape.
A semi-ring shaped region, roughly coincident with knots A2, A3, and A4, is located just behind the nucleus.
Three more bright spots identify the two wings. The first emission coincides with C1 on the left wing, the second, on the right wing,
extends to an area corresponding to B2 and B3 in the optical, and the third to B5 and B6 on the right wing as well.
Finally, fainter and diffuse emission is detected on the D1, D2 and D3 knots.

In addition to the three HH\,53 knots already known (A, B, and C), we detect several more emissions coming from this region.
Three more knots, mainly emitting in the H$\alpha$ line (Fig.~\ref{HH52-3-4Hfig:fig}), are found along the main flow, forming an elongated S-shape chain of E-W
knots. We label C1 the emission between C and A, and B1 and B2 those close to B. These objects are not detected in our
H$_2$ images and two of them (knots C1 and B1) are just barely visible on the [\ion{S}{ii}] images.
More diffuse emission, detected in all three filters, appears superimposed on the HH\,53 main flow, connected
to the HH\,52 streamer. HH\,53\,I lies SW, towards HH\,52, while H and E are situated NE in the opposite direction.
Two brighter structures are located farther NE along the streamer. Group F, with a jet-like  appearance in the optical filters,
is approximately elongated towards NE, also visible in H$_2$. Knot G, only visible in the optical, is approximatively located
at the origin of the continuous emission of the streamer.


\subsection{HH\,54}
\label{appendix_HH54:sec}

\begin{figure*}
 \centering
\fbox{\includegraphics [width=8.05 cm] {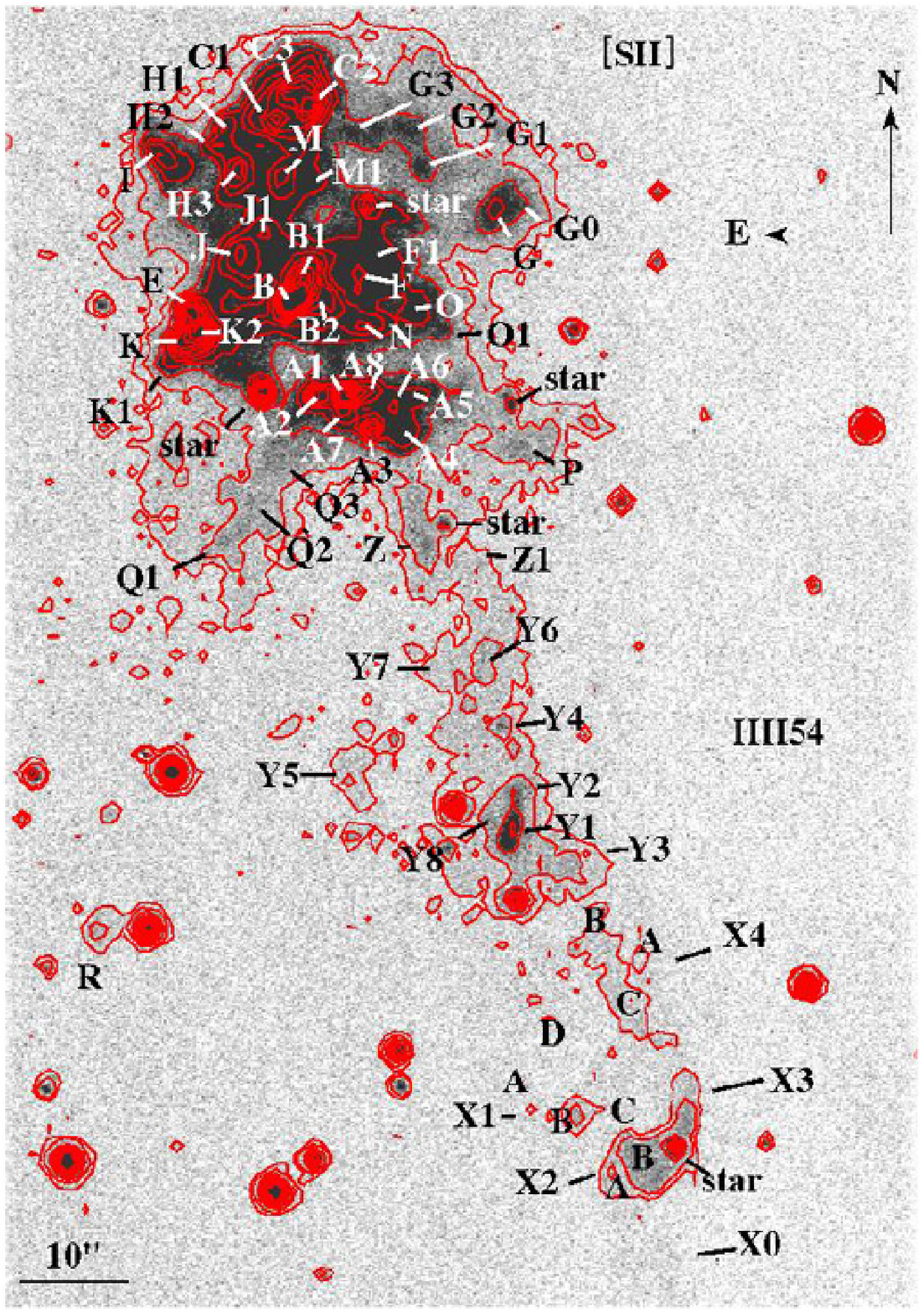}}
\fbox{\includegraphics [width=7.73 cm] {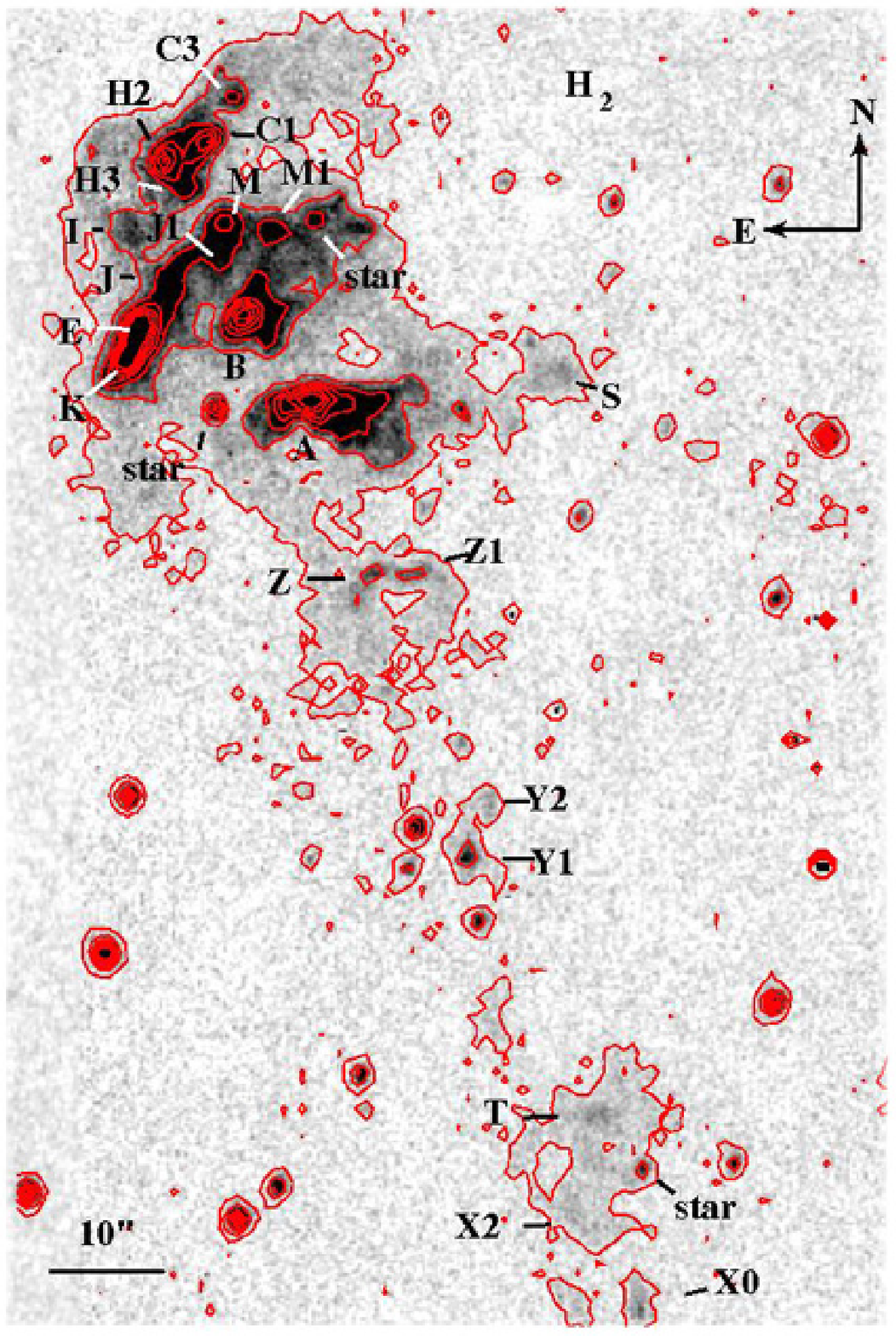}}\\
   \caption{ HH\,54 regions with [\ion{S}{ii}] (EMMI 2006) ({\bf left}) and H$_2$ (SofI 1999) ({\bf right}) filters. The labels indicate the position of the knots, including the newly detected ones. [\ion{S}{ii}] contour levels are 3, 10, 20, 30, 40, 50, 60, 70, 80, 90, 100$\times$
   the standard deviation to the mean background ($\sim$~4~$\times$~10$^{-17}$~erg~s$^{-1}$~cm$^{-2}$~arcsec$^{-2}$).
    H$_2$ contour levels are 3, 10, 20, 30, 40, 50, 60$\times$ the standard deviation to the mean background ($\sim$10$^{-16}$~erg~s$^{-1}$~cm$^{-2}$~arcsec$^{-2}$).
\label{HH54-fig:fig}}
\end{figure*}

In Figures~\ref{HH52-3-4Hfig:fig} (left panel) and \ref{HH54-fig:fig} we show the
HH\,54 images and contours in H$\alpha$, [\ion{S}{ii}] (left panel
of Fig.~\ref{HH54-fig:fig}) and H$_2$ (right panel of Fig.~\ref{HH54-fig:fig}).
The morphology of the object
appears extremely complex. Our images reveal several
substructures. For the sake of simplicity, we divide the
description of the region in two parts, the streamer, composed
of groups X, Y, and Z, and the main body of HH\,54, made up
of the remaining knots.

HH\,54 streamer shows an uncommon shape of a double helix, clearly
visible in H$\alpha$. The brightest helix is also well delineated
in [\ion{S}{ii}], appearing as a wiggling jet fragmented in
several knots. The morphology of the streamer closely resembles
that of the HH\,46/47 outflow (see e.\,g. Eisl\"{o}ffel \& Mundt~\cite{eis94}, Heathcote et al.\cite{heathcote}),
except for the presence of a second
faint helix. The base of the
streamer is X0, with knots X2 and X3 delineating the first bend of
the bright helix. The path follows with X4. Y1 and Y2 delineate
the second bend of the jet, that continues with Y4. The
faint helix is delineated by X1, Y3, Y8, Y7 and Y6. Both the
bright and faint structures end in two shocks (Z and Z1,
respectively), that barely appear as a bow shock feature in H$_2$ (Fig.~\ref{HH54-fig:fig}, right panel).
Molecular hydrogen emission is also detected in knots X0, X2, Y1, and
Y2. A further knot (T) complementary to the ionic emission is
observed along the jet. Finally, some filaments (X4\,A, Y5, R), detected at
optical wavelengths, are located aside, detached from the main flow.

The morphology of the main body appears even more complicated than the streamer.
In addition to the original knots (from A to K), observed in previous papers (Schwartz~\cite{schwartz},
Schwartz \& Dopita~\cite{SD80}, Sandell et al.~\cite{sandell}), we detect more features (labelled L to S).
Moreover, several knots present sub-structures, labelled here with numbers.

Group A at the bottom appears as bow-shaped in the optical,
with a P.A. of about 75$\degr$. In H$_2$ the group exhibits emission only
in the northern part of the structure, only partially overlapping the atomic emission.
A fainter arc-shaped emission (group Q), mainly visible in H$\alpha$, is located east.
Knots P (detected in the optical) and S (detected in H$_2$) are located west, and together
with knots O (farther north, visible in the optical) form the farthest point of the HH\,52 streamer.

Towards north, more groups are detected. From east to west, almost
coincident in all three wavelengths, we detect groups K and E, slightly elongated northward,
then group B, and, finally group F, detected only in the optical.
Farther NE an arc-like structure is identified by groups J and M in both optical and NIR.

In the north-eastern region, knot I and group H appear as optical jets with P.A.s of $\sim$45$\degr$
and $\sim$20$\degr$, respectively.
An arc emission in H$_2$ is delineated ahead of I.

In the north-western region, we detect group G, composed of five subsequent knots only detected in the optical
(mainly in the H$\alpha$ filter, see also Fig.~\ref{HH52-3-4Hfig:fig}), with a P.A. of $\sim$55$\degr$.
At the end of this flow another group of knots (C) is observed, with C1 and C2 already known, and the new knot C3
located on a straight line connecting knots G and C2.

It is worth to note that the identification in the NIR of knots C accomplished by Gredel~(\cite{gredel})
(and then used in Caratti o Garatti et al.~\cite{caratti06b} and in Giannini et al.~\cite{giannini06}) is erroneous.
Indeed the knots that were labelled in H$_2$ as C1 and C2 have a similar appearance in the optical, but are situated about 6$\arcsec$ SW.
As a consequence, the knot previously known as C1 in the NIR coincides with knot H2 in the optical, while C2 roughly corresponds
to C1 in the optical (see Fig.~\ref{HH52-3-4Hfig:fig} and \ref{HH54-fig:fig}).
In this paper we use this nomenclature to avoid further confusion.

\end{appendix}

\begin{appendix}
\section{Proper motions}
\label{appendixB:sec}
\subsection{HH\,52 - 53}

Knot position angles along HH\,52 flow have a quite wide spread in their values (30$\degr$-80$\degr$).
The bulk of HH\,52 and other knots, as HH\,53 C1, I, and H, have a P.A. of $\sim$50$\degr$.
On the other hand HH\,52\,A3, A4 and a few more knots as HH\,53\,F and G, proceed towards HH\,54 with different P.A.s between
60$\degr$ and 80$\degr$.
A completely different behaviour is found for the features located on HH\,52 wings in the rear of the flow. Although following the bulk of the
flow, they show a high degree of variability during the 20 years of observations, as reported in Appendix~\ref{appendixC:sec}.
Such a variability, both due to local instabilities inside the flow and fast moving shocks overtaking HH\,52 from behind,
makes their identification in different epochs and a correct evaluation of the P.M.s and P.A.s sometimes difficult
(see also Fig~\ref{HH52-3-pms:fig} and \ref{HH54-pms:fig}).

In Fig.~\ref{HH52-H2-pms:fig} H$_2$ proper motions are reported (see also Tab.~\ref{PM52_h2:tab} and \ref{PM54_h2:tab}).
Significant measurements in HH\,52 were possible only for the three bright spots located in the wings.
B2-B3 and C1 follow the flow as in the optical, with similar tangential velocities and position angles.

HH\,53\,A, B and C, in the HH\,53 outflow, exhibit small P.M.s with P.A.s around 270$\degr$ also in the NIR.
Here, however, the measurements are less significant due to the larger errors.

\subsection{HH\,54}

Proper motions in the HH\,54 streamer range between $\sim$0.01 and 0.07$\arcsec$\,yr$^{-1}$
(i.\,e. $v_\mathrm{tan}$ 8 and 61\,km\,s$^{-1}$), where the highest velocities (between $\sim$30 and 60\,km\,s$^{-1}$)
are detected in the inner part of the flow and exhibit alternating P.A.s (322$\degr$-69$\degr$), likely due to precession,
with directions that oscillate around an average value of $\sim$20$\degr$.

Some features, located on the edge of the streamer, are not collimated and are drifting apart (P.A.$\sim$90$\degr$) at a lower velocity.
At the tip of the streamer group Z appears almost stationary in H$\alpha$ ($v_\mathrm{tan}\sim$10\,km\,s$^{-1}$), while in [\ion{S}{ii}]
has a higher value of $\sim$30\,km\,s$^{-1}$.

Moving towards NE we observe a decrease in the tangential velocities from groups Q ($\sim$40\,km\,s$^{-1}$) to K and E
($\sim$20 and 15\,km\,s$^{-1}$, respectively). From the P.A.s and inclinations (see Sect.~\ref{inclination:sec})
we can reasonably associate these knots with the HH\,54 flow.

\begin{figure*}
 \centering
  \fbox{\includegraphics [height=8.8 cm] {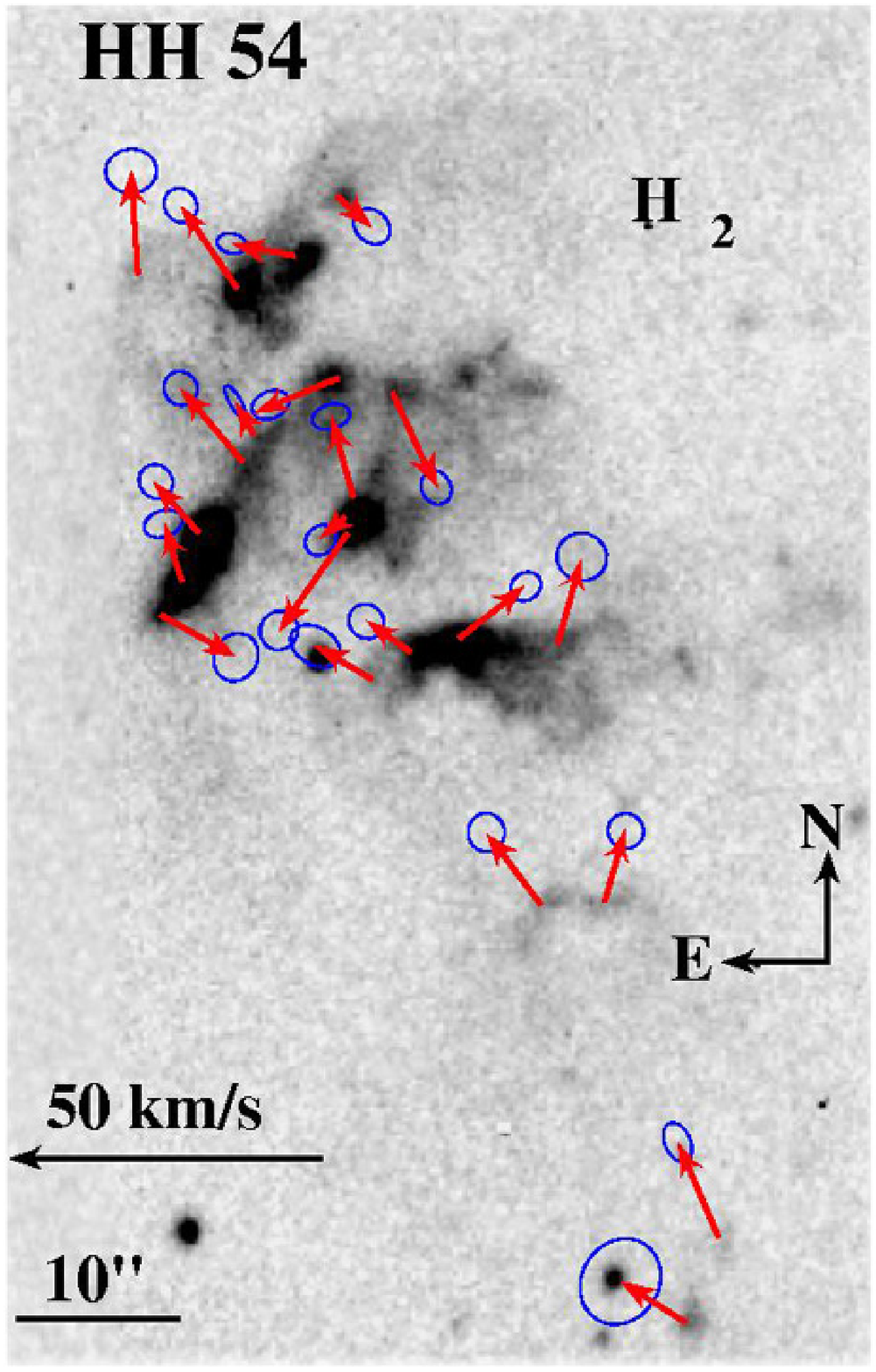}}
  \fbox{\includegraphics [height=6.1 cm] {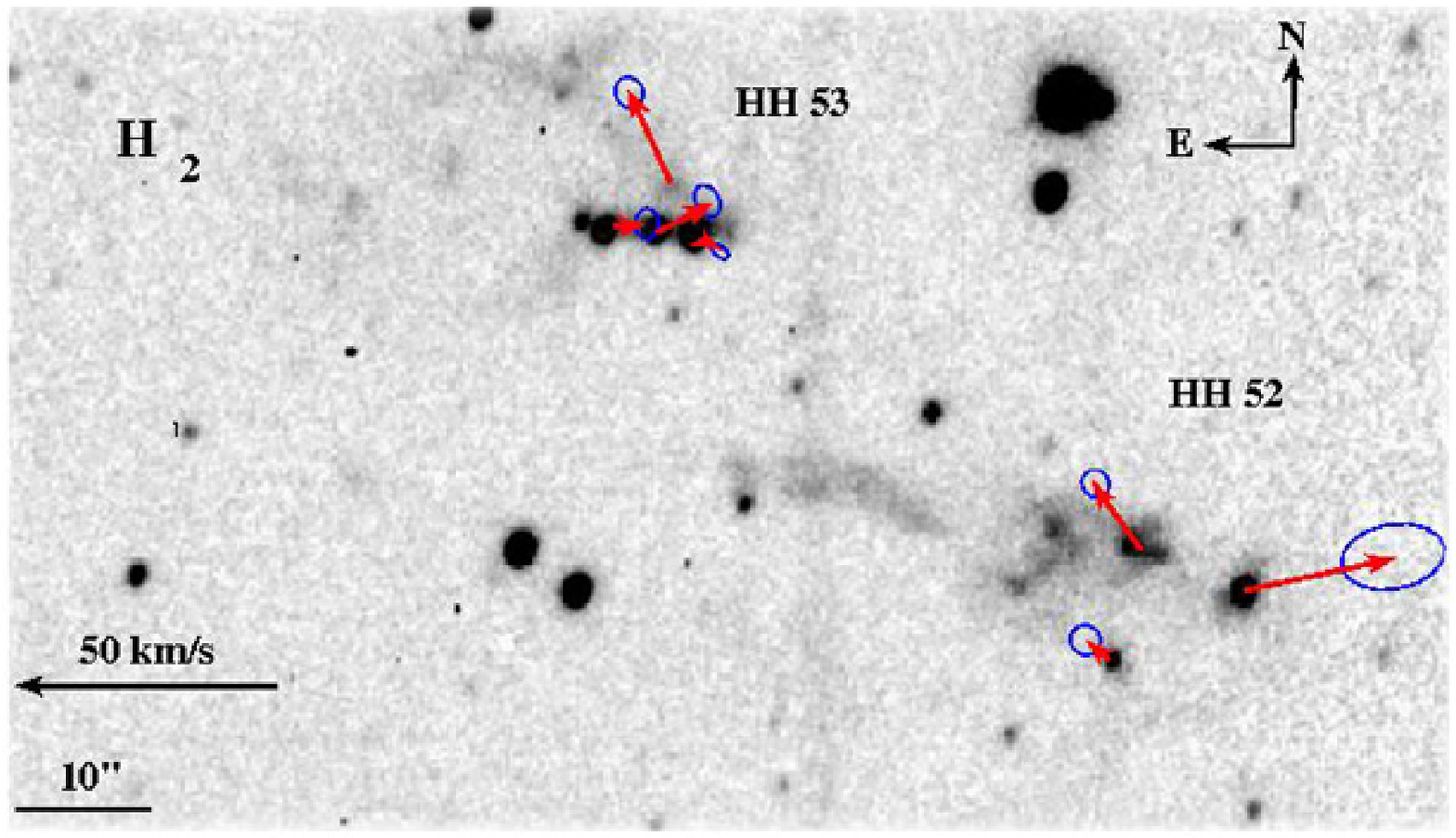}}
    \caption{ Flow charts with error bars of HH\,54 ({\bf left panel}), HH\,52 and 53 ({\bf right panel}) in H$_2$ filter.
    Proper motions and their error bars are indicated by arrows and ellipses, respectively.
\label{HH52-H2-pms:fig}}
\end{figure*}

Group A appears as an expanding bow shock, towards ENE with an average direction of about 80$\degr$ and a mean $v_\mathrm{tan}$ of $\sim$30\,km\,s$^{-1}$. Knots A1, A2, A6 and A8 exhibit some variability, as reported in Appendix~\ref{appendixC:sec}.
Because of the morphology, the direction, and the average inclination (see Sect.~\ref{inclination:sec}), the structure appears to be connected to knot P and to the HH\,52 flow. Indeed part of the flow along the HH\,52 streamer curves and shocks this region, producing group O as well (see also Fig.~\ref{slit-positions:fig}).
H$_2$ emission in HH\,54\,A (Fig.~\ref{HH54-fig:fig}) appears as an expanding arc, and its motion ($v_\mathrm{tan}\sim$30\,km\,s$^{-1}$) has a different orientation depending on the position along the structure (Fig.~\ref{HH52-H2-pms:fig}). Proceeding from east to west, we separately measured four regions, indicated in Tab.~\ref{PM54_h2:tab} as A$_1$, A$_2$, A$_3$, and A$_4$. The directions of the motions have different P.A.s with respect to the optical counterparts and it is not clear if the molecular emission comes from material swept up by the
bow shock on one side (see e.\,g. HH\,219, Caratti o Garatti et al.~\cite{caratti04}) or from the HH\,54 streamer.

The faintest component of the HH\,52 streamer follows a straight trajectory (of $\sim$55$\degr$) and impacts ahead,
producing groups HH\,54\,G and then C.
We observe a sudden deceleration of the flow along the HH\,52 streamer (from $\sim$65 to 30\,km\,s$^{-1}$)
as it collides with the leading material, that is partially deflected sideways (P.A.$\sim$80$\degr$).
G0 is part of such a fast flow, that shocks at the end of a slow moving component (knot G), changing its direction (see also Sect.~\ref{variability:sec}).
This dynamic is also visible in knots G1 and G3, that are not aligned with the flow, moving with a P.A. of $\sim$80$\degr$ and $\sim$65$\degr$,
respectively.
Group C ahead of G is proceeding in the same direction at similar tangential velocities (25-30\,km\,s$^{-1}$).
Surprisingly, knot C3, at the tip of the flow, slightly moves in the opposite direction in both optical and NIR filters
(P.A.$\sim$250$\degr$ and $v_\mathrm{tan}\sim$15\,km\,s$^{-1}$). This behaviour is observed also in knot K1, on the opposite side of the HH object,
showing nearly a reverse motion with respect to the main flows.

Velocities of knots M are quite low (about 20\,km\,s$^{-1}$) with a position angle around 80$\degr$. The direction and the velocities
suggest that they could be part of the same outflow, generated by the same deflection mechanism.

Groups H, I and J, on the northeast side, are moving NE (P.A.$\sim$45$\degr$) with a $v_\mathrm{tan}$ between 10 and 50\,km\,s$^{-1}$.

Group N, in the central region of the main body, could be part of the HH\,54 flow, with a $v_\mathrm{tan}\sim$20-25\,km\,s$^{-1}$ and a P.A.$\sim$320$\degr$.

It is difficult to determine the membership of several knots (as B or F groups) in the central part of the HH object.
The bulk of knot B possibly moves slightly towards ESE, while B1, few arcseconds north, seems to proceed in the opposite direction.
Also F and F1 show a similar behaviour.

The molecular component of knot B presents distinct internal motions. We detect three components, reported in
Tab.~\ref{PM54_h2:tab} (B$_1$, B$_2$, B$_3$, from north to south). B$_1$ appears as a bright arc in the high spatial resolution ISAAC images and it is moving towards NNE ($\sim$20$\degr$). B$_2$ and B$_3$ are proceeding towards ESE (130$\degr$-140$\degr$) as in the optical.
In this case we are possibly observing the motion of the two different outflows.

\begin{table*}
\caption[]{ Proper Motions in HH\,52 \& HH\,53 - H$\alpha$ \& [\ion{S}{ii}] - Parameters derived from the linear
fit \it{to be continued} \label{PM52:tab}}
\begin{center}
\begin{tabular}{ccccccccccc}
\hline\hline\\[-5pt]
Knot ID   &  \multicolumn{3}{c}{$\alpha$(2000.0)} & \multicolumn{3}{c}{$\delta$(2000.0)} &  P.M.  &  P.A.  &  $v_{tan}$  & Filter \\
          &   ($^{h}$ & $^{m}$ & $^{s}$) & ($\degr$ & $\arcmin$ & $\arcsec$)     & ($\arcsec$\,yr$^{-1}$) &  ($\degr$)  & (km\,s$^{-1}$) \\
\hline\\[-5pt]
HH\,52\,A1 &12&55&07.6&-76&57&52  & 0.055$\pm$0.002        &  42$\pm$2  & 48$\pm$1 & [\ion{S}{ii}] \\
           &12&55&07.6&-76&57&52  & 0.052$\pm$0.002        &  52$\pm$2  & 44$\pm$2 & H$\alpha$ \\
HH\,52\,A2 &12&55&06.7&-76&57&52  &    --                  &  -- & --  &  [\ion{S}{ii}],H$\alpha$ \\
HH\,52\,A3 &12&55&06.7&-76&57&55  &    --                  &  -- & --  &  [\ion{S}{ii}] \\
           &12&55&06.7&-76&57&55  & 0.047$\pm$0.004        &  70$\pm$4 & 40$\pm$3 & H$\alpha$ \\
HH\,52\,A4 &12&55&07.4&-76&57&54  & 0.059$\pm$0.009        &  70$\pm$9  & 50$\pm$8 & [\ion{S}{ii}] \\
           &12&55&07.5&-76&57&55  & 0.048$\pm$0.002        &  71$\pm$3  & 40$\pm$2 & H$\alpha$ \\
HH\,52\,A5 &12&55&09.1&-76&57&54  & 0.028$\pm$0.002$^{a}$  &  62$\pm$3  & 24$\pm$1$^{a}$ & [\ion{S}{ii}] \\
           &12&55&09.1&-76&57&54  & 0.056$\pm$0.003$^{a}$  &  67$\pm$3 & 48$\pm$2$^{a}$ & H$\alpha$ \\
HH\,52\,A6 &12&55&09.5&-76&57&52  &    --                  &  --       & --  & [\ion{S}{ii}] \\
           &12&55&09.5&-76&57&52  & 0.035$\pm$0.003        &  78$\pm$4  & 29$\pm$3 & H$\alpha$ \\
HH\,52\,B1 &12&55&05.6&-76&57&54  & 0.046$\pm$0.002$^{a}$  &  116$\pm$3 & 39$\pm$2$^{a}$  & [\ion{S}{ii}] \\
           &12&55&05.2&-76&57&55  & 0.031$\pm$0.004$^{a}$  &  111$\pm$6   & 26$\pm$3$^{a}$ & H$\alpha$ \\
HH\,52\,B2 &12&55&05.3&-76&57&53  & 0.053$\pm$0.002$^{a}$  &  2$\pm$3   & 45$\pm$2$^{a}$ & [\ion{S}{ii}] \\
           &12&55&05.2&-76&57&53  & 0.062$\pm$0.002$^{a}$  &  1$\pm$3   & 53$\pm$2$^{a}$& H$\alpha$ \\
HH\,52\,B3 &12&55&04.6&-76&57&54  & 0.038$\pm$0.002$^{a}$  &  71$\pm$3  & 32$\pm$2$^{a}$  & [\ion{S}{ii}] \\
           &12&55&04.6&-76&57&54  & 0.044$\pm$0.003        &  80$\pm$3  & 37$\pm$2        & H$\alpha$ \\
HH\,52\,B4 &12&55&03.0&-76&57&57  & 0.017$\pm$0.002$^{a}$  &  63$\pm$7  & 14$\pm$2$^{a}$  & [\ion{S}{ii}] \\
           &12&55&03.0&-76&57&58  & --  &  --  & -- & H$\alpha$ \\
HH\,52\,B5 &12&55&02.9&-76&57&58  & 0.011$\pm$0.002$^{a}$  & 309$\pm$10 & 10$\pm$2$^{a}$  & [\ion{S}{ii}] \\
           &12&55&02.7&-76&57&58  & 0.097$\pm$0.003        & 195$\pm$2  & 82$\pm$2         & H$\alpha$ \\
HH\,52\,B6 &12&55&02.2&-76&57&55  & 0.034$\pm$0.002        & 257$\pm$3  & 29$\pm$1        & [\ion{S}{ii}] \\
           &12&55&02.1&-76&57&55  & 0.033$\pm$0.001        & 249$\pm$2  & 28$\pm$1        & H$\alpha$ \\
HH\,52\,B7 &12&55&03.4&-76&58&00  & 0.033$\pm$0.003        & 157$\pm$6  & 28$\pm$3        & [\ion{S}{ii}] \\
           &12&55&03.2&-76&58&00  & 0.062$\pm$0.003        & 152$\pm$3  & 52$\pm$2        & H$\alpha$ \\
HH\,52\,B8 &12&55&02.4&-76&57&57  & 0.019$\pm$0.005$^{a}$  & 288$\pm$11 & 16$\pm$4$^{a}$  & [\ion{S}{ii}] \\
           &12&55&02.5&-76&57&56  & --                     &  --         & -- & H$\alpha$ \\
HH\,52\,B9 &12&55&03.6&-76&57&57  & 0.045$\pm$0.011        &  82$\pm$15  & 38$\pm$9        & H$\alpha$ \\
HH\,52\,C1 &12&55&05.4&-76&58&01  & 0.037$\pm$0.002$^{a}$  &  229$\pm$3  & 31$\pm$1$^{a}$& [\ion{S}{ii}] \\
           &12&55&05.4&-76&58&01  & 0.009$\pm$0.002        &  47$\pm$14  & 8$\pm$2       & H$\alpha$ \\
HH\,52\,C2 &12&55&05.5&-76&57&58  & 0.056$\pm$0.002$^{a}$  &  38$\pm$2  & 48$\pm$2$^{a}$ & [\ion{S}{ii}] \\
           &12&55&05.4&-76&57&58  & 0.046$\pm$0.004        &  56$\pm$5  & 39$\pm$3       & H$\alpha$ \\
HH\,52\,C3 &12&55&06.2&-76&57&59  &    --                  &  -- & --  &  [\ion{S}{ii}],H$\alpha$ \\
HH\,52\,C4 &12&55&06.7&-76&57&58  & 0.056$\pm$0.002$^{a}$  &   4$\pm$2  & 47$\pm$2$^{a}$& [\ion{S}{ii}] \\
           &12&55&06.7&-76&57&58  & 0.068$\pm$0.003$^{a}$  &   4$\pm$3  & 58$\pm$2$^{a}$ & H$\alpha$ \\
HH\,52\,C5 &12&55&08.0&-76&58&01  &    --                  &  --       & --  & [\ion{S}{ii}] \\
HH\,52\,C6 &12&55&05.3&-76&58&01  &    --                  &  --       & --  & [\ion{S}{ii}], H$\alpha$ \\
HH\,52\,D1 &12&55&13.9&-76&57&47  & --  &  -- & -- &[\ion{S}{ii}] \\
           &12&55&13.9&-76&57&47  & 0.060$\pm$0.006        & 35$\pm$5  & 50$\pm$5 & H$\alpha$ \\
HH\,52\,D2 &12&55&12.8&-76&57&49  & 0.046$\pm$0.002        & 32$\pm$2  & 38$\pm$2 &[\ion{S}{ii}] \\
           &12&55&12.7&-76&57&49  & 0.049$\pm$0.002        & 23$\pm$3  & 41$\pm$2 & H$\alpha$ \\
HH\,52\,D3 &12&55&12.1&-76&57&49  & 0.046$\pm$0.002        & 49$\pm$2  & 39$\pm$2 & [\ion{S}{ii}] \\
           &12&55&12.0&-76&57&48  & 0.047$\pm$0.003        & 22$\pm$3  & 40$\pm$2 & H$\alpha$ \\
HH\,52\,D4 &12&55&10.0&-76&57&46  & 0.044$\pm$0.002$^{a}$  & 47$\pm$2  & 37$\pm$1$^{a}$ & [\ion{S}{ii}] \\
           &12&55&09.7&-76&57&47  & 0.043$\pm$0.003        & 57$\pm$4  & 36$\pm$3 & H$\alpha$ \\
HH\,52\,D5 &12&55&06.7&-76&57&43  & 0.036$\pm$0.002$^{a}$  & 53$\pm$3  & 30$\pm$2$^{a}$ & [\ion{S}{ii}] \\
           &12&55&06.6&-76&57&44  & 0.040$\pm$0.003        & 24$\pm$4  & 34$\pm$3 & H$\alpha$ \\
HH\,53\,A  &12&55&15.4&-76&57&30  & 0.019$\pm$0.002$^{a}$  & 303$\pm$5 & 16$\pm$1$^{a}$  & [\ion{S}{ii}] \\
           &12&55&15.4&-76&57&30  & 0.021$\pm$0.003        & 275$\pm$6 & 18$\pm$2   & H$\alpha$ \\
HH\,53\,B  &12&55&14.5&-76&57&31  & 0.007$\pm$0.002$^{a}$  & 235$\pm$14 &  6$\pm$2$^{a}$  & [\ion{S}{ii}] \\
           &12&55&14.5&-76&57&31  & 0.013$\pm$0.003        & 252$\pm$11  & 11$\pm$2   & H$\alpha$ \\
HH\,53\,B1 &12&55&13.6&-76&57&31  & 0.081$\pm$0.002        & 262$\pm$1   & 69$\pm$2   & H$\alpha$ \\
HH\,53\,B2 &12&55&14.1&-76&57&31  & --  & -- &  --  & [\ion{S}{ii}],H$\alpha$ \\
HH\,53\,C  &12&55&16.5&-76&57&30  & 0.013$\pm$0.002  & 239$\pm$9 &  11$\pm$2  & [\ion{S}{ii}] \\
           &12&55&16.5&-76&57&30  & 0.014$\pm$0.003  & 232$\pm$12  & 12$\pm$3   & H$\alpha$ \\
HH\,53\,C1 &12&55&16.1&-76&57&30  & 0.060$\pm$0.006  & 58$\pm$6 &  50$\pm$5  & H$\alpha$ \\
\hline \hline
\end{tabular}
\end{center}
Notes:\\ $^{a}$ Mean value, variable in velocity, see text and Table~\ref{acc:tab} \\
$^{b}$ Below 3$\sigma$.\\
\end{table*}

\addtocounter{table}{-1}
\begin{table*}
\caption[]{ Proper Motions in HH\,52 \& HH\,53 - H$\alpha$ \& [\ion{S}{ii}]  - Parameters derived from the linear
fit \it{continued} }
\begin{center}
\begin{tabular}{ccccccccccc}
\hline\hline\\[-5pt]
Knot ID   &  \multicolumn{3}{c}{$\alpha$(2000.0)} & \multicolumn{3}{c}{$\delta$(2000.0)} &  P.M.  &  P.A.  &  $v_{tan}$  & Filter \\
          &   ($^{h}$ & $^{m}$ & $^{s}$) & ($\degr$ & $\arcmin$ & $\arcsec$)     & ($\arcsec$\,yr$^{-1}$) &  ($\degr$)  & (km\,s$^{-1}$) \\
\hline\\[-5pt]
HH\,53\,E  &12&55&19.4&-76&57&21  & 0.016$\pm$0.002  & 240$\pm$6 & 14$\pm$2  & [\ion{S}{ii}] \\
           &12&55&19.3&-76&57&21  & --  & -- & --   & H$\alpha$ \\
HH\,53\,E1 &12&55&19.9&-76&57&24  & 0.052$\pm$0.004  & 63$\pm$4   & 44$\pm$3  & [\ion{S}{ii}] \\
           &12&55&19.8&-76&57&24  & 0.049$\pm$0.003  & 97$\pm$3   & 41$\pm$3   & H$\alpha$ \\
HH\,53\,F  &12&55&23.0&-76&57&23  & 0.031$\pm$0.002  & 64$\pm$3   & 26$\pm$1  & [\ion{S}{ii}] \\
           &12&55&23.0&-76&57&23  & 0.049$\pm$0.003  & 66$\pm$3   & 41$\pm$3   & H$\alpha$ \\
HH\,53\,F1 &12&55&24.6&-76&57&20  & 0.079$\pm$0.002$^{a}$  & 64$\pm$2  & 67$\pm$2$^{a}$  & [\ion{S}{ii}] \\
           &12&55&24.5&-76&57&20  & 0.049$\pm$0.003$^{a}$  & 89$\pm$3  & 42$\pm$2$^{a}$   & H$\alpha$ \\
HH\,53\,F2 &12&55&22.1&-76&57&28  & 0.051$\pm$0.008        & 52$\pm$8 & 43$\pm$6   & H$\alpha$ \\
HH\,53\,G  &12&55&30.3&-76&57&11  & 0.063$\pm$0.002$^{a}$  & 67$\pm$2  & 53$\pm$2$^{a}$   & [\ion{S}{ii}] \\
           &12&55&30.3&-76&57&11  & 0.037$\pm$0.003$^{a}$  & 61$\pm$4  & 31$\pm$2$^{a}$   & H$\alpha$ \\
HH\,53\,H  &12&55&15.4&-76&57&25  & --  &  -- & --   & [\ion{S}{ii}] \\
           &12&55&15.3&-76&57&25  & 0.065$\pm$0.004  & 52$\pm$3 & 55$\pm$3        & H$\alpha$ \\
HH\,53\,I  &12&55&13.1&-76&57&34  & 0.063$\pm$0.002  & 51$\pm$2 & 53$\pm$2  & [\ion{S}{ii}] \\
           &12&55&12.2&-76&57&36  & 0.066$\pm$0.003  & 53$\pm$2 & 56$\pm$2   & H$\alpha$ \\
\hline \hline
\end{tabular}
\end{center}
Notes:\\ $^{a}$ Mean value, variable in velocity, see text and Table~\ref{acc:tab} \\
\end{table*}


\begin{table*}
\caption[]{ Proper Motions in HH\,54 - H$\alpha$ \& [\ion{S}{ii}] - Parameters derived from the linear
fit \it{to be continued}
\label{PM54:tab}}
\begin{center}
\begin{tabular}{ccccccccccc}
\hline\hline\\[-5pt]
Knot ID   &  \multicolumn{3}{c}{$\alpha$(2000.0)} & \multicolumn{3}{c}{$\delta$(2000.0)} &  P.M.  &  P.A.  &  $v_{tan}$  & Filter \\
            &   ($^{h}$ & $^{m}$ & $^{s}$) & ($\degr$ & $\arcmin$ & $\arcsec$)     & ($\arcsec$\,yr$^{-1}$) &  ($\degr$)  & (km\,s$^{-1}$) \\
\hline\\[-5pt]
HH\,54\,A1 &12&55&49.4&-76&56&30  & 0.036$\pm$0.004  & 70$\pm$6   & 31$\pm$3   & [\ion{S}{ii}] \\
           &12&55&49.4&-76&56&30  & 0.035$\pm$0.002  & 85$\pm$2   & 29$\pm$2   & H$\alpha$ \\
HH\,54\,A2 &12&55&50.2&-76&56&30  & 0.036$\pm$0.004  & 81$\pm$6   & 30$\pm$3   & [\ion{S}{ii}] \\
           &12&55&50.2&-76&56&30  & 0.034$\pm$0.002  & 73$\pm$3   & 29$\pm$2   & H$\alpha$ \\
HH\,54\,A3 &12&55&48.7&-76&56&33  & 0.013$\pm$0.004  & 166$\pm$17 & 11$\pm$3   & [\ion{S}{ii}] \\
           &12&55&48.7&-76&56&33  & 0.015$\pm$0.002  & 121$\pm$6  & 13$\pm$3   & H$\alpha$ \\
HH\,54\,A4 &12&55&47.7&-76&56&33  & 0.030$\pm$0.004  & 225$\pm$7  & 25$\pm$3   & [\ion{S}{ii}] \\
           &12&55&47.8&-76&56&34  & --  & -- & --    & H$\alpha$ \\
HH\,54\,A5 &12&55&47.4&-76&56&30  & 0.066$\pm$0.006  & 125$\pm$6  & 56$\pm$5   & [\ion{S}{ii}] \\
           &12&55&47.4&-76&56&31  & --  & -- & --   & H$\alpha$ \\
HH\,54\,A6 &12&55&48.1&-76&56&31  & 0.031$\pm$0.007  & 53$\pm$13  & 27$\pm$6   & [\ion{S}{ii}] \\
           &12&55&48.1&-76&56&31  & 0.025$\pm$0.005  & 40$\pm$11  & 21$\pm$4   & H$\alpha$ \\
HH\,54\,A7 &12&55&49.4&-76&56&32  & 0.017$\pm$0.005  & 88$\pm$27  & 14$\pm$4   & [\ion{S}{ii}] \\
           &12&55&49.4&-76&56&32  & 0.016$\pm$0.002  & 23$\pm$6   & 14$\pm$1   & H$\alpha$ \\
HH\,54\,A8 &12&55&48.7&-76&56&29  & 0.039$\pm$0.009  & 28$\pm$20  & 33$\pm$8   & [\ion{S}{ii}] \\
           &12&55&48.7&-76&56&29  & 0.034$\pm$0.002  & 66$\pm$3   & 29$\pm$2   & H$\alpha$ \\
HH\,54\,B  &12&55&50.9&-76&56&21  & 0.008$\pm$0.004$^{b}$  & 110$\pm$30 & 6$\pm$3$^{b}$   & [\ion{S}{ii}] \\
           &12&55&50.9&-76&56&21  & --  & -- & --   & H$\alpha$ \\
HH\,54\,B1 &12&55&50.6&-76&56&19  & 0.010$\pm$0.005$^{b}$ & 282$\pm$32 & 8$\pm$4$^{b}$   & [\ion{S}{ii}] \\
           &12&55&50.6&-76&56&19  & --  & -- & --   & H$\alpha$ \\
HH\,54\,B2 &12&55&49.8&-76&56&21  & --                    & -- & --   & [\ion{S}{ii}] \\
           &12&55&49.8&-76&56&22  & 0.017$\pm$0.002       & 313$\pm$7  & 15$\pm$2   & H$\alpha$ \\
HH\,54\,B3 &12&55&50.5&-76&56&24  & --                    & -- & --   & H$\alpha$ \\	
HH\,54\,C1 &12&55&51.3&-76&56&05  & 0.032$\pm$0.004       & 73$\pm$8   & 27$\pm$3   & [\ion{S}{ii}] \\
           &12&55&51.3&-76&56&05  & 0.017$\pm$0.002       & 67$\pm$5   & 14$\pm$1   & H$\alpha$ \\
HH\,54\,C2 &12&55&50.4&-76&56&03  & 0.041$\pm$0.004       & 42$\pm$5   & 34$\pm$3   & [\ion{S}{ii}] \\
           &12&55&50.4&-76&56&03  & 0.017$\pm$0.002       & 70$\pm$5   & 14$\pm$1   & H$\alpha$ \\
HH\,54\,C3 &12&55&51.0&-76&56&02  & 0.017$\pm$0.006$^{b}$ & 239$\pm$19 & 14$\pm$5$^{b}$   & [\ion{S}{ii}] \\
           &12&55&51.0&-76&56&02  & 0.016$\pm$0.002$^{a}$ & 267$\pm$6  & 13$\pm$2$^{a}$   & H$\alpha$ \\
HH\,54\,E  &12&55&53.7&-76&56&22  & 0.017$\pm$0.005       & 35$\pm$15 &14$\pm$4   & [\ion{S}{ii}] \\
           &12&55&53.7&-76&56&22  & 0.018$\pm$0.002       & 16$\pm$6  & 15$\pm$2   & H$\alpha$ \\
HH\,54\,F  &12&55&49.0&-76&56&19  & 0.018$\pm$0.003       & 125$\pm$11 & 15$\pm$3    & [\ion{S}{ii}] \\
           &12&55&49.0&-76&56&19  & 0.013$\pm$0.002       & 129$\pm$7 &  11$\pm$2   & H$\alpha$ \\
HH\,54\,F1 &12&55&48.4&-76&56&17  & 0.015$\pm$0.002$^{a}$ & 251$\pm$9 & 13$\pm$2$^{a}$   & H$\alpha$ \\
HH\,54\,G  &12&55&45.4&-76&56&13  & 0.048$\pm$0.004$^{a}$ & 78$\pm$4   & 41$\pm$3$^{a}$   & [\ion{S}{ii}] \\
           &12&55&45.4&-76&56&13  & 0.035$\pm$0.001$^{a}$ & 71$\pm$3   & 29$\pm$1$^{a}$   & H$\alpha$ \\
HH\,54\,G0 &12&55&44.8&-76&56&13  & 0.036$\pm$0.005$^{a}$ & 36$\pm$7   & 31$\pm$4$^{a}$   & [\ion{S}{ii}] \\
           &12&55&44.8&-76&56&13  & 0.055$\pm$0.003$^{a}$ & 52$\pm$3   & 46$\pm$3$^{a}$   & H$\alpha$ \\
HH\,54\,G1 &12&55&47.3&-76&56&09  & 0.096$\pm$0.006$^{a}$ & 81$\pm$3  & 81$\pm$5$^{a}$  & [\ion{S}{ii}] \\
           &12&55&47.3&-76&56&09  & 0.073$\pm$0.006$^{a}$ & 72$\pm$6  & 62$\pm$5$^{a}$   & H$\alpha$ \\
HH\,54\,G2 &12&55&47.6&-76&56&05  & 0.017$\pm$0.005$^{b}$ & 58$\pm$17  & 14$\pm$5$^{b}$   & [\ion{S}{ii}] \\
           &12&55&47.8&-76&56&05  & 0.079$\pm$0.003$^{a}$ & 68$\pm$2   & 67$\pm$2$^{a}$  & H$\alpha$ \\
HH\,54\,G3 &12&55&49.1&-76&56&06  & 0.079$\pm$0.006$^{a}$ & 63$\pm$4   & 67$\pm$5$^{a}$   & [\ion{S}{ii}] \\
           &12&55&49.3&-76&56&06  & 0.080$\pm$0.002$^{a}$ & 66$\pm$2   & 68$\pm$2$^{a}$   & H$\alpha$ \\
HH\,54\,H1 &12&55&52.8&-76&56&05  & 0.026$\pm$0.004       & 36$\pm$8   & 22$\pm$3   & [\ion{S}{ii}] \\
           &12&55&52.8&-76&56&05  & 0.033$\pm$0.002$^{a}$ & 53$\pm$4   & 28$\pm$2$^{a}$   & H$\alpha$ \\
HH\,54\,H2 &12&55&53.1&-76&56&07  & 0.035$\pm$0.004       & 23$\pm$6   & 29$\pm$3   & [\ion{S}{ii}] \\
           &12&55&53.1&-76&56&07  & 0.051$\pm$0.002$^{a}$ & 38$\pm$3   & 43$\pm$2$^{a}$   & H$\alpha$ \\
HH\,54\,H3 &12&55&52.4&-76&56&10  & 0.025$\pm$0.003       & 187$\pm$9  & 21$\pm$3   & [\ion{S}{ii}] \\
           &12&55&52.6&-76&56&11  & 0.043$\pm$0.003$^{a}$ & 127$\pm$4  & 37$\pm$3$^{a}$   & H$\alpha$ \\
HH\,54\,I  &12&55&54.4&-76&56&08  & 0.031$\pm$0.006       & 48$\pm$11  & 26$\pm$5   & [\ion{S}{ii}] \\
           &12&55&54.5&-76&56&08  & 0.031$\pm$0.002       & 41$\pm$4   &  26$\pm$2   & H$\alpha$ \\
HH\,54\,J  &12&55&52.3&-76&56&17  & 0.012$\pm$0.005$^{b}$ & 45$\pm$25  & 10$\pm$4$^{b}$   & [\ion{S}{ii}] \\
           &12&55&52.3&-76&56&18  &    --  &  -- & --  & [\ion{S}{ii}],H$\alpha$ \\
HH\,54\,J1 &12&55&51.4&-76&56&16  &    --  &  -- & --  & [\ion{S}{ii}],H$\alpha$ \\
\hline \hline
\end{tabular}
\end{center}
Notes:\\ $^{a}$ Mean value, variable in velocity, see text and Table~\ref{acc:tab} \\
$^{b}$ Below 3$\sigma$.\\
\end{table*}

\addtocounter{table}{-1}
\begin{table*}
\caption[]{ Proper Motions in HH\,54 - H$\alpha$ \& [\ion{S}{ii}] - Parameters derived from the linear
fit \it{continued}}
\begin{center}
\begin{tabular}{ccccccccccc}
\hline\hline\\[-5pt]
Knot ID   &  \multicolumn{3}{c}{$\alpha$(2000.0)} & \multicolumn{3}{c}{$\delta$(2000.0)} &  P.M.  &  P.A.  &  $v_{tan}$  & Filter \\
            &   ($^{h}$ & $^{m}$ & $^{s}$) & ($\degr$ & $\arcmin$ & $\arcsec$)     & ($\arcsec$\,yr$^{-1}$) &  ($\degr$)  & (km\,s$^{-1}$) \\
\hline\\[-5pt]
HH\,54\,K  &12&55&53.8&-76&56&25  & 0.017$\pm$0.003   & 2$\pm$14 & 14$\pm$3    & [\ion{S}{ii}] \\
           &12&55&53.8&-76&56&25  & 0.021$\pm$0.003  & 329$\pm$11     & 18$\pm$3   & H$\alpha$ \\
HH\,54\,K1 &12&55&54.1&-76&56&27  & 0.042$\pm$0.004  & 207$\pm$5  & 36$\pm$3   & [\ion{S}{ii}] \\
           &12&55&54.2&-76&56&27  & 0.048$\pm$0.005  & 233$\pm$7  & 41$\pm$4   & H$\alpha$ \\
HH\,54\,K2 &12&55&53.3&-76&56&24  & 0.022$\pm$0.005  & 45$\pm$14  & 18$\pm$4   & [\ion{S}{ii}] \\
           &12&55&53.4&-76&56&24  & 0.016$\pm$0.002  & 345$\pm$10 & 13$\pm$1   & H$\alpha$ \\
HH\,54\,M  &12&55&51.2&-76&56&08  & 0.026$\pm$0.004  & 76$\pm$9   & 22$\pm$4    & [\ion{S}{ii}] \\
           &12&55&51.0&-76&56&09  & 0.019$\pm$0.002$^{a}$  & 81$\pm$7 & 16$\pm$2$^{a}$   & H$\alpha$ \\
HH\,54\,M1 &12&55&50.2&-76&56&11  &    --  &  -- & --  & [\ion{S}{ii}],H$\alpha$ \\
HH\,54\,N  &12&55&48.8&-76&56&23  & 0.029$\pm$0.005   & 312$\pm$9 & 25$\pm$4   & [\ion{S}{ii}] \\
           &12&55&49.0&-76&56&23  & 0.021$\pm$0.002   & 326$\pm$6 & 22$\pm$2   & H$\alpha$ \\
HH\,54\,O  &12&55&47.4&-76&56&22  & 0.013$\pm$0.004  & 328$\pm$16 & 11$\pm$3   & [\ion{S}{ii}] \\
           &12&55&47.3&-76&56&22  & 0.051$\pm$0.005  & 58$\pm$6 & 43$\pm$4   & H$\alpha$ \\
HH\,54\,O1 &12&55&46.7&-76&56&24  & 0.046$\pm$0.004  & 15$\pm$7 & 39$\pm$3   & [\ion{S}{ii}] \\
           &12&55&46.6&-76&56&24  & 0.037$\pm$0.003$^{a}$  & 47$\pm$5 & 31$\pm$3$^{a}$   & H$\alpha$ \\
HH\,54\,P  &12&55&45.2&-76&56&34  &    --  &  -- & --  & [\ion{S}{ii}],H$\alpha$ \\
HH\,54\,Q1 &12&55&47.2&-76&56&43  & 0.044$\pm$0.006  & 23$\pm$8   & 37$\pm$5   & [\ion{S}{ii}] \\
           &12&55&52.5&-76&56&43  & 0.041$\pm$0.002$^{a}$  & 37$\pm$3   & 34$\pm$2$^{a}$   & H$\alpha$ \\
HH\,54\,Q2 &12&55&51.8&-76&56&40  & 0.046$\pm$0.013  & 34$\pm$17 & 39$\pm$11   & [\ion{S}{ii}] \\
           &12&55&51.8&-76&56&40  & 0.043$\pm$0.003  & 79$\pm$4  & 37$\pm$3   & H$\alpha$ \\
HH\,54\,Q3 &12&55&51.3&-76&56&36  & --  & --  & --   & [\ion{S}{ii}] \\
           &12&55&51.3&-76&56&36  & 0.043$\pm$0.003$^{a}$  & 80$\pm$4 & 37$\pm$2$^{a}$   & H$\alpha$ \\
HH\,54\,R  &12&55&56.3&-76&57&20  &    --  &  -- & --  & [\ion{S}{ii}],H$\alpha$ \\
HH\,54\,X0 &12&55&40.3&-76&57&50  & 0.022$\pm$0.003  & 325$\pm$6  & 19$\pm$2   & H$\alpha$ \\
HH\,54\,X1\,A&12&55&44.6&-76&57&35  & 0.035$\pm$0.003$^{a}$  & 101$\pm$4  & 29$\pm$2$^{a}$   & H$\alpha$ \\
HH\,54\,X1\,B&12&55&42.8&-76&57&36  & 0.013$\pm$0.002        & 313$\pm$11  & 11$\pm$2        & [\ion{S}{ii}] \\
             &12&55&42.8&-76&57&37  & 0.012$\pm$0.003  & 275$\pm$8  & 10$\pm$3   & H$\alpha$ \\
HH\,54\,X1\,C&12&55&41.6&-76&57&35  & 0.036$\pm$0.002$^{a}$  & 340$\pm$4  & 30$\pm$2$^{a}$   & H$\alpha$ \\
HH\,54\,X1\,D&12&55&43.8&-76&57&27  & 0.037$\pm$0.004   & 33$\pm$7 & 31$\pm$4      & H$\alpha$ \\
HH\,54\,X2\,A&12&55&40.7&-76&57&42  & 0.018$\pm$0.006  & 204$\pm$23  & 15$\pm$5   & [\ion{S}{ii}] \\
             &12&55&40.9&-76&57&39  & 0.021$\pm$0.002  & 55$\pm$6 & 18$\pm$2   & H$\alpha$ \\
HH\,54\,X2\,B&12&55&41.3&-76&57&39  & 0.013$\pm$0.002   & 314$\pm$11  & 11$\pm$2    & [\ion{S}{ii}] \\
             &12&55&41.0&-76&57&41  & 0.036$\pm$0.002        & 253$\pm$5 & 30$\pm$2   & H$\alpha$ \\
HH\,54\,X3 &12&55&39.8&-76&57&36  & --  & --  & --   & [\ion{S}{ii}] \\
           &12&55&39.8&-76&57&36  & 0.021$\pm$0.003  & 256$\pm$13 & 18$\pm$4   & H$\alpha$ \\
HH\,54\,X4\,A&12&55&40.5&-76&57&28  & 0.068$\pm$0.007  & 45$\pm$6  & 57$\pm$7   & H$\alpha$  \\
HH\,54\,X4\,B&12&55&42.0&-76&57&22  & 0.043$\pm$0.003  & 22$\pm$5  & 37$\pm$3   & H$\alpha$  \\
HH\,54\,X4\,C&12&55&41.8&-76&57&25  & 0.067$\pm$0.006  & 49$\pm$5   & 57$\pm$5   & [\ion{S}{ii}] \\
             &12&55&41.8&-76&57&25  & 0.047$\pm$0.003  & 44$\pm$3 & 39$\pm$2   & H$\alpha$ \\
HH\,54\,Y1 &12&55&44.9&-76&57&11  & 0.033$\pm$0.005  & 33$\pm$8 & 28$\pm$4   & [\ion{S}{ii}] \\
           &12&55&44.9&-76&57&11  & 0.018$\pm$0.002  & 356$\pm$5 & 16$\pm$2   & H$\alpha$ \\
HH\,54\,Y2 &12&55&44.7&-76&57&06  & 0.038$\pm$0.005  & 18$\pm$9 & 32$\pm$4   & [\ion{S}{ii}] \\
           &12&55&44.6&-76&57&07  & 0.037$\pm$0.002$^{a}$  & 26$\pm$3 & 31$\pm$2$^{a}$   & H$\alpha$ \\
HH\,54\,Y3 &12&55&43.2&-76&57&13  & 0.042$\pm$0.004        & 63$\pm$5 & 36$\pm$4   & [\ion{S}{ii}] \\
           &12&55&43.2&-76&57&13  & 0.068$\pm$0.003$^{a}$  & 43$\pm$3 & 58$\pm$3   & H$\alpha$ \\
HH\,54\,Y4 &12&55&45.0&-76&57&00 & 0.043$\pm$0.008        & 62$\pm$17 & 36$\pm$7   & [\ion{S}{ii}] \\
           &12&55&45.0&-76&57&00 & 0.067$\pm$0.008        & 41$\pm$7  & 57$\pm$7   & H$\alpha$ \\
HH\,54\,Y5 &12&55&49.1&-76&57&04  & 0.013$\pm$0.004$^{b}$  & 14$\pm$19   & 11$\pm$3$^{b}$   & [\ion{S}{ii}] \\
           &12&55&49.0&-76&57&05  & 0.022$\pm$0.008$^{b}$  & 90$\pm$20 & 19$\pm$7$^{b}$   & H$\alpha$ \\
HH\,54\,Y6 &12&55&45.5&-76&56&53  & 0.059$\pm$0.006       & 69$\pm$6   & 50$\pm$5   & [\ion{S}{ii}] \\
           &12&55&45.5&-76&56&53  & 0.059$\pm$0.010       & 16$\pm$18  & 50$\pm$9   & H$\alpha$ \\
HH\,54\,Y7 &12&55&46.1&-76&57&00  & 0.072$\pm$0.017      & 322$\pm$15 & 61$\pm$14   & H$\alpha$ \\
HH\,54\,Y8 &12&55&45.3&-76&57&09  & 0.026$\pm$0.010$^{b}$  & 22$\pm$11 & 22$\pm$8$^{b}$   & H$\alpha$ \\
HH\,54\,Z  &12&55&47.2&-76&56&43  & 0.037$\pm$0.004  & 344$\pm$7  & 31$\pm$3   & [\ion{S}{ii}] \\
           &12&55&47.2&-76&56&45  & 0.010$\pm$0.002  & 325$\pm$13 & 8$\pm$2   & H$\alpha$ \\
HH\,54\,Z1 &12&55&46.1&-76&56&45  & 0.013$\pm$0.002  & 292$\pm$9 & 11$\pm$2   & H$\alpha$ \\
\hline \hline
\end{tabular}
\end{center}
Notes:\\ $^{a}$ Mean value, variable in velocity, see text and Table~\ref{acc:tab} \\
$^{b}$ Below 3$\sigma$.\\
\end{table*}

\begin{table*}
\caption[]{ Proper Motions in HH\,52 \& HH\,53 - H$_2$  \label{PM52_h2:tab}}
\begin{center}
\begin{tabular}{ccccccccccc}
\hline\hline\\[-5pt]
Knot ID   &  \multicolumn{3}{c}{$\alpha$(2000.0)} & \multicolumn{3}{c}{$\delta$(2000.0)} &  P.M.  &  P.A.  &  $v_{tan}$  & Filter \\
            &   ($^{h}$ & $^{m}$ & $^{s}$) & ($\degr$ & $\arcmin$ & $\arcsec$)     & ($\arcsec$\,yr$^{-1}$) &  ($\degr$)  & (km\,s$^{-1}$) \\
\hline\\[-5pt]
HH\,52\,B2/B3 &12&55&05.0&-76&57&54  & 0.026$\pm$0.004$^{a}$  &  40$\pm$9 & 22$\pm$4$^{a}$      & H$_2$ \\
HH\,52\,B5/B6 &12&55&02.7&-76&57&57  & 0.041$\pm$0.014$^{b}$  &  281$\pm$12  & 35$\pm$12$^{b}$      &  H$_2$  \\
HH\,52\,C1 &12&55&05.4&-76&58&01  & 0.008$\pm$0.004$^{b}$  &  51$\pm$26  & 7$\pm$3$^{b}$    & H$_2$ \\
HH\,53\,A  &12&55&15.4&-76&57&30  & 0.016$\pm$0.003  & 298$\pm$15 & 14$\pm$3  & H$_2$ \\
HH\,53\,B  &12&55&14.5&-76&57&31  & 0.007$\pm$0.003$^{b}$  & 231$\pm$34 &  6$\pm$3$^{b}$  & H$_2$ \\
HH\,53\,C  &12&55&16.5&-76&57&30  & 0.010$\pm$0.003  & 270$\pm$26 & 9$\pm$2             & H$_2$ \\
HH\,53\,H  &12&55&15.2&-76&57&26  & 0.024$\pm$0.004  &  27$\pm$8  & 21$\pm$4    & H$_2$ \\
\hline \hline
\end{tabular}
\end{center}
Notes:\\ $^{a}$ Mean value, variable in velocity, see text and Table~\ref{acc:tab} \\
$^{b}$ Below 3$\sigma$.\\
\end{table*}

\begin{table*}
\caption[]{ Proper Motions in HH\,54 - H$_2$  \label{PM54_h2:tab}}
\begin{center}
\begin{tabular}{ccccccccccc}
\hline\hline\\[-5pt]
Knot ID   &  \multicolumn{3}{c}{$\alpha$(2000.0)} & \multicolumn{3}{c}{$\delta$(2000.0)} &  P.M.  &  P.A.  &  $v_{tan}$  & Filter \\
           &   ($^{h}$ & $^{m}$ & $^{s}$) & ($\degr$ & $\arcmin$ & $\arcsec$)     & ($\arcsec$\,yr$^{-1}$) &  ($\degr$)  & (km\,s$^{-1}$) \\
\hline\\[-5pt]
HH\,54\,A$_1$ &12&55&47.3&-76&56&29   & 0.018$\pm$0.005   & 343$\pm$16  & 15$\pm$4    &    H$_2$ \\
HH\,54\,A$_2$ &12&55&50.0&-76&56&29  & 0.011$\pm$0.003  & 51$\pm$17  & 9$\pm$3   &    H$_2$ \\
HH\,54\,A$_3$ &12&55&49.2&-76&56&29  & 0.018$\pm$0.004  & 300$\pm$10 & 16$\pm$3     & H$_2$ \\
HH\,54\,A$_4$ &12&55&50.9&-76&56&30  & 0.014$\pm$0.006$^{b}$   & 62$\pm$16  & 12$\pm$5$^{b}$   &    H$_2$ \\
HH\,54\,B$_1$&12&55&51.0&-76&56&20  & 0.016$\pm$0.003$^{a}$   & 15$\pm$13 & 14$\pm$2$^{a}$   & H$_2$ \\
HH\,54\,B$_2$&12&55&51.1&-76&56&21& 0.008$\pm$0.003$^{b}$ & 131$\pm$23 & 6$\pm$3$^{b}$   & H$_2$ \\
HH\,54\,B$_3$&12&55&51.2&-76&56&22  & 0.022$\pm$0.003$^{a}$   & 140$\pm$9 & 18$\pm$3$^{a}$  & H$_2$ \\
HH\,54\,C1  &12&55&51.9&-76&56&05  & 0.015$\pm$0.003  & 77$\pm$9  & 13$\pm$3   & H$_2$ \\
HH\,54\,C3  &12&55&51.3&-76&56&01  & 0.009$\pm$0.004$^{b}$  & 239$\pm$19 & 8$\pm$3$^{b}$   & H$_2$ \\
HH\,54\,E   &12&55&53.9&-76&56&22  & 0.012$\pm$0.003$^{a}$  & 48$\pm$14  & 11$\pm$3   & H$_2$ \\
HH\,54\,H2  &12&55&53.2&-76&56&07  & 0.020$\pm$0.003  & 39$\pm$9   & 17$\pm$3   & H$_2$ \\
HH\,54\,I  &12&55&55.2&-76&56&06  & 0.016$\pm$0.004$^{a}$   & 3$\pm$14 & 14$\pm$3$^{a}$   & H$_2$ \\
HH\,54\,J  &12&55&53.0&-76&56&17  & 0.019$\pm$0.003   & 45$\pm$10 & 16$\pm$3   & H$_2$ \\
HH\,54\,J1  &12&55&52.6&-76&56&16  & 0.008$\pm$0.004$^{b}$   & 32$\pm$36 & 8$\pm$4$^{b}$   & H$_2$ \\
HH\,54\,K   &12&55&54.1&-76&56&25  & 0.012$\pm$0.003  & 19$\pm$17  & 10$\pm$2   & H$_2$ \\
HH\,54\,K1   &12&55&54.6&-76&56&27  & 0.019$\pm$0.005 & 243$\pm$16 & 16$\pm$4   & H$_2$ \\
HH\,54\,M  &12&55&51.5&-76&56&12  & 0.020$\pm$0.007$^{b}$   & 110$\pm$13 & 17$\pm$6$^{b}$   & H$_2$ \\
HH\,54\,M1 &12&55&50.3&-76&56&13  & 0.021$\pm$0.003   & 211$\pm$8 & 18$\pm$3   & H$_2$ \\
HH\,54\,T &12&55&41.6&-76&57&34  &    --  &  -- & --  &  H$_2$ \\
HH\,54\,Y1 &12&55&45.0&-76&57&10  & 0.016$\pm$0.008$^{b}$  & 67$\pm$25 & 14$\pm$7$^{b}$   & H$_2$ \\
HH\,54\,Y2 &12&55&44.4&-76&57&06  & 0.020$\pm$0.005$^{a}$  & 21$\pm$7 & 17$\pm$4$^{a}$   & H$_2$ \\
HH\,54\,Z  &12&55&47.5&-76&56&45  & 0.018$\pm$0.004   & 39$\pm$12 & 15$\pm$3   & H$_2$ \\
HH\,54\,Z1 &12&55&46.3&-76&56&45  & 0.015$\pm$0.004    & 317$\pm$13 & 13$\pm$3    & H$_2$ \\
\hline \hline
\end{tabular}
\end{center}
\end{table*}

\end{appendix}

\begin{appendix}
\section{Flux and velocity variability}
\label{appendixC:sec}


\begin{table*}
\caption[]{ Accelerated Motions in HH\,52, HH\,53, and  HH\,54 - H$\alpha$, [\ion{S}{ii}] and H$_2$ - Parameters derived from the quadratic
fit  \label{acc:tab}}
\begin{center}
\begin{tabular}{cccccccc}
\hline\hline\\[-5pt]
Knot ID   & $P_\mathrm{a}$($\chi^2/\nu$) &  P.M.     & Accelerated motion   & P.A.($\vec{v}$)    & P.A.($\vec{a}$)    & Filter  & Note\\
          &  $\times$100 (\%) &  ($\arcsec$\,yr$^{-1}$) &  ($\arcsec$\,yr$^{-2}$)  & ($\degr$)  & ($\degr$)  &      &  Reliable  \\
\hline\\[-5pt]
HH\,52\,A5   & 0.10     &  0.071$\pm$0.012   & 0.0067$\pm$0.0013          &  110$\pm$8 &  133$\pm$11   &  [\ion{S}{ii}]  & no \\
             & 1.00$^a$ &  0.137$\pm$0.086$^b$  & 0.018$\pm$0.008$^b$     &  5$\pm$33   &  340$\pm$26   &  H$\alpha$ &  \\
HH\,52\,B1   & 0.05     &  0.132$\pm$0.017      & 0.0226$\pm$0.0018  &  116$\pm$8    &  299$\pm$5   &  [\ion{S}{ii}] & yes \\
             & 1.00$^a$ &  0.093$\pm$0.068$^b$  & 0.02$\pm$0.01$^b$  &  111$\pm$20   &  285$\pm$16   &  H$\alpha$ & \\
HH\,52\,B2   & 0.99     &  0.099$\pm$0.021   & 0.0060$\pm$0.0021$^b$  &  7$\pm$11 &  12$\pm$23   &  [\ion{S}{ii}] & yes \\
             & 1.00     &  0.131$\pm$0.023   & 0.0109$\pm$0.0027      &  9$\pm$16 &  16$\pm$23   &  H$\alpha$ & \\
HH\,52\,B3   & 0.34     &  0.122$\pm$0.017   & 0.0121$\pm$0.0019  &  112$\pm$7 &  128$\pm$9   &  [\ion{S}{ii}] & no \\
HH\,52\,B2/B3 & 1.00$^a$     &  0.194$\pm$0.077$^b$   & 0.061$\pm$0.015  &  357$\pm$22 &  348$\pm$23   &  H$_2$ & no \\
HH\,52\,B4   & 0.22     &  0.124$\pm$0.016   & 0.0134$\pm$0.0019  &    90$\pm$6 &  94$\pm$7   &  [\ion{S}{ii}] & no\\
HH\,52\,B5   & 0.05     &  0.106$\pm$0.041$^b$  & 0.0117$\pm$0.0030  &  243$\pm$13 &  238$\pm$13   &  [\ion{S}{ii}] & no\\
             & 0.10     &  0.146$\pm$0.045     & 0.0076$\pm$0.0052$^b$  &  202$\pm$12   &  215$\pm$36   &  H$\alpha$ & \\
HH\,52\,B8   & 1.00     &  0.164$\pm$0.067$^b$   & 0.020$\pm$0.007$^b$  &  290$\pm$16 &  291$\pm$19   &  [\ion{S}{ii}] & yes  \\
HH\,52\,C1   & 0.05     &  0.052$\pm$0.012        & 0.0105$\pm$0.0014  &  69$\pm$13 &  229$\pm$10   &  [\ion{S}{ii}] & yes\\
HH\,52\,C4   & 0.28     &  0.144$\pm$0.025     & 0.0105$\pm$0.0018  &  347$\pm$7 &  338$\pm$11   &  [\ion{S}{ii}] &  yes\\
             & 1.00$^a$ &  0.166$\pm$0.047    & 0.0163$\pm$0.0050$^b$       &  348$\pm$14   &  338$\pm$21   &  H$\alpha$ & \\
HH\,52\,D4   & 0.18     &  0.073$\pm$0.014   & 0.0038$\pm$0.0013$^b$  &  33$\pm$9 &  16$\pm$21   &  [\ion{S}{ii}] & no \\
	     & 0.53     &  0.093$\pm$0.050$^b$   & 0.011$\pm$0.0050$^b$  &  12$\pm$22 &  347$\pm$28   &  H$\alpha$ \\
HH\,53\,A    & 0.93   &  0.034$\pm$0.012$^b$   & 0.0040$\pm$0.0012  &  230$\pm$18 &  197$\pm$20   &  [\ion{S}{ii}] & no \\
HH\,53\,B    & 0.96   &  0.038$\pm$0.022$^b$   & 0.0047$\pm$0.0012    &  332$\pm$18 &  343$\pm$18   &  [\ion{S}{ii}] & no \\
HH\,53\,G    & 0.07   &  0.039$\pm$0.013    & 0.0049$\pm$0.0014  &  104$\pm$16 &  212$\pm$18   &  [\ion{S}{ii}]  & no\\
             & 1.00$^a$  &  0.212$\pm$0.069$^b$  & 0.0276$\pm$0.0089       & 105$\pm$15   &  113$\pm$18   &  H$\alpha$ \\
HH\,54\,B$_1$& 0.34     &  0.04$\pm$0.01   & 0.0088$\pm$0.0024  &  40$\pm$12 &  61$\pm$15   &  H$_2$ & no\\
HH\,54\,B$_3$& 0.58     &  0.067$\pm$0.013   & 0.0145$\pm$0.0023  &  173$\pm$8 &  185$\pm$10   &  H$_2$ & no \\
HH\,54\,C3   & 0.85   &  0.046$\pm$0.020$^b$  & 0.054$\pm$0.014       & 184$\pm$21   &  165$\pm$20   &  H$\alpha$ & no\\
HH\,54\,E    & 0.86     &  0.027$\pm$0.008   & 0.0039$\pm$0.0015$^b$  &  28$\pm$19 &  16$\pm$41   &  H$_2$ & yes \\
HH\,54\,F1   & 0.09   &  0.072$\pm$0.024  & 0.084$\pm$0.019       &  312$\pm$12   &  324$\pm$13   &  H$\alpha$ & no\\
HH\,54\,G    & 1.00$^a$   &  0.134$\pm$0.057$^b$   & 0.013$\pm$0.006$^b$  &  67$\pm$21 &  61$\pm$33   &  [\ion{S}{ii}] & yes  \\
             & 0.08   &  0.155$\pm$0.015  & 0.0146$\pm$0.0017       &  85$\pm$7   &  89$\pm$9   &  H$\alpha$ \\
HH\,54\,G0   & 1.00$^a$   &  0.2$\pm$0.1$^b$   & 0.020$\pm$0.007$^b$  &  28$\pm$16 &  220$\pm$20   &  [\ion{S}{ii}] & yes \\
             & 0.12   &  0.05$\pm$0.02$^b$  & 0.012$\pm$0.002       &  40$\pm$20   &  230$\pm$13   &  H$\alpha$ \\
HH\,54\,G1   & 1.00$^a$   &  0.239$\pm$0.067   & 0.02$\pm$0.009$^b$  &  79$\pm$15 &  76$\pm$27   &  [\ion{S}{ii}] & yes \\
             & 0.06       &  0.214$\pm$0.031  & 0.0177$\pm$0.0039   &  85$\pm$12   &  85$\pm$19   &  H$\alpha$ \\
HH\,54\,G2   & 1.00$^a$   &  0.12$\pm$0.11$^b$   & 0.017$\pm$0.0077$^b$  &  70$\pm$35 &  316$\pm$26   &  [\ion{S}{ii}] & no  \\
             & 0.37       &  0.125$\pm$0.019  & 0.0097$\pm$0.0026       & 70$\pm$13   &  143$\pm$19   &  H$\alpha$ \\
HH\,54\,G3   & 1.00$^a$   &  0.291$\pm$0.059   & 0.0299$\pm$0.0077        &  76$\pm$11 &  80$\pm$15   &  [\ion{S}{ii}] & yes \\
             & 0.10       &  0.134$\pm$0.025  & 0.0068$\pm$0.0025$^b$       & 66$\pm$16   &  67$\pm$38   &  H$\alpha$ \\
HH\,54\,H1   & 0.49       &  0.033$\pm$0.013$^b$  & 0.0063$\pm$0.0017       & 55$\pm$30   &  199$\pm$17   &  H$\alpha$ & yes \\
HH\,54\,H2   & 0.10       &  0.158$\pm$0.017  & 0.0228$\pm$0.0019       &  41$\pm$5   &  256$\pm$4   &  H$\alpha$ & yes \\
HH\,54\,H3   & 1.00$^a$   &  0.062$\pm$0.023$^b$  & 0.012$\pm$0.004       &  127$\pm$4   &  273$\pm$16   &  H$\alpha$ & no \\
             & 0.50       &  0.079$\pm$0.017    & 0.007$\pm$0.003$^b$     &  66$\pm$12   &  140$\pm$35   &  H$_2$ \\
HH\,54\,M    & 1.00       &  0.071$\pm$0.019  & 0.069$\pm$0.021       & 70$\pm$14   &  67$\pm$19   &  H$\alpha$ & yes\\
HH\,54\,Q1   & 1.00$^a$   &  0.165$\pm$0.103$^b$  & 0.0171$\pm$0.0078$^b$       &  12$\pm$20   &  8$\pm$28   &  [\ion{S}{ii}] & yes\\
             & 0.97       &  0.093$\pm$0.023  & 0.0084$\pm$0.0023               &  7$\pm$22   &  349$\pm$30   &  H$\alpha$ \\
HH\,54\,Q3   & 0.99       &  0.097$\pm$0.018  & 0.0069$\pm$0.023       &  79$\pm$12   &  79$\pm$22   &  H$\alpha$ & yes\\
HH\,54\,X1\,A  & 0.57     &  0.079$\pm$0.015  & 0.055$\pm$0.018     &  105$\pm$9   &  107$\pm$16   &  H$\alpha$ & yes\\
HH\,54\,X1\,C  & 0.38     &  0.025$\pm$0.012$^b$   & 0.0065$\pm$0.0015         &  225$\pm$30   &  186$\pm$14   &  H$\alpha$ & no \\
HH\,54\,Y2     & 1.00    &  0.082$\pm$0.014      & 0.0057$\pm$0.0013       &  25$\pm$7   &  23$\pm$12   &  H$\alpha$ & yes \\
               & 0.59   &  0.071$\pm$0.026$^b$   & 0.0098$\pm$0.0037$^b$  &  19$\pm$16 &  40$\pm$20   &  H$_2$ \\
HH\,54\,Y3    & 0.06   &  0.145$\pm$0.026     & 0.0097$\pm$0.0026       &  37$\pm$9   &  32$\pm$18   &  H$\alpha$ & yes\\
\hline \hline
\end{tabular}
\end{center}
Notes:\\ $^{a}$ Only 3 points in the fit.\\
$^{b}$ Below 3$\sigma$.\\
\end{table*}

\subsection{HH\,52}

\label{52:sec}

\begin{figure*}
\centering
   \includegraphics [width=9 cm] {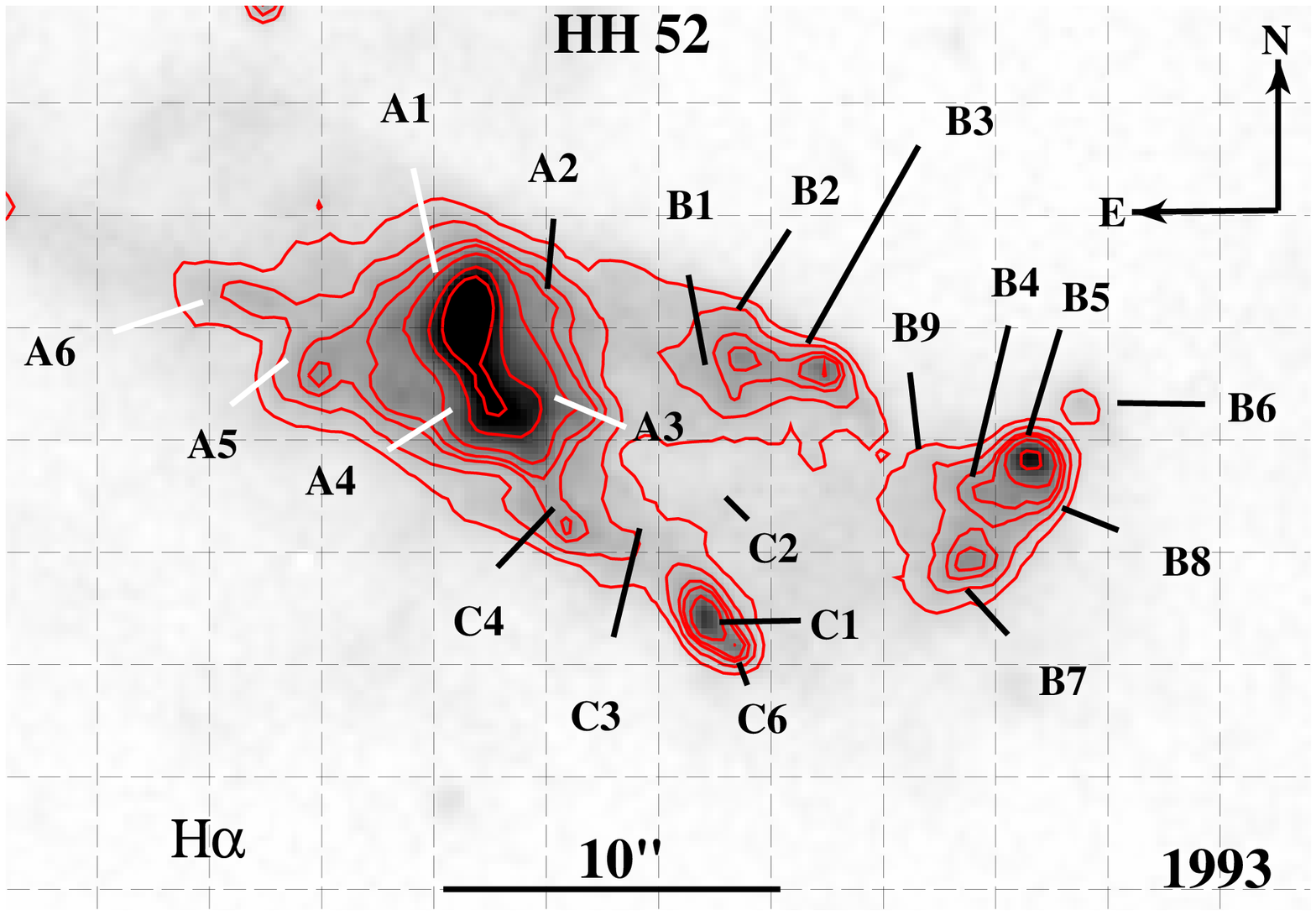}
   \includegraphics [width=9 cm] {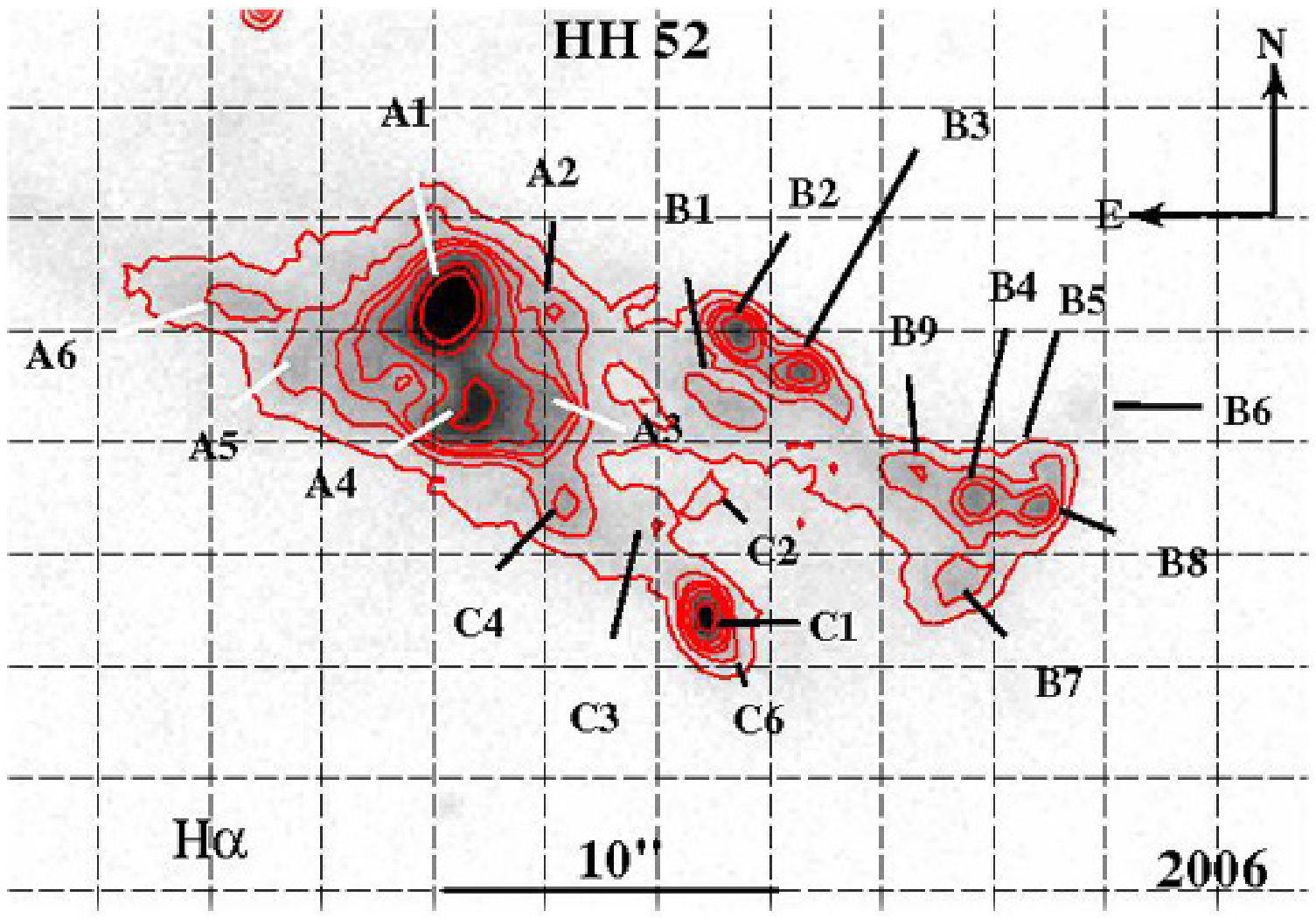}
   \caption{ Variability in morphology and flux of HH\,52. {\bf Left panel:} close up of HH\,52 from 1993 H$\alpha$ calibrated image (labels refer to 2006 image). {\bf Right panel:} close up of HH\,52 from 2006 H$\alpha$ calibrated image. Both images have the same contour levels.
\label{HH52var-flux:fig}}
\end{figure*}

\subsubsection{HH\,52\,B and C}

\begin{figure}
   \includegraphics [width=9.2 cm] {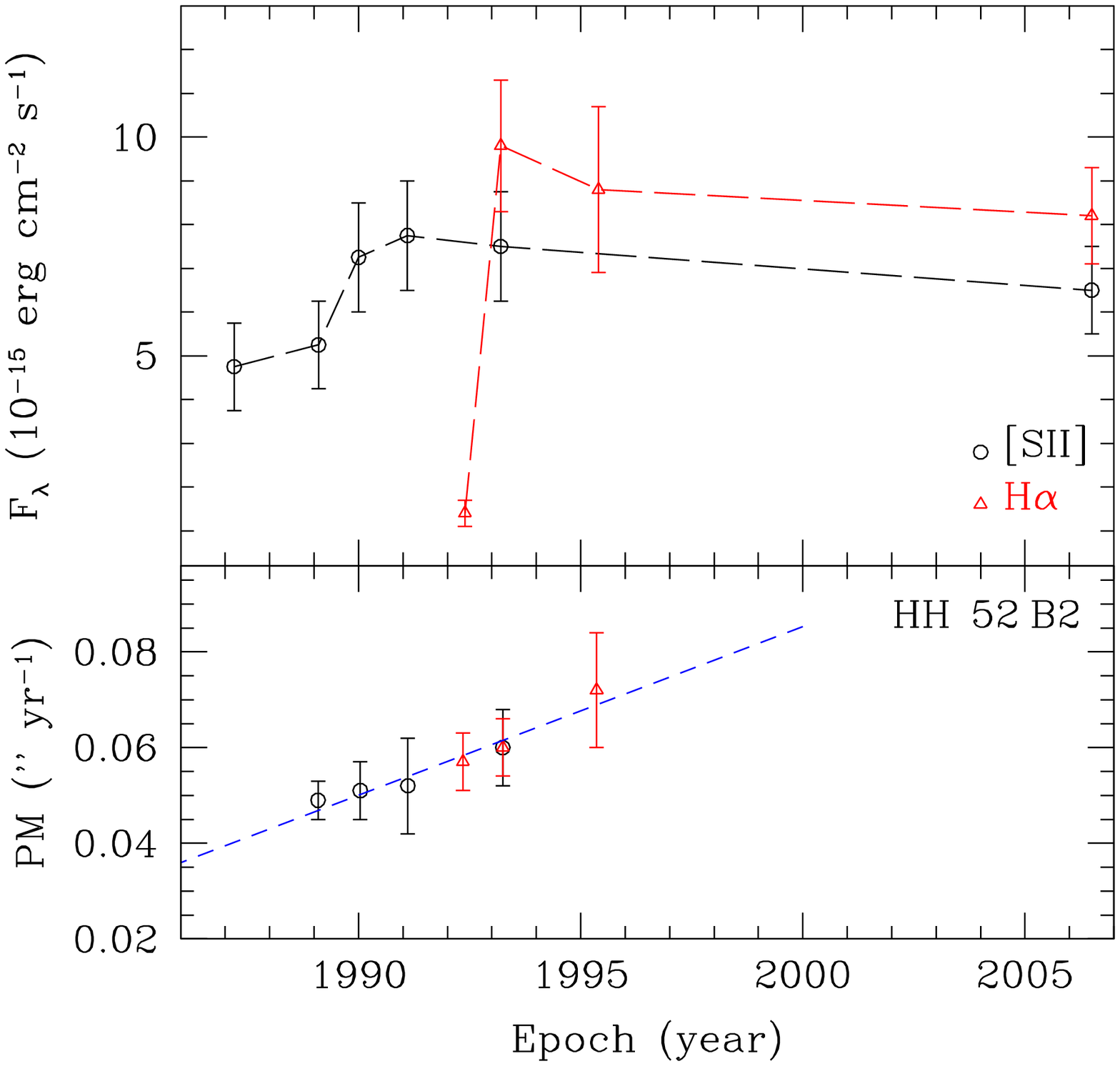}
   \caption{ Variability in HH\,52\,B2. {\bf Bottom panel} shows the P.M.s as a function of time in [\ion{S}{ii}] (circles)and H$\alpha$ (triangles). The dashed line is the best fit to the data points.
   The slope of the fit gives the acceleration of the knot.
   In the {\bf top panel} of the figure the measured fluxes (uncorrected for the extinction) in the two filters are reported.
\label{HH52B2-flux:fig}}
\end{figure}

\begin{figure}
   \includegraphics [width=9.2 cm] {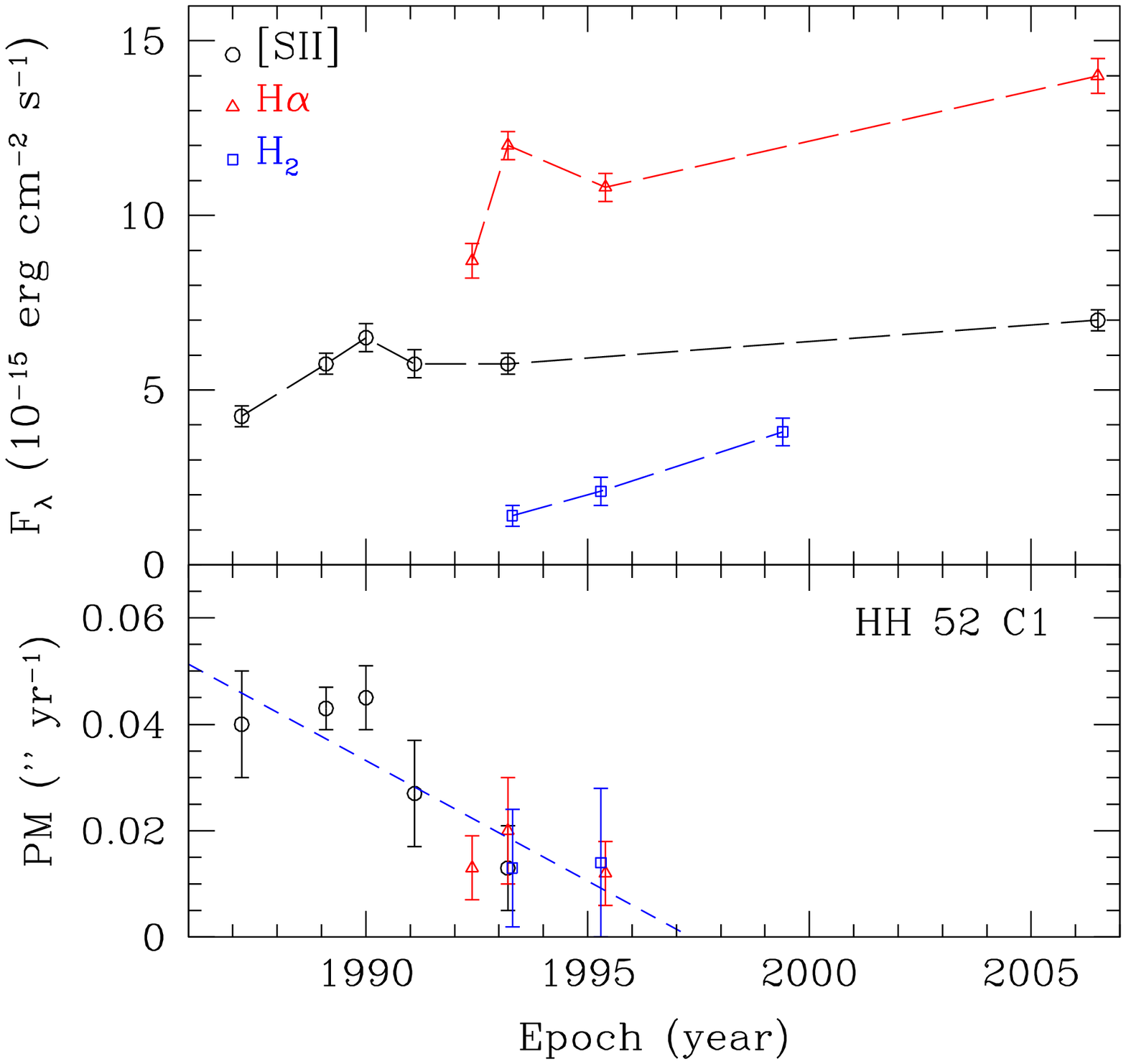}
   \caption{ Variability in HH\,52\,C1. {\bf Bottom panel} shows the P.M.s as a function of time in [\ion{S}{ii}] (circles)and H$\alpha$ (triangles). The dashed line is the best fit to the data points. The slope of the fit gives the acceleration of the knot.
   In the {\bf top panel} of the figure the measured fluxes (uncorrected for the extinction) in the two filters are reported.
\label{HH52C1-flux:fig}}
\end{figure}

Both wings of HH\,52 bow-shock show a high degree of variability.
As an example of such a variability, we show in Fig.~\ref{HH52var-flux:fig} a close up of HH\,52
in 1993 and 2006 H$\alpha$ calibrated images. Changes in morphology and flux are easy to recognise, especially for the knots
of group B.

HH\,52 B1 and B2 come from the fragmentation of a single structure around 1987. The first knot is moving ESE with a
P.A.$\sim$110$\degr$, while the second is moving northward (P.A.$\sim$5$\degr$). A third knot B3, clearly visible from 1989, likely
causes the break, proceeding from SE to NE and pushing forward the structure. The direction of the feature considerably changes with
time in both optical filters, accordingly, from $\sim$55$\degr$ in 1989 to 85$\degr$ in 1993.
Also B1 and B2 show variability (see Fig.~\ref{HH52var-flux:fig}, Fig.~\ref{HH52B1-fig:fig}, and Fig.~\ref{HH52B2-flux:fig}), exhibiting evidence of deceleration and acceleration, respectively.

Knots HH\,52\,B4-B9, in the outskirts of the wing, exhibit large variability and complicated motions as well.
The presence of such a high variability and the large gap between the second to last (1993 for [\ion{S}{ii}] and 1995 for H$\alpha$) and
the last epoch image (2006) produce uncertainties in the identification of the knots and in the P.M. derivation.
A large structure, B5, appears in 1989 (see Fig.~\ref{HH52B5-flux:fig}, top panel and also Fig.~\ref{HH52var-flux:fig}), rapidly increasing its brightness and then almost vanishing in 2006, apparently fragmented in several knots (B4, B5, B6, and B8), that move in different directions.
The flux variation of the object has different timing in the three filters,
with a delay of some years between the flux variations of [\ion{S}{ii}] and H$\alpha$, and then between H$\alpha$ and H$_2$
(Fig~\ref{HH52B5-flux:fig}, top panel).
In [\ion{S}{ii}] it slightly varies its brightness, reaching a maximum possibly around 1991-1993. In H$\alpha$ and H$_2$
the variations are more considerable, especially in the molecular component, where the flux increases an order of magnitude
during six years.

Knots C in the left wing of HH\,52 present a peculiar behaviour as well.
C1 (see Fig.~\ref{HH52B5-flux:fig}, central panel and Fig.~\ref{HH52C1-flux:fig}) shows a large variability similar to B5, with a large brightness increase in H$\alpha$, due to the presence of a fast shock colliding against the target.
This is clearly visible in 1993 and 1995 H$\alpha$ images (Fig.~\ref{HH52var-flux:fig}), where a high velocity knot (C6) appears behind C1.
In the 2006 image the bullet is not visible anymore, but an increase in C1 brightness is detected, in both atomic emissions. As in B5, also the molecular emission increases with time. Here the increment is not so pronounced as in the previous knot.
HH\,52\,C2, located behind HH\,52 nucleus, starts forming in 1989 as a [\ion{S}{ii}] emission (Fig~\ref{HH52B5-flux:fig}, bottom panel)
and increases its luminosity until 2006. On the other hand the H$\alpha$ emission is visible above a 3$\sigma$ limit only in the 2006 image.

\subsection{HH\,54}

\subsubsection{HH\,54 streamer}
\label{streamer54:sec}

\begin{figure}
   \includegraphics [width=9.2 cm] {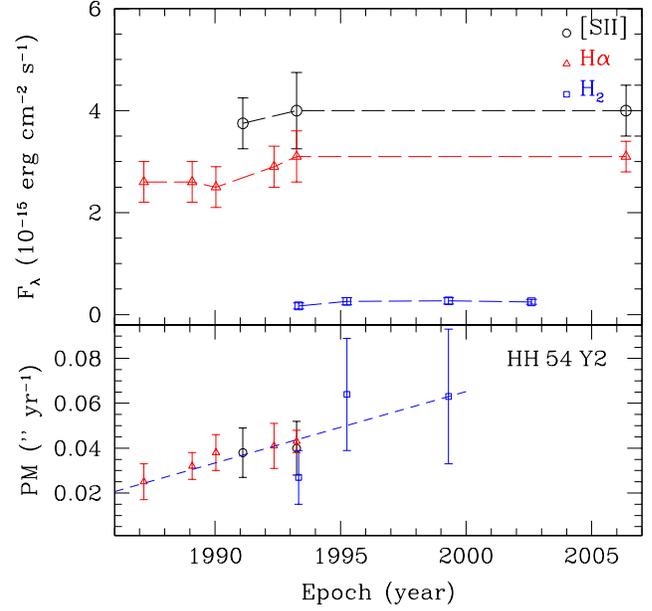}
   \caption{ Variability in HH\,54\,Y2. {\bf Bottom panel} shows the P.M.s as a function of time in [\ion{S}{ii}] (circles), H$\alpha$ (triangles), and H$_2$ (squares). The dashed line is the best fit to the data points.
   The slope of the fit gives the acceleration of the knot.
   In the {\bf top panel} of the figure the measured fluxes (uncorrected for the extinction) in the three filters are reported.
\label{HH52Y2-flux:fig}}
\end{figure}

Figure~\ref{HH52Y2-flux:fig} (bottom panel) displays the peculiar kinematics of HH\,54\,Y2, located in the middle of the streamer.
The proper motion measurements reveal an apparent acceleration of the knot, detected in all three filters.
However, due to the faint emission, errors in H$_2$ appear remarkably larger. Here, the P.M. measurements are referred to epoch 2002, since in H$_2$ 2005 image Y2 is out of the FoV. We measure an acceleration of 0.006$\pm$0.0002$\arcsec$\,yr$^{-2}$ and  0.01$\pm$0.004$\arcsec$\,yr$^{-2}$
in H$\alpha$ and H$_2$, respectively (see Table~\ref{acc:tab}). The average value obtained including both optical (H$\alpha$ and [\ion{S}{ii}]) and molecular P.M.s is 0.0031$\pm$0.0008$\arcsec$\,yr$^{-2}$.
Thus a mechanism must exist there that is accelerating the streamer or, conversely,
the observed motion could be a reminiscence of the original acceleration experienced by the flow.
The upper panel of Fig.~\ref{HH52Y2-flux:fig} reports the measured flux for the three filters. For this object, all the
emissions appear almost constant with time (inside the error bars).
This is indeed due to the high degree of ionisation along the HH\,54 streamer.

Also along the streamer it is possible to observe the formation of new condensations. A new H$\alpha$ knot is forming
$\sim$1$\arcsec$.6 ENE of HH\,54\,Y3. This is probably an instability in the flow created by the material compressed and pushed ahead of Y3.

\subsubsection{HH\,54\,A}
\label{54A:sec}

\begin{figure}
   \includegraphics [width=9.2 cm] {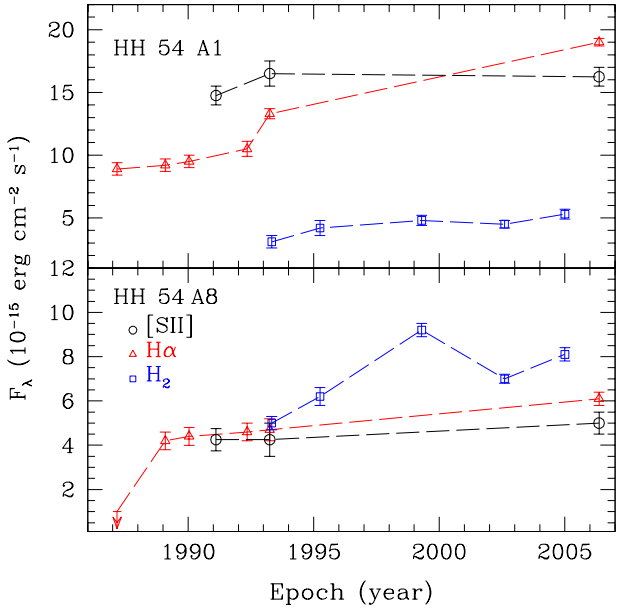}
   \caption{ Measured fluxes (uncorrected for the extinction) in knots HH\,54\,A1 ({\bf bottom panel}), and A8 ({\bf top panel})
   in [\ion{S}{ii}] (circles), H$\alpha$ (triangles), and H$_2$ (squares) filters.
\label{HH54A-flux:fig}}
\end{figure}

Figure~\ref{HH54A-flux:fig} illustrates flux variability in HH\,54\,A group.
Here, at variance with other knots, the measured velocities are constant (inside the error bars) in all the three filters.
The brightest feature (knot A1, bottom panel) doubles its H$\alpha$ luminosity, after 1990,.
Variability is, however, not observed in [\ion{S}{ii}], which appears constant
during the considered period. The molecular emission, that does not spatially coincide with the atomic one but it is located
about 1$\arcsec$.6 forward, from 1993 shows an increase with time.
Knot A8, next to A1 along the flow, appears in H$\alpha$ images in 1989 (Fig.~\ref{HH54A-flux:fig}, top panel), and increases its brightness
in all the three filters, and it is extremely luminous in the molecular component. In both
[\ion{S}{ii}] and H$_2$ it has a diffuse
emission, probably originating from moving gas cooling behind the H$\alpha$ emission.
The feature has a different direction (P.A.$\sim$35$\degr$) with respect to the bulk of the flow (P.A.$\sim$80$\degr$). This deflection could be generated in the interaction with knot A6 (see also Fig.~\ref{HH54-fig:fig}), a faint emission following the group, that seems to push aside A8, or, conversely, it could indicate that the knot is not part of the same flow.

\subsubsection{HH\,54\,G}
\label{54G:sec}
\begin{figure}
   \includegraphics [width=9.2 cm] {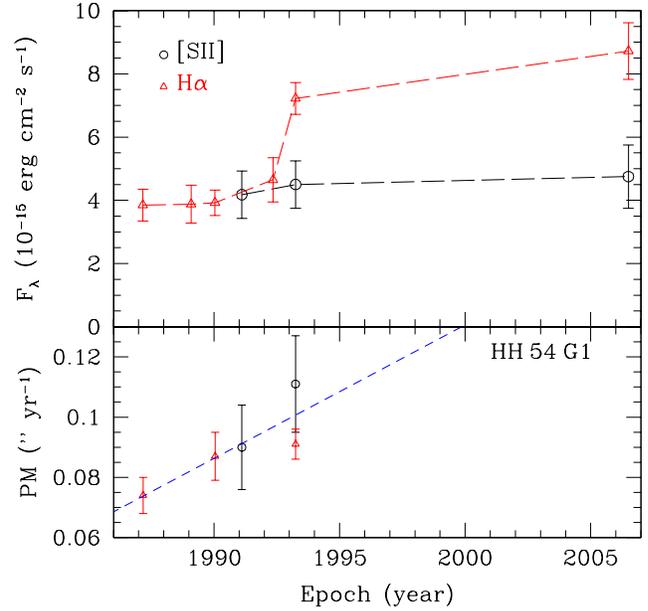}
   \caption{ Variability in HH\,54\,G1. {\bf Bottom panel} shows the P.M.s as a function of time in [\ion{S}{ii}] (circles)and H$\alpha$ (triangles). The dashed line is the best fit to the data points. The slope of the fit gives the acceleration of the knot. In the {\bf top panel} of the figure the measured fluxes (uncorrected for the extinction) in the two filters are reported.
\label{HH54G1-flux:fig}}
\end{figure}

\begin{figure}
   \includegraphics [width=9.2 cm] {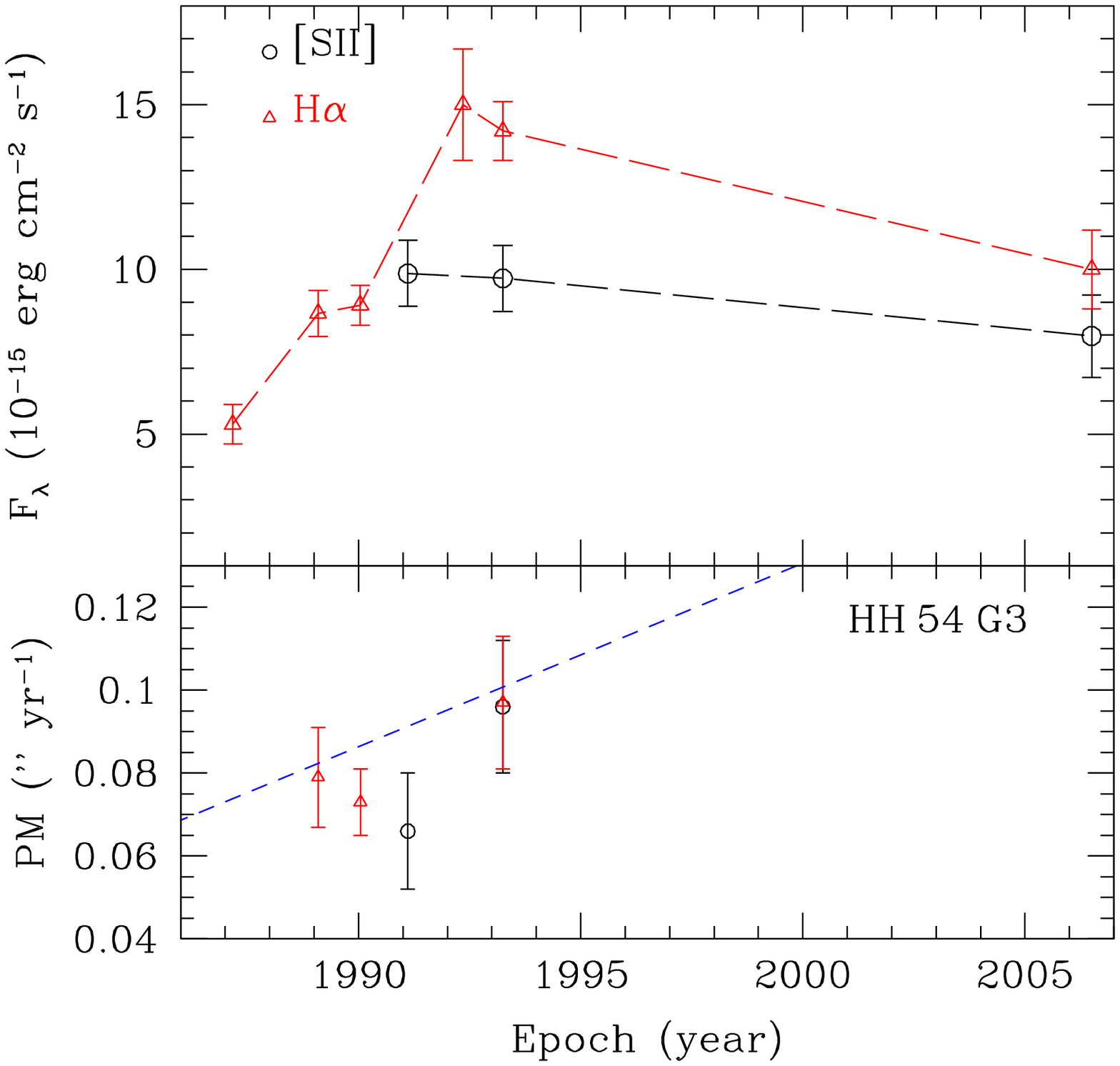}
   \caption{ Variability in HH\,54\,G3. {\bf Bottom panel} shows the P.M.s as a function of time in [\ion{S}{ii}] (circles)and H$\alpha$ (triangles). The dashed line is the best fit to the data points. The slope of the fit gives the acceleration of the knot. In the {\bf top panel} of the figure the measured fluxes (uncorrected for the extinction) in the two filters are reported.
\label{HH54G3-flux:fig}}
\end{figure}

Few arcseconds ahead of knots G and G0 (see Section~\ref{variability:sec}, also knots
HH\,54 G1 and G3 are deflected (P.A.$\sim$60-80$\degr$) and accelerated (see Fig.~\ref{HH54G1-flux:fig}, \ref{HH54G3-flux:fig}),
likely by the same mechanism of HH\,54\,G/G0.

\subsubsection{HH\,54\,H }
\label{54H:sec}

\begin{figure}
   \includegraphics [width=9.2 cm] {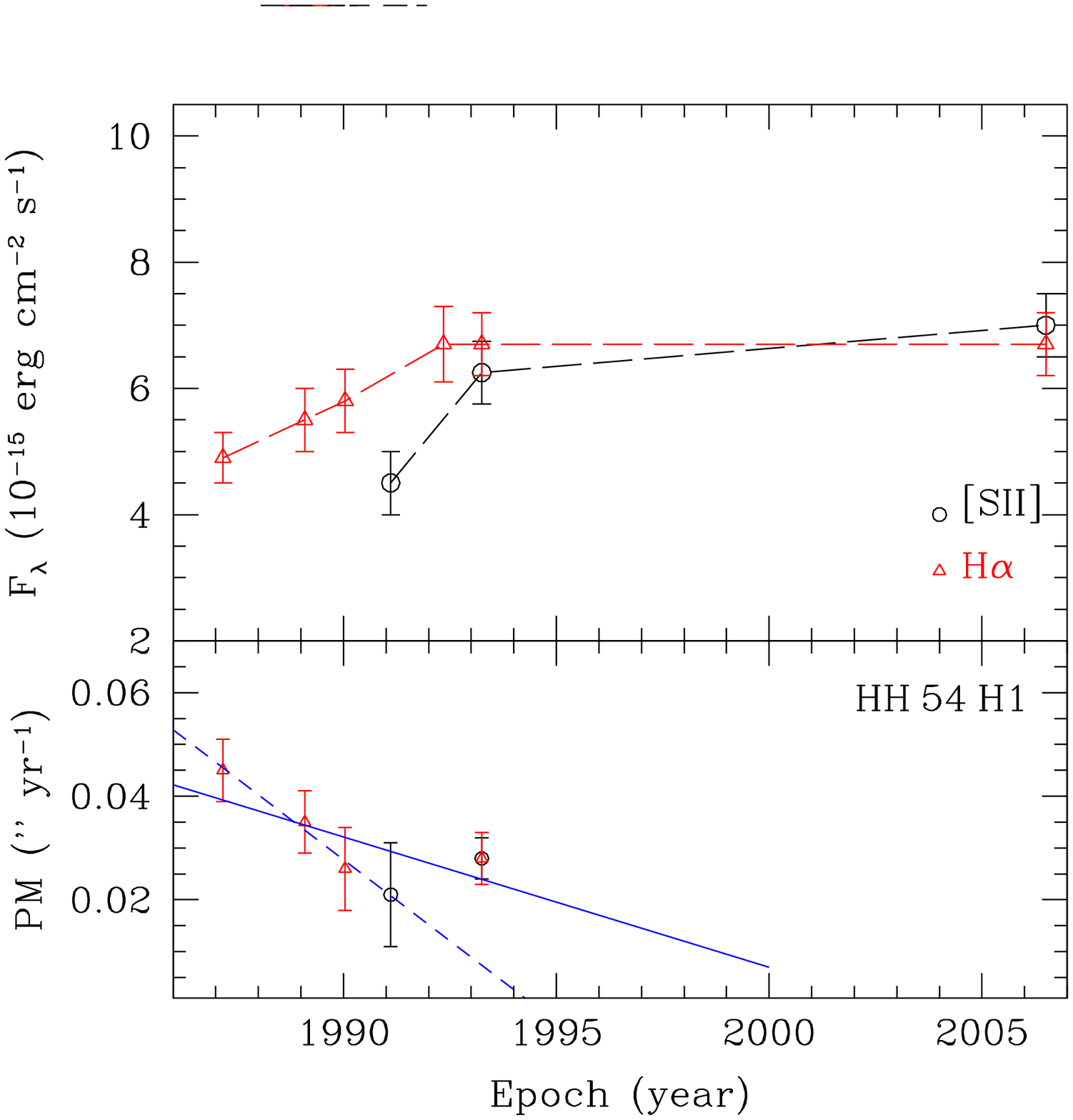}
   \includegraphics [width=9.2 cm] {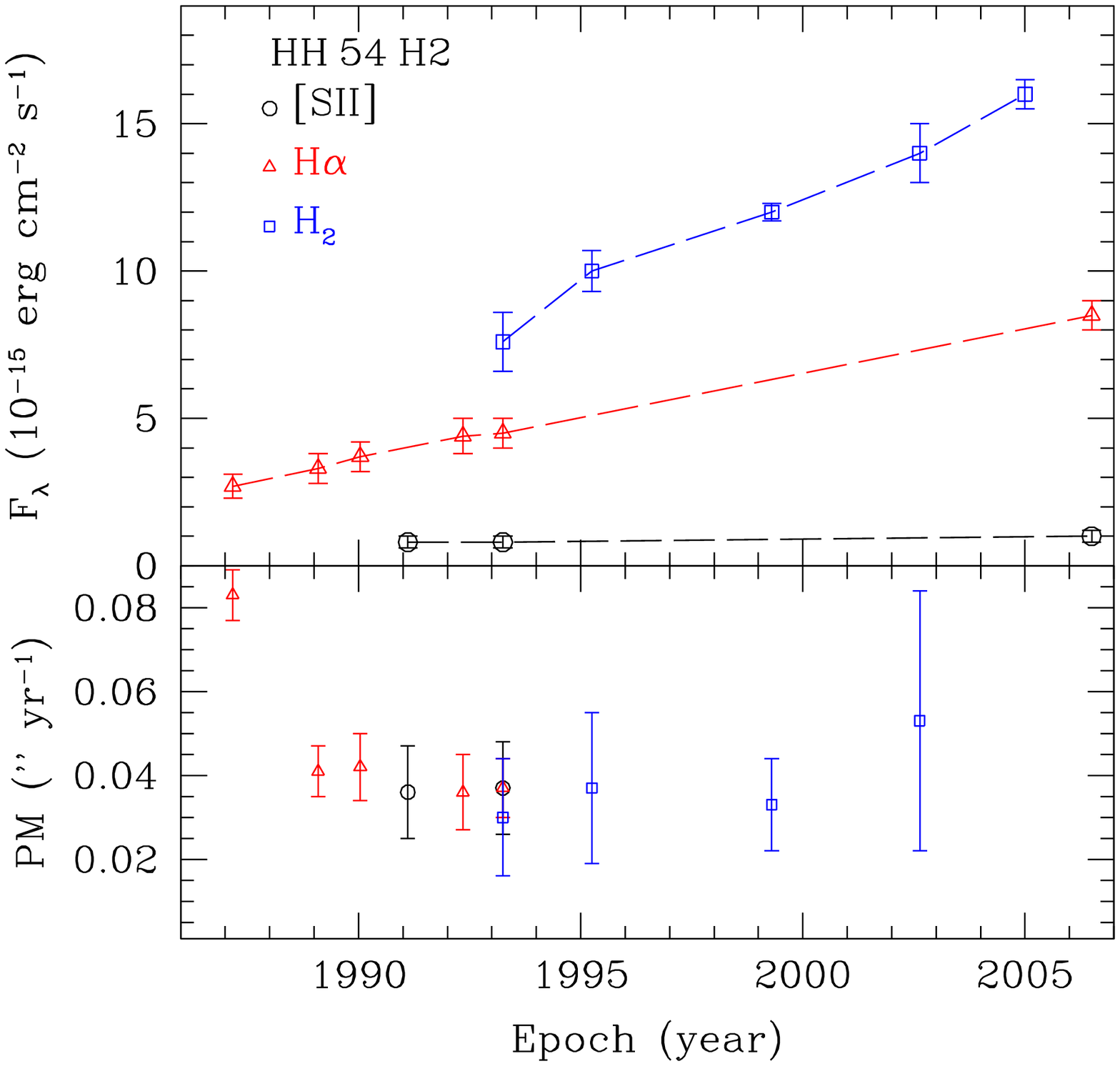}
   \caption{ Variability in HH\,54\,H1 {\bf upper Figure} and H2 {\bf bottom Figure}
   The {\bf bottom panels} show the P.M.s as a function of time in both [\ion{S}{ii}] (circles) and H$\alpha$ (triangles). The continuous and dashed lines in the top figure are the best fits
   of the data points with and without 1993 data points (see text), respectively.
   The slope of the fits gives the deceleration of the knot.
   In the {\bf top panel} of the figures the measured fluxes (uncorrected for the extinction) in the two filters are reported.
\label{HH54H:fig} }
\end{figure}

Group HH\,54\,H is a very active and fast variable region of the flow.

Between 1987 and 1993 H1 has decreased the speed in the optical bands from 40 to 20\,km\,s$^{-1}$
with a P.M. deceleration of 0.02$\pm$0.01$\arcsec$\,yr$^{-2}$ in H$\alpha$ (see Table~\ref{acc:tab}), and a value of 0.0025$\pm$0.0012$\arcsec$\,yr$^{-2}$,
or 0.0063$\pm$0.0005$\arcsec$\,yr$^{-2}$ without considering 1993 data, combining all the optical P.M.s
(see upper Fig.~\ref{HH54H:fig}, bottom panel).
During this period the H$\alpha$ and [\ion{S}{ii}] fluxes gradually increase from 4.9$\times$10$^{-15}$\,erg\,s$^{-1}$\,cm$^{-2}$
to 7.2$\times$10$^{-15}$\,erg\,s$^{-1}$\,cm$^{-2}$, and 1.8$\times$10$^{-15}$\,erg\,s$^{-1}$\,cm$^{-2}$
to 2.8$\times$10$^{-15}$\,erg\,s$^{-1}$\,cm$^{-2}$, respectively (top panel).

Knot H2 (see lower Fig.~\ref{HH54H:fig}, bottom panel) apparently drops in velocity between 1987 and 1989 H$\alpha$, changing the P.M. value from $\sim$0.08 to 0.04\,$\arcsec$\,yr$^{-1}$ (decelerating of 0.022$\pm$0.002$\arcsec$\,yr$^{-2}$ in H$\alpha$, see Table~\ref{acc:tab}). The knot has also a relevant flux variability at least in the H$\alpha$ and H$_2$ emissions,
that have a constant increase of the flux from 2.7$\times$10$^{-15}$\,erg\,s$^{-1}$\,cm$^{-2}$ (1987) to
8.5$\times$10$^{-15}$\,erg\,s$^{-1}$\,cm$^{-2}$ (2006) in H$\alpha$, and from 7.6$\times$10$^{-15}$\,erg\,s$^{-1}$\,cm$^{-2}$ (1993)
to 16$\times$10$^{-15}$\,erg\,s$^{-1}$\,cm$^{-2}$ (2005) in H$_2$. On the other hand the measured flux in the three [\ion{S}{ii}] images
is constant.

Finally H3 emission appears only in 2006 optical images, whereas in H$_2$ it appears as a faint featureless
emission and no significant motion is detectable. Again, the gap between the second to last and the last optical images
makes it hard to derive the initial position of the condensation. We have tried to correlate it with a feature close to H2 in 1993,
but the interpretation is not certain.

\end{appendix}

\begin{appendix}
\section{Radial velocity}
\label{appendixD:sec}

\subsection{HH\,52 flow }

Along the HH\,52 flow, knots have radial velocities from about -70 to -90\,km\,s$^{-1}$, with lower values detected at the end of the flow,
in HH\,54, where the velocity decreases moving from G to C, indicating that the flow has been decelerated.
The lowest value is detected in HH\,54\,C2 (-18$\pm$4\,km\,s$^{-1}$) at the very end of the flow. C3 is following C2 at a higher
velocity (-45$\pm$4\,km\,s$^{-1}$) (see Tab.~\ref{HH54vrad1:tab}). Along the jet, in knots G1-G3, we observe two distinct velocity components of -87 and -21\,km\,s$^{-1}$. In the EMMI spectral image the emissions are well separated, delineating two distinct lines of similar intensities for each detected species (see Tab.~\ref{HH54vrad1:tab}). At the base of the jet, where G0 and G are located, the fast component, extending few arcseconds from the HH\,52 streamer to G0, shows an abrupt deceleration (from -78 to -18\,km\,s$^{-1}$) and then ahead, along the flow, a second slower velocity component is detected. Also in consideration of the proper motion analysis, where a portion of the flow appears to be deflected in this region, the two velocities could be ascribed to the two flow components, rather than a bow shock structure. Moreover, due to the profile of the observed emission lines, a hypothetical bow shock would lie close the plane of the sky
(see e.\,g. Hartigan et al.~\cite{hart87}), but, as the inclination angle suggests (see also Sect.~\ref{inclination:sec}),
this is not the case.

Two velocities components, not completely separated, are also detected in HH\,52 knots located in front (D4) and behind (A3-A4)
the nucleus (A1, v$_{rad}$=-79$\pm$6\,km\,s$^{-1}$) (see Tab.~\ref{HH52vrad1:tab}).
Velocities in A3-A4 are slightly higher (-98$\pm$5\,km\,s$^{-1}$ and -35$\pm$6\,km\,s$^{-1}$,) with respect to D4
(-85$\pm$3\,km\,s$^{-1}$ and -30$\pm$3\,km\,s$^{-1}$). Line fluxes of the low velocity component are considerably fainter
than the other component. The velocity profiles of the H$\alpha$ line in both structures closely match the theoretical
profile of a bow shock with an inclination angle of $\sim$60$\degr$ with respect to the plane of the sky
(see Hartigan et al.~\cite{hart87}) and a shock velocity of 70-80\,km\,s$^{-1}$ (observed from the {\it full width zero intensity - FWZI}, measured on the spectrum where the flux reaches a 2$\sigma$ background noise level, see e.\,g. Davis et al.~\cite{davis01}).

\subsection{HH\,53 flow }

The highest $v_{rad}$ values are observed in the three knots of HH\,53 (see Tab.~\ref{HH53vrad1:tab}) outflow, A (-110$\pm$10\,km\,s$^{-1}$),
B (-113$\pm$9\,km\,s$^{-1}$), and C (-96$\pm$14\,km\,s$^{-1}$), derived averaging the radial velocities of the different species from
B\&C slit~1 and 2. In the EMMI spectrum (Tab.~\ref{HH53vrad1:tab}) it is not possible to spatially disentangle the emission coming from knots
C and C1. Here, however, we measure two different velocities of -104$\pm$7\,km\,s$^{-1}$ and -55$\pm$18\,km\,s$^{-1}$.
Considering that the two knots are from two separate flows, the higher velocity should be ascribed to knot C and the lower to C1.

\subsection{HH\,54 flow }

Along the portion of the HH\,54 streamer encompassed by our B\&C slits, we measure an increase in the velocity
moving towards the main body of HH\,54 (see also Tab.~\ref{HH54vrad1:tab}). The values rise from about -50\,km\,s$^{-1}$, roughly at the base
of the streamer (X1A-X3), up to -100\,km\,s$^{-1}$, close to the middle (X4A-X4B). Such a behaviour was also observed by
Graham \& Hartigan~(\cite{GH88}), but the value measured in the middle (considering also the errors) is lower than ours of
$\sim$10-20\,km\,s$^{-1}$. Such a difference in their values could be due to a not perfect positioning of the slit,
due to the sinuous geometry of the HH\,54 streamer.

\begin{table*}
\caption[]{ Radial velocities and line intensities for different
optical lines of individual knots in HH\,52 from EMMI and B\&C spectra. Radial
velocities are corrected for the cloud speed with respect to the
LSR (v$_{LSR}$=2\,km\,s$^{-1}$, Knee~\cite{knee}). When detected, two
velocity components are reported.
 \label{HH52vrad1:tab}}
\begin{center}
\begin{tabular}{ccccc}
\multicolumn{5}{c}{HH\,52}\\ [+5pt] \hline \hline
Knot  & Line & $\lambda$ & $v_{rad}$      & $F\pm\Delta~F$      \\
      & Id.     &   ($\AA$)    & (km\,s$^{-1}$) & ($10^{-15}$erg\,s$^{-1}$\,cm$^{-2}$)\\
\hline
\multicolumn{5}{c}{EMMI}\\ [+1pt]
\hline
A1    &  [\ion{O}{i}]  & 6300.3   & -78$\pm$4      & 8.1$\pm$0.1  \\
      &  [\ion{O}{i}]  & 6363.8   & -77$\pm$4      & 2.9$\pm$0.1  \\
      &  [\ion{N}{ii}] & 6548.1  & -78$\pm$4      & 2.2$\pm$0.1  \\
      &  H$\alpha$     & 6562.8   & -70$\pm$4      &16.7$\pm$0.2  \\
      &  [\ion{N}{ii}] & 6583.4  & -79$\pm$4      & 5.8$\pm$0.1  \\
      &  [\ion{S}{ii}] & 6716.4  & -87$\pm$4      &12.0$\pm$0.1  \\
      &  [\ion{S}{ii}] & 6730.8  & -86$\pm$4      &12.4$\pm$0.1  \\
A3-A4 &  [\ion{O}{i}]  & 6300.3   & -89$\pm$4 ; -43$\pm$4 & 1.5$\pm$0.1 ; 1.4$\pm$0.1 \\
      &  [\ion{O}{i}]  & 6363.8   &   -     & 1.0$\pm$0.1  \\
      &  [\ion{N}{ii}] & 6548.1  & -96$\pm$4 ; -27$\pm$4 & 1.5$\pm$0.1 ; 0.5$\pm$0.1 \\
      &  H$\alpha$     & 6562.8  & -101$\pm$4 ; -40$\pm$4 &   2.3$\pm$0.2 ; 1.9$\pm$0.2  \\
      &  [\ion{N}{ii}] & 6583.4  &      -      & 0.3$\pm$0.1  \\
      &  [\ion{S}{ii}] & 6716.4  & -100$\pm$4 ; -34$\pm$4 &   2.1$\pm$0.1  ; 1.3$\pm$0.1  \\
      &  [\ion{S}{ii}] & 6730.8  & -102$\pm$4 ; -31$\pm$4 &   1.8$\pm$0.2  ; 1.0$\pm$0.1  \\
D4    &  [\ion{O}{i}]  & 6300.3   & -84$\pm$4 ; -30$\pm$4     & 0.7$\pm$0.1 ; 0.3$\pm$0.1 \\
      &  [\ion{N}{ii}] & 6548.1   & -84$\pm$4 ; -33$\pm$4     & 0.8$\pm$0.1 ; 0.4$\pm$0.1 \\
      &  H$\alpha$     & 6562.8   & -73$\pm$4$^{a}$      &2.4$\pm$0.2  \\
      &  [\ion{N}{ii}] & 6583.4  &    -      & 0.5$\pm$0.2$^{b}$  \\
      &  [\ion{S}{ii}] & 6716.4  & -87$\pm$4 ; -27$\pm$4     & 1.4$\pm$0.1 ; 0.5$\pm$0.1 \\
      &  [\ion{S}{ii}] & 6730.8  & -87$\pm$4 ; -29$\pm$4    &1.1$\pm$0.1 ; 0.5$\pm$0.1 \\
\hline
\multicolumn{5}{c}{B\&C}\\ [+1pt]
\hline
A$^{c}$ &  [\ion{O}{i}]  & 6300.3 & -102$\pm$20     & 40.0$\pm$0.2  \\
      &  [\ion{O}{i}]  & 6363.8   & -103$\pm$20    & 14.3$\pm$0.2  \\
      &  [\ion{N}{ii}] & 6548.1   & -94$\pm$20      & 10.3$\pm$0.2  \\
      &  H$\alpha$     & 6562.8   & -101$\pm$20     & 95.6$\pm$0.3  \\
      &  [\ion{N}{ii}] & 6583.4   & -89$\pm$20      & 31.9$\pm$0.2  \\
      &  [\ion{S}{ii}] & 6716.4   & -81$\pm$20      & 62.8$\pm$0.3  \\
      &  [\ion{S}{ii}] & 6730.8   & -78$\pm$20      & 58.7$\pm$0.3  \\
B$^{d}$ &  [\ion{O}{i}]  & 6300.3 & -77$\pm$20      &  5.3$\pm$0.4  \\
      &  [\ion{O}{i}]  & 6363.8   & -83$\pm$20      &  2.2$\pm$0.3  \\
      &  [\ion{N}{ii}] & 6548.1   & -85$\pm$20     &  2.6$\pm$0.2  \\
      &  H$\alpha$     & 6562.8   & -85$\pm$20    &  17.2$\pm$0.2  \\
      &  [\ion{N}{ii}] & 6583.4   & -85$\pm$20      &  8.1$\pm$0.2  \\
      &  [\ion{S}{ii}] & 6716.4   & -63$\pm$20     &   6.5$\pm$0.2  \\
      &  [\ion{S}{ii}] & 6730.8   & -67$\pm$20     &  6.8$\pm$0.2  \\
D2-D3 &    [\ion{O}{i}]  & 6300.3 & -112$\pm$20     &  7.7$\pm$0.2  \\
      &  [\ion{O}{i}]  & 6363.8   & -132$\pm$20    &  3.5$\pm$0.2  \\
      &  [\ion{N}{ii}] & 6548.1   & -112$\pm$20    &  1.4$\pm$0.2  \\
      &  H$\alpha$     & 6562.8   & -94$\pm$20      & 25.3$\pm$0.3  \\
      &  [\ion{N}{ii}] & 6583.4   & -103$\pm$20     &  5.3$\pm$0.3 \\
      &  [\ion{S}{ii}] & 6716.4   & -103$\pm$20     & 13.1$\pm$0.3  \\
      &  [\ion{S}{ii}] & 6730.8   & -99$\pm$20      & 10.6$\pm$0.2 \\
\hline\hline
\end{tabular}
\end{center}
Notes:\\ $^{a}$ Not possible to deblend.\\
$^{b}$S/N between 2 and 3.\\
$^{c}$ Blend of knots A1,A3,A4,A6,A7.\\
$^{d}$ Blend of knots B5,B8,B9.\\
\end{table*}

\begin{table*}
\caption[]{ Radial velocities and line intensities for different
optical lines of individual knots in HH\,53 from EMMI and B\&C spectra. Radial
velocities are corrected for the cloud speed with respect to the
LSR (v$_{LSR}$=2\,km\,s$^{-1}$, Knee~\cite{knee}). When detected, two
velocity components are reported.
 \label{HH53vrad1:tab}}
\begin{center}
\begin{tabular}{ccccc}
\multicolumn{5}{c}{HH\,53}\\ [+5pt] \hline \hline
Knot  & Line & $\lambda$ & $v_{rad}$      & $F\pm\Delta~F$      \\
      & Id.     &   ($\AA$)    & (km\,s$^{-1}$) & ($10^{-15}$erg\,s$^{-1}$\,cm$^{-2}$)\\
\hline
\multicolumn{5}{c}{EMMI}\\ [+1pt]
\hline
C-C1  &  [\ion{O}{i}]  & 6300.3   & -101$\pm$4 ;  -32$\pm$4 &   1.9$\pm$0.1 ; 0.9$\pm$0.2\\
      &  [\ion{O}{i}]  & 6363.8   & --     & 1.0$\pm$0.4  \\
      &  H$\alpha$     & 6562.8   & -97$\pm$4 ; -47$\pm$4  & 9.8$\pm$0.1 ; 2.4$\pm$0.2 \\
      &  [\ion{N}{ii}] & 6583.4   & -100$\pm$4 ; -51$\pm$4  & 1.5$\pm$0.1 ; 0.4$\pm$0.2 \\
      &  [\ion{S}{ii}] & 6716.4   & -111$\pm$4 ; -72$\pm$4  & 2.2$\pm$0.1 ; 0.7$\pm$0.1 \\
      &  [\ion{S}{ii}] & 6730.8   & -112$\pm$4 ; -74$\pm$4  & 2.0$\pm$0.1 ; 0.7$\pm$0.1 \\
E1    &  H$\alpha$     & 6562.8  & -65$\pm$4 &   0.9$\pm$0.1   \\
      &  [\ion{S}{ii}] & 6716.4  & -74$\pm$4 &   0.7$\pm$0.1   \\
      &  [\ion{S}{ii}] & 6730.8  & -76$\pm$4 &   0.7$\pm$0.1   \\
\hline
\multicolumn{5}{c}{B\&C slit~1}\\ [+1pt]
\hline
A     &  [\ion{O}{i}]  & 6300.3   & -140$\pm$30     & 4.2$\pm$0.8  \\
      &  H$\alpha$     & 6562.8   & -114$\pm$30     & 25.0$\pm$0.6  \\
      &  [\ion{N}{ii}] & 6583.4   & -104$\pm$30     & 5.3$\pm$0.8  \\
      &  [\ion{S}{ii}] & 6716.4   & -117$\pm$30    & 6.6$\pm$0.4  \\
      &  [\ion{S}{ii}] & 6730.8   & -128$\pm$30    & 6.4$\pm$0.5  \\
B     &  [\ion{O}{i}]  & 6300.3   & -130$\pm$30    & 6.3$\pm$0.6  \\
      &  [\ion{N}{ii}] & 6548.1   & -141$\pm$30    & 2.5$\pm$0.6  \\
      &  H$\alpha$     & 6562.8   & -124$\pm$30    & 34.3$\pm$0.6  \\
      &  [\ion{N}{ii}] & 6583.4   & -103$\pm$30    & 3.6$\pm$0.5  \\
      &  [\ion{S}{ii}] & 6716.4   & -139$\pm$30    & 10.5$\pm$0.5  \\
      &  [\ion{S}{ii}] & 6730.8   & -127$\pm$30    & 9.0$\pm$0.6  \\
C     &  H$\alpha$     & 6562.8   & -102$\pm$30    & 8.8$\pm$0.6  \\
      &  [\ion{N}{ii}] & 6583.4   & -103$\pm$30    & 2.7$\pm$0.5  \\
      &  [\ion{S}{ii}] & 6716.4   & -101$\pm$30    & 2.1$\pm$0.6  \\
\hline
\multicolumn{5}{c}{B\&C slit~2}\\[+1pt]
\hline
A     &  [\ion{O}{i}]  & 6300.3   & -96$\pm$20      & 6.7$\pm$0.1  \\
      &  [\ion{O}{i}]  & 6363.8   & -112$\pm$20     & 2.2$\pm$0.1  \\
      &  [\ion{N}{ii}] & 6548.1   & -76$\pm$20      & 1.6$\pm$0.1 \\
      &  H$\alpha$     & 6562.8   & -106$\pm$20     & 26.3$\pm$0.1  \\
      &  [\ion{N}{ii}] & 6583.4   & -82$\pm$20      & 5.1$\pm$0.1  \\
      &  [\ion{S}{ii}] & 6716.4   & -115$\pm$20     & 7.2$\pm$0.1  \\
      &  [\ion{S}{ii}] & 6730.8   & -111$\pm$20     & 7.8$\pm$0.1 \\
B     &  [\ion{O}{i}]  & 6300.3   & -97$\pm$20     & 9.7$\pm$0.1 \\
      &  [\ion{O}{i}]  & 6363.8   & -96$\pm$20     & 2.9$\pm$0.1 \\
      &  [\ion{N}{ii}] & 6548.1   & -94$\pm$20     & 2.1$\pm$0.1 \\
      &  H$\alpha$     & 6562.8   & -98$\pm$20     & 33.3$\pm$0.1 \\
      &  [\ion{N}{ii}] & 6583.4   & -94$\pm$20     & 6.9$\pm$0.1 \\
      &  [\ion{S}{ii}] & 6716.4   & -106$\pm$20    & 11.5$\pm$0.1 \\
      &  [\ion{S}{ii}] & 6730.8   & -112$\pm$20    & 11.3$\pm$0.1 \\
C     &  [\ion{O}{i}]  & 6300.3   & -109$\pm$20     & 3.1$\pm$0.1 \\
      &  [\ion{O}{i}]  & 6363.8   & -125$\pm$20    & 1.1$\pm$0.1 \\
      &  [\ion{N}{ii}] & 6548.1   & -120$\pm$20     & 0.6$\pm$0.1 \\
      &  H$\alpha$     & 6562.8   & -122$\pm$20    & 14.2$\pm$0.1  \\
      &  [\ion{N}{ii}] & 6583.4   & -120$\pm$20    & 2.0$\pm$0.1 \\
      &  [\ion{S}{ii}] & 6716.4   & -116$\pm$20   & 4.5$\pm$0.1 \\
      &  [\ion{S}{ii}] & 6730.8   & -111$\pm$20   & 3.9$\pm$0.1 \\
F2    &  [\ion{O}{i}]  & 6300.3   & -87$\pm$20    & 1.5$\pm$0.1 \\
      &  [\ion{O}{i}]  & 6363.8   & --            & 0.8$\pm$0.4 \\
      &  [\ion{N}{ii}] & 6548.1   & --            & 1.0$\pm$0.2 \\
      &  H$\alpha$     & 6562.8   & -97$\pm$20     & 7.3$\pm$0.2  \\
      &  [\ion{N}{ii}] & 6583.4   & -85$\pm$20     & 1.7$\pm$0.2 \\
      &  [\ion{S}{ii}] & 6716.4   & -81$\pm$20     & 2.1$\pm$0.2 \\
      &  [\ion{S}{ii}] & 6730.8   & -95$\pm$20   & 1.7$\pm$0.2 \\
\hline\hline
\end{tabular}
\end{center}
Notes:\\ $^{a}$ Not possible to deblend.\\
$^{b}$S/N between 2 and 3.\\
\end{table*}

\begin{table*}
\caption[]{ Radial velocities and line intensities for different
optical lines of individual knots in HH\,54 from EMMI and B\&C spectra. Radial
velocities are corrected for the cloud speed with respect to the
LSR (v$_{LSR}$=2\,km\,s$^{-1}$, Knee~\cite{knee}). When detected, two
velocity components are reported.
 \label{HH54vrad1:tab}}
\begin{center}
\begin{tabular}{ccccc}
\multicolumn{5}{c}{HH\,54}\\ [+5pt] \hline \hline
Knot  & Line & $\lambda$ & $v_{rad}$      & $F\pm\Delta~F$      \\
      & Id.     &   ($\AA$)    & (km\,s$^{-1}$) & ($10^{-15}$erg\,s$^{-1}$\,cm$^{-2}$)\\
\hline
\multicolumn{5}{c}{EMMI}\\[+1pt]
\hline
C2    &  [\ion{O}{i}]  & 6300.3   & -40$\pm$4      & 6.4$\pm$0.3  \\
      &  [\ion{O}{i}]  & 6363.8   & -43$\pm$4      & 2.8$\pm$0.3  \\
      &  [\ion{N}{ii}] & 6548.1  & -47$\pm$4      & 2.2$\pm$0.2  \\
      &  H$\alpha$     & 6562.8   & -52$\pm$4      &13.9$\pm$0.3  \\
      &  [\ion{N}{ii}] & 6583.4  & -44$\pm$4      & 6.5$\pm$0.2  \\
      &  [\ion{S}{ii}] & 6716.4  & -46$\pm$4      &8.9$\pm$0.2  \\
      &  [\ion{S}{ii}] & 6730.8  & -46$\pm$4      &8.9$\pm$0.2  \\
C3    &  [\ion{O}{i}]  & 6300.3   & -12$\pm$4      & 1.5$\pm$0.1  \\
      &  [\ion{O}{i}]  & 6363.8   & -17$\pm$4      & 0.7$\pm$0.2  \\
      &  [\ion{N}{ii}] & 6548.1  & -16$\pm$4      & 0.4$\pm$0.1  \\
      &  H$\alpha$     & 6562.8  & -19$\pm$4      & 2.8$\pm$0.1  \\
      &  [\ion{N}{ii}] & 6583.4  & -20$\pm$4      & 1.6$\pm$0.1  \\
      &  [\ion{S}{ii}] & 6716.4  & -17$\pm$4      & 1.7$\pm$0.1  \\
      &  [\ion{S}{ii}] & 6730.8  & -22$\pm$4      & 1.8$\pm$0.1  \\
G-G0  &  [\ion{O}{i}]  & 6300.3   & -71$\pm$4 ; -24$\pm$4 & 0.4$\pm$0.1 ; 0.6$\pm$0.1 \\
      &  [\ion{N}{ii}] & 6548.1  & -- & 0.5$\pm$0.2 ; -- \\
      &  H$\alpha$     & 6562.8   & -80$\pm$4 ; -15$\pm$4 & 3.0$\pm$0.1 ; 0.5$\pm$0.1 \\
      &  [\ion{N}{ii}] & 6583.4  & -73$\pm$4 ; -15$\pm$4 &   0.8$\pm$0.1 ; 0.6$\pm$0.1  \\
      &  [\ion{S}{ii}] & 6716.4  & -78$\pm$4 ; -17$\pm$4 &   1.0$\pm$0.2 ; 0.6$\pm$0.2  \\
      &  [\ion{S}{ii}] & 6730.8  & -79$\pm$4 ; -21$\pm$4 &   1.0$\pm$0.2 ; 0.5$\pm$0.2  \\
G1-G3$^{a}$ &  [\ion{O}{i}]  & 6300.3   & -86$\pm$4 ; -18$\pm$4 & 1.9$\pm$0.1 ; 2.5$\pm$0.1 \\
      &  [\ion{O}{i}]  & 6363.8   &   --     & 0.6$\pm$0.2 ; 0.8$\pm$0.2 \\
      &  [\ion{N}{ii}] & 6548.1  & -- & 0.6$\pm$0.2 ; 1.0$\pm$0.3 \\
      &  H$\alpha$     & 6562.8  & -82$\pm$4 ; -24$\pm$4 &   5.6$\pm$0.3 ; 5.0$\pm$0.3  \\
      &  [\ion{N}{ii}] & 6583.4  & -86$\pm$4 ; -20$\pm$4 &   1.2$\pm$0.2 ; 2.1$\pm$0.2  \\
      &  [\ion{S}{ii}] & 6716.4  & -89$\pm$4 ; -20$\pm$4 &   2.6$\pm$0.2 ; 2.3$\pm$0.2  \\
      &  [\ion{S}{ii}] & 6730.8  & -91$\pm$4 ; -22$\pm$4 &   1.6$\pm$0.2 ; 2.0$\pm$0.1  \\
\hline
\multicolumn{5}{c}{B\&C slit~3}\\[+1pt]
\hline
X1D-X4C & [\ion{O}{i}] & 6300.3   & -34$\pm$30     & 6.5$\pm$1.3  \\
      &  H$\alpha$     & 6562.8   & -70$\pm$30     & 40.5$\pm$2.8  \\
      &  [\ion{N}{ii}] & 6583.4   & -58$\pm$30     & 5.4$\pm$2.0  \\
      &  [\ion{S}{ii}] & 6716.4   & -61$\pm$30     & 6.6$\pm$0.6  \\
      &  [\ion{S}{ii}] & 6730.8   & -67$\pm$30     & 6.3$\pm$2.5  \\
X1A   &  H$\alpha$     & 6562.8   & -70$\pm$20     & 4.5$\pm$0.1  \\
X3    &  [\ion{O}{i}]  & 6300.3   & -40$\pm$20     & 1.6$\pm$0.1  \\
      &  H$\alpha$     & 6562.8   & -42$\pm$20     & 9.9$\pm$0.1  \\
      &  [\ion{N}{ii}] & 6583.4   & -65$\pm$20     & 1.1$\pm$0.1  \\
      &  [\ion{S}{ii}] & 6716.4   & -53$\pm$20     & 2.0$\pm$0.2  \\
      &  [\ion{S}{ii}] & 6730.8   & -34$\pm$20     & 1.3$\pm$0.2  \\
X4A-X4B &  H$\alpha$   & 6562.8   & -115$\pm$20    & 11.8$\pm$0.2  \\
      &  [\ion{N}{ii}] & 6583.4   & -110$\pm$20    & 0.8$\pm$0.2  \\
      &  [\ion{S}{ii}] & 6716.4   & -92$\pm$20     & 1.8$\pm$0.2  \\
      &  [\ion{S}{ii}] & 6730.8   & -82$\pm$20     & 1.9$\pm$0.2  \\
\hline\hline
\end{tabular}
\end{center}
Notes:\\ $^{a}$ Blend of knots G1,G2 and G3.\\
\end{table*}

\begin{table*}
\caption[]{ [\ion{Fe}{ii}] radial velocities of the individual knots of HH\,54
obtained from ISAAC high resolution spectroscopy, corrected for the cloud speed with respect to
the LSR (v$_{LSR}$=2\,km\,s$^{-1}$, Knee~\cite{knee}). When detected, two velocities components are reported.
 \label{FeIIvrad:tab}}
\vspace {0.5 cm}
\begin{center}
\begin{tabular}{cc}
\multicolumn{2}{c}{HH\,54 - ISAAC slit~1}\\ \hline \hline
knot  & $v$$_{rad}$  (km\,s$^{-1}$)   \\
\hline
A1    & -52$\pm$3 ; -105$\pm$3 \\
A3    & -34$\pm$3  \\
B     & -14$\pm$3 ; -109$\pm$3 \\
B3    & -49$\pm$3 ; -105$\pm$3 \\
J-J1  &  -9$\pm$3 ; -91$\pm$3 \\
H2    & -36$\pm$3 ; -90$\pm$3 \\
H3    & -20$\pm$3 ; -45$\pm$3 \\
Z     & -100$\pm$3  \\
\hline
\multicolumn{2}{c}{HH\,54 - ISAAC slit~2}\\   \hline \hline
knot  & $v$$_{rad}$  (km\,s$^{-1}$)   \\
\hline
C1    &  -5$\pm$3 ; -90$\pm$3 \\
C3    &  -9$\pm$3 ; -85$\pm$3 \\
E     & -52$\pm$3 \\
J-J1  & -18$\pm$3 ; -62$\pm$3\\
H3    &  -40$\pm$3 \\
K     & -62$\pm$3 \\
K1    & -45$\pm$3 \\
\hline\hline
\end{tabular}
\end{center}
\end{table*}

\end{appendix}

\begin{appendix}
\section{Inclination and spatial velocity}
\label{appendixE:sec}

In HH\,52 bow shock we obtain an average inclination with
respect to the sky plane of 58$\degr\pm$3$\degr$, excluding D2-D3 measurement, that shows a value of
67$\degr\pm$4$\degr$. HH\,53\,E1 and F2, as already deduced from the P.M. analysis, are part of the same outflow with
inclinations of 57$\degr$$\pm$3$\degr$ and 63$\degr$$\pm$4$\degr$, respectively.
The association of groups HH\,54\,G and C to the HH\,52 outflow is also confirmed.
HH\,54\,G0 has an inclination of 55$\degr\pm$3$\degr$, at the end of the streamer.
Moreover, if we assume that along HH\,54\,G1-G3 the high and low radial velocities are associated with the high
and low tangential velocities observed along the jet, we obtain inclinations ranging between
50$\degr\pm$4$\degr$ and 56$\degr\pm$5$\degr$, respectively.
Values in HH\,54 C2 and C3 are 61$\degr\pm$3$\degr$ and 52$\degr\pm$8$\degr$, respectively.
Spatial velocity along the flow are around 100\,km\,s$^{-1}$ (see Tab.~\ref{incl:tab}) and decreases
in group G and C (20-50\,km\,s$^{-1}$).

HH\,53 A, B, and C have similar inclinations (83$\degr\pm$2$\degr$) and spatial velocities (around 110\,km\,s$^{-1}$).
HH\,53\,C1 is superimposed on the second flow and is moving accordingly with the HH\,52 streamer, with an inclination angle of
$\sim$49$\degr\pm$9$\degr$, considering the low velocity component measured in HH\,53\,C-C1 spectrum (see Tab.~\ref{HH53vrad1:tab}).

In the HH\,54 streamer we measure an average inclination angle of 67$\pm$3$\degr$, but the errors of the
single data points are too large to derive variations of inclination along the wiggling jet.

\begin{table*}
\caption[]{ Inclination and spatial velocities of individual knots of HH\,52, 53, and 54 inferred from the
kinematical analysis.
 \label{incl:tab}}
\vspace {0.5 cm}
\begin{center}
\begin{tabular}{ccc}\\ [+5pt] \hline \hline
knot ID & i ($\degr$) & $v_{tot}$  (km\,s$^{-1}$)   \\
\hline
HH\,52\,A1     & 58$\pm$4 & 93$\pm$6 \\
HH\,52\,A3-A4  & 57$\pm$2 & 83$\pm$6 \\
HH\,52\,A      & 62$\pm$4 & 106$\pm$14 \\
HH\,52\,B      & 63$\pm$3 & 88$\pm$22 \\
HH\,52\,D2-D3  & 67$\pm$4 & 112$\pm$20 \\
HH\,52\,D4     & 61$\pm$3 & 83$\pm$11 \\
HH\,53\,A      & 81$\pm$1 & 111$\pm$11 \\
HH\,53\,B      & 85$\pm$1 & 113$\pm$10 \\
HH\,53\,C      & 83$\pm$1 & 105$\pm$8 \\
HH\,53\,C1     & 49$\pm$9 &  76$\pm$21 \\
HH\,53\,E1     & 57$\pm$3 &  85$\pm$9 \\
HH\,53\,F2     & 63$\pm$3 & 100$\pm$15 \\
HH\,54\,C2     & 61$\pm$3 &  51$\pm$11 \\
HH\,54\,C3     & 52$\pm$8 &  24$\pm$9 \\
HH\,54\,G-G0   & 55$\pm$8 &  60-95$\pm$20 \\
HH\,54\,G1-G3  & 53$\pm$8 &  70$\pm$20 \\
HH\,54\,X1A    & 66$\pm$6 &   77$\pm$21 \\	
HH\,54\,X1D-X4C& 64$\pm$7 &   65$\pm$24 \\	
HH\,54\,X3     & 67$\pm$10 &   50$\pm$15 \\	
HH\,54\,X4A-X4B& 68$\pm$4 &  124$\pm$23 \\	
\hline\hline
\end{tabular}
\end{center}
\end{table*}

\end{appendix}

\end{document}